\documentclass[letterpaper, 12pt]{article}[2023/08/08]
\usepackage[english]{babel}
\usepackage{amsfonts,amsmath,amssymb,amsthm,latexsym,amscd,mathrsfs}
\usepackage{ifthen,cite}
\usepackage[bookmarksnumbered=true]{hyperref}
\usepackage{times,fancyhdr}

\usepackage{amsmath}
\numberwithin{equation}{section}

\usepackage{graphicx}

\hypersetup{pdfpagetransition={Split}}

\newcommand{\evenhead}{Author \ name}
\newcommand{\oddhead}{Article \ name}
\newcommand{\theArticleName}{Article \ name}

\newcommand{\FirstPageHeading}[1]{\thispagestyle{empty}%
\noindent\raisebox{0pt}[0pt][0pt]{\makebox[\textwidth]{\protect\footnotesize \sf }}\par}

\newcommand{\ArticleName}[1]{\renewcommand{\theArticleName}{#1}\vspace{-2mm}\par\noindent {\LARGE\bf  #1\par}}
\newcommand{\Author}[1]{\vspace{5mm}\par\noindent {\Large  #1\par} \par\vspace{2mm}\par}
\newcommand{\Address}[1]{\vspace{2mm}\par\noindent {\it #1} \par}
\newcommand{\Email}[1]{\ifthenelse{\equal{#1}{}}{}{\par\noindent {\rm E-mail: }{\it  #1} \par}}
\newcommand{\URLaddress}[1]{\ifthenelse{\equal{#1}{}}{}{\par\noindent {\rm URL: }{\tt  #1} \par}}
\newcommand{\EmailD}[1]{\ifthenelse{\equal{#1}{}}{}{\par\noindent {$\phantom{\dag}$~\rm E-mail: }{\it  #1} \par}}
\newcommand{\URLaddressD}[1]{\ifthenelse{\equal{#1}{}}{}{\par\noindent {$\phantom{\dag}$~\rm URL: }{\tt  #1} \par}}

\newcommand{\Abstract}[1]{\vspace{6mm}\par\noindent\hspace*{10mm}
\parbox{140mm}{\small {\bf Abstract.} #1}\par}
\newcommand{\Keywords}[1]{\vspace{3mm}\par\noindent\hspace*{10mm}
\parbox{140mm}{\small {\bf Key words:} \rm #1}\par}
\newcommand{\Classification}[1]{\vspace{3mm}\par\noindent\hspace*{10mm}
\parbox{140mm}{\small {\it 2020 Mathematics Subject Classification:} \rm #1}\vspace{3mm}\par}
\newcommand{\ShortArticleName}[1]{\renewcommand{\oddhead}{#1}}
\newcommand{\AuthorNameForHeading}[1]{\renewcommand{\evenhead}{#1}}

\long\def\@makecaption#1#2{
  \sbox\@tempboxa{\small \textbf{#1.}\ \ #2}%
  \ifdim \wd\@tempboxa >\hsize
    {\small \textbf{#1.}\ \ #2}\par \else
    \global \@minipagefalse
    \hb@xt@\hsize{\hfil\box\@tempboxa\hfil}%
  \fi \vskip\belowcaptionskip}

\def\numberwithin#1#2{\@ifundefined{c@#1}{\@nocounterr{#1}}{%
  \@ifundefined{c@#2}{\@nocnterr{#2}}{%
  \@addtoreset{#1}{#2}%
  \toks@\@xp\@xp\@xp{\csname the#1\endcsname}%
  \@xp\xdef\csname the#1\endcsname
    {\@xp\@nx\csname the#2\endcsname.\the\toks@}}}}
\def\E^#1{{\buildrel #1 \over\vee}}

{\theoremstyle{definition} 

\newtheorem*{criterion*}{Criterion}
\newtheorem*{theorem*}{Theorem}
\newtheorem*{proposition*}{Proposition}

}

\usepackage{tikz}

\begin{document}

\FirstPageHeading{V.I. Gerasimenko, I.V. Gapyak}

\ShortArticleName{Evolution equations of colliding particles}

\AuthorNameForHeading{V.I. Gerasimenko, I.V. Gapyak}

\ArticleName{\textcolor{blue!55!black}{Advances in theory of evolution equations \\ of many colliding particles}}

\Author{V.I. Gerasimenko$^\ast$\footnote{E-mail: \emph{gerasym@imath.kiev.ua}}
        and I.V. Gapyak$^\ast$$^\ast$\footnote{E-mail: \emph{gapjak@ukr.net}}}

\Address{$^\ast$\hspace*{2mm}Institute of Mathematics of NAS of Ukra\"{\i}ne,\\
    \hspace*{4mm}3, Tereshchenkivs'ka Str.,\\
    \hspace*{4mm}01601, Ky\"{\i}v-4, Ukra\"{\i}ne}

\Address{$^\ast$$^\ast$Taras Shevchenko National University of Ky\"{\i}v,\\
    \hspace*{4mm}Department of Mechanics and Mathematics,\\
    \hspace*{4mm}2, Academician Glushkov Av.,\\
    \hspace*{4mm}03187, Ky\"{\i}v, Ukra\"{\i}ne}

\bigskip

\Abstract{The review presents rigorous results of the theory of fundamental equations of evolution
of many-particle systems with collisions and also considers their connection with nonlinear
kinetic equations describing the collective behavior of particles in scaling approximations.

This work is dedicated to the 160th anniversary of the birth of Dmytro Oleksandrovych Grave,
the first academician of the Ukraine Academy of Sciences in mathematics and the founder of the
Institute of Mathematics in 1920.
}

\bigskip

\Keywords{hierarchy of evolution equations; kinetic equation; cumulant; semigroup of operators; colliding particles}
\vspace{2pc}
\Classification{35C10; 35Q20; 82C05; 82D05; 70F45}

\makeatletter
\renewcommand{\@evenhead}{
\hspace*{-3pt}\raisebox{-7pt}[\headheight][0pt]{\vbox{\hbox to \textwidth {\thepage \hfil \evenhead}\vskip4pt \hrule}}}
\renewcommand{\@oddhead}{
\hspace*{-3pt}\raisebox{-7pt}[\headheight][0pt]{\vbox{\hbox to \textwidth {\oddhead \hfil \thepage}\vskip4pt\hrule}}}
\renewcommand{\@evenfoot}{}
\renewcommand{\@oddfoot}{}
\makeatother

\newpage
\vphantom{math}

\protect\textcolor{blue!55!black}{\tableofcontents}
\vspace{0.7cm}

\newpage
\textcolor{blue!55!black}{\section{Preface}}

This review presents the modern theory of evolution equations for systems of many particles
with collisions. The traditional approach to describing the evolution of both finitely and
infinitely many classical particles is based on the description of the evolution of all
possible states by means of the reduced distribution functions governed by the BBGKY
hierarchy (Bogolyubov--Born--Green--Kirkwood--Yvon), which for finitely many particles
is an equivalent to the Liouville equation for the probability distribution function
\cite{CGP97},\cite{PG83},\cite{PG90}.

As is known, in the basis of the description of many-particle systems, there are notions
of the state and the observable. The functional of the mean value of observables defines
a duality between observables and states, and as a consequence, there exist two approaches
to the description of evolution. Thus, an equivalent approach to the description of evolution
is to describe the evolution by means of observables governed by the so-called dual BBGKY
hierarchy for reduced observables \cite{BGer},\cite{GG18}.

In certain situations \cite{CGP97}, the collective behavior of many-particle systems can be
adequately described by the kinetic equations. The conventional philosophy of the description
of kinetic evolution consists of the following: if the initial state is specified by a
one-particle reduced distribution function, then the evolution of the state can be effectively
described by means of a one-particle reduced distribution function governed by the nonlinear
kinetic equation in a suitable scaling limit. A well-known historical example of a kinetic
equation is the Boltzmann equation, which describes the process of particle collisions in
rarefied gases \cite{CC}.

Nowadays, a number of papers have appeared that discuss possible approaches to the rigorous
description of the evolution of many colliding particles
\cite{DS},\cite{GS-RT},\cite{GG21},\cite{PS16},\cite{PS17},\cite{Ch16},\cite{S}.
In particular, this is related to the problem of the rigorous derivation of the Boltzmann
kinetic equation from the underlying hierarchies of fundamental evolution equations. The
conventional method of deriving the Boltzmann equation consists of constructing the
Boltzmann--Grad scaling asymptotics of the BBGKY hierarchy solution represented by series
expansions within the framework of perturbation theory. The most advanced and rigorous
results to date have been obtained for systems of colliding particles, which is why the
motivation for writing this work is to discuss the theory of evolution equations for such
many-particle systems \cite{BGS-RS23},\cite{GG21},\cite{V02}.


In what follows, we mainly consider three challenges are left open until recently \cite{GG21}.

One of them is related to the construction of solutions to the Cauchy problem for hierarchies
of fundamental evolution equations for systems of many particles with collisions, using the
example of hard spheres with elastic collisions. It is established that the cluster expansions
of groups of operators for the Liouville equations for observables and a state of many hard
spheres underlie the classification of possible non-perturbative solution representations of
the Cauchy problem for the dual BBGKY hierarchy and the BBGKY hierarchy, respectively. As a
consequence, these solutions are represented in the form of series expansions whose generating
operators are the cumulants of the groups of operators for the Liouville equations. In a particular
case, the non-perturbative solutions of these hierarchies are represented in the form of the
perturbation (iteration) series as a result of applying analogs of the Duhamel equation to their
generating operators. The paper also formulated the Liouville hierarchy of evolution equations
for correlation functions of the state and established that the dynamics of correlations underlie
the description of the evolution of an infinitely many hard spheres that are governed by the BBGKY
hierarchy for the reduced distribution functions or the hierarchy of nonlinear evolution equations
for the reduced correlation functions \cite{GG22}.

Another challenge considered below is an approach to the description of the kinetic evolution
within the framework of the evolution of the observables of many colliding particles \cite{GG18}.
The problem of a rigorous description of the kinetic evolution of hard sphere observables is
considered by giving the example of the Boltzmann–Grad asymptotics of the non-perturbative
solution of the dual BBGKY hierarchy. One of the advances of  this approach is the opportunity
to construct kinetic equations, taking into account the correlations of particles in the initial
state and also the description of the process of propagation of initial correlations in scaling
approximations.

In addition, this paper discusses the approach to describing the evolution of a state by means
of the state of a typical particle of a system of many hard spheres, or, in other words, we
consider the origin of the description of the evolution of the state of hard spheres by the
Enskog-type kinetic equation \cite{GG12}. One of the applications of  the method  is related
to the challenge of the rigorous derivation of kinetic equations of the non-Markovian type
based on the dynamics of correlations, which allow us to describe the memory effects in
complex systems.

Thus, this review presents rigorous results in the theory of fundamental evolution equations for
many-particle systems with collisions, as well as nonlinear kinetic equations describing their
collective behavior in scaling approximations.


\bigskip
\subsection{A chronological overview of the theory of evolution equations\\ for many colliding particles}
The theory of kinetic equations begins with the work of L.~Boltzmann \cite{B72}, where an evolution
equation for collision dynamics was formulated based on phenomenological models of kinetic phenomena.
Later, to generalize the Boltzmann equation for the case of dense gases or fluids, D.~Enskog \cite{E22}
formulated a kinetic equation for a system of many hard spheres, now known as the Enskog equation.

The idea that equations formulated on the basis of phenomenological models of phenomena, such as
hydrodynamics equations or kinetic equations, should be derived from fundamental evolutionary equations
for systems of many particles, namely the Liouville equations, apparently goes back to the works of
D.~Hilbert \cite{H12} and H.~Poincar{\'e} \cite{Poi1905}. At the Second International Congress of
Mathematicians, held in Paris at the beginning of the 20th century, D. Hilbert formulated this idea
in his list of open questions as follows: "Boltzmann's work on the principles of mechanics suggests
the problem of developing mathematically the limiting processes that lead from the atomistic view to
the laws of motion of continua".

The approach to describe the evolution of the state of many-particle systems in a way equivalent to
the Liouville equation for the probability distribution function based on the hierarchy of evolution
equations for reduced distribution functions, known in our time as the BBGKY hierarchy, was most
consistently formulated in the work of M.~M.~Bogolyubov \cite{B1}, and independently by M.~Born and
H.~S.~Green \cite{BG46}, J.~G.~Kirkwood \cite{K47}, J.~Yvon \cite{Y35}.

In his famous monograph "Problems of the Dynamical Theory in Statistical Physics" \cite{B1}, which
was actually the manuscript of a 1945 report at the Institute of Mathematics in Kyiv, M.~M.~Bogolyubov
also formulated a consistent approach to the problem of deriving kinetic equations from the dynamics
of many particles. Using the methods of perturbation theory, an approach was developed to construct
a generalization of the Boltzmann equation, known as the Bogolyubov kinetic equation, as well as
justify the Vlasov and Landau kinetic equations for the first time. Thanks to this work, the
irreversibility mechanism of the evolution of systems of many particles, whose dynamics are described
as reversible in time by the fundamental equations of motion, became clear. A little later, in the
Proceedings of the Institute of Mathematics, M.~M.~Bogolyubov published a paper on the derivation of
the equations of hydrodynamics from the BBGKY hierarchy \cite{Bh}. These works became widely known as
a result of G.~E.~Uhlenbeck's lectures \cite{ChUh}. Bogolyubov's ideas became the cradle of modern
kinetic theory, as M.~H.~Ernst noted in his review \cite{E98}. The Bogolyubov method of deriving the
Boltzmann kinetic equation directly from the BBGKY hierarchy is presented in modern terminology in
the book \cite{CGP97}.

Since these lines are written in the work dedicated to the 160th anniversary of the birth of
Dmytro Oleksandrovych Grave, the first academician of the Ukraine Academy of Sciences in mathematics
and the founder of the Institute of Mathematics in 1920, it should be reminded that M.~M.~Bogolyubov
was one of the students of D.~O.~Grave, thanks to whom he became an outstanding scientist in the
field of mathematical physics. It is known that in 1922 at the age of thirteen, M.~Bogolyubov
became a participant in Grave's famous  mathematician seminar. In 1925, at the request of professor
D.~O.~Grave, the Small Presidium of Ukrgolovnauka made a decision: "In view of his phenomenal
abilities in mathematics, to consider M.~Bogolyubov as a post-graduate student of the research
department of mathematics in Kyiv from June, 1925", and already in 1928 he defended his doctoral
thesis. By the way, this historical precedent convincingly illustrates the significance of a
scientific school for the development of mathematics.

Rigorous methods for the description of the equilibrium state by the Gibbs distribution functions
\cite{WG}, i.e., by solutions of the steady BBGKY hierarchy, originate from the works of
M.~M.~Bogolyubov and D.~Ya.~Petrina within the framework of a canonical ensemble \cite{BPKh} and
D.~Ruelle within the framework of a grand canonical ensemble \cite{R69} and were investigated in
numerous works as a new direction of the progress of modern mathematical physics in the 70-80s.
In our time, mention above work of M.~M.~Bogolyubov, D.~Ya.~Petrina and B.~I.~Khatset was included
in the special issue of the Ukrainian Journal of Physics dedicated to the 90th Academy of Sciences
of Ukraine, which was was republished the most significant works of Ukrainian physicists over the
entire period of the Academy's existence, in other words, works that contributed to the golden fund
of world physical science (Golden Pages of Ukrainian Physics \cite{UJP}).

Note that the mathematical description of the Gibbs equilibrium states for infinitely many particles
forms the principal part of modern statistical mechanics. The main rigorous results about the
equilibrium Gibbs states were presented in the book \cite{GPM}.

The mathematical theory of the BBGKY hierarchy originates from the works of D.~Ya.~Petrina and
V.~I.~Gerasimenko \cite{GP85s},\cite{GP85_D},\cite{GP85},\cite{P79},\cite{PG83} in the early 80s.
The dual BBGKY hierarchy for reduced functions of observables was introduced by V.~I.~Gerasimenko
in the middle of the 1980s, and the theory of these evolution equations began to develop in the
last two decades (see \cite{GG21} and references therein).

Mathematical methods for deriving nonlinear kinetic equations from the BBGKY hierarchy began to
develop intensively in the early 1980s \cite{C72},\cite{L75}. One of the achievements of this
period was the formal derivation of the Boltzmann equation from the dynamics of an infinite
number of hard spheres in the Boltzmann--Grad limit.

In the approach to the problem of deriving kinetic equations from particle dynamics, which was
formulated by H.~Grad \cite{G49} and has now become generally accepted, the philosophy of the
description of kinetic evolution looks like this: if the initial state is specified by a
one-particle distribution function, then the evolution of the state of many particles can be
effectively described by means of a one-particle distribution function governed by a nonlinear
kinetic equation in a certain scaling approximation.

The Boltzmann–Grad asymptotics of a solution of the BBGKY hierarchy, represented as an iteration
series for infinitely many hard spheres, was first constructed by C.~Cercignani \cite{C72} and
O.~E.~Lanford \cite{L75} and rigorously justified in a series of papers
\cite{GP87},\cite{GP88},\cite{GP90},\cite{PG90} by D.~Ya.~Petrina and V.~I.~Gerasimenko
(some details are given in sections 2.2 and 3.1). Incidentally, it should be noted that the results
of the papers \cite{GP85_D},\cite{GP87} were discussed with Academician M.~M.~Bogolyubov at that
time and were submitted by him for publication.

Rigorous results of the theory of evolution equations for hard spheres and the derivation of the
Boltzmann equation from the BBGKY hierarchy in the Boltzmann--Grad limit were summarized in monographs
C.~Cercignani, V.~I.~Gerasimenko and D.~Ya.~Petrina \cite{CGP97}, C.~Cercignani, R.~Illner and
M.~Pulvirenti \cite{CIP}, H.~Spohn \cite{Sp91} at the end of the 90th.

The last two decades of progress in solving the problem of rigorous derivation of kinetic equations
from the collisional dynamics of particles are represented in numerous recent works \cite{BGS-RS23},\cite{D16},\cite{DS},\cite{G19},\cite{GS-RT},\cite{PSS},\cite{PS16},\cite{PS17},\cite{Ch16},\cite{S}.
The challenges of this area of contemporary mathematical physics are also discussed in
the latest review \cite{GG21}. With respect to the modern progress in the theory of evolution
equations of quantum many-particle systems, we refer to the overview \cite{G20}.

In these notes, recent advances in the theory of evolutionary equations for many colliding particles
will be considered; more precisely, we focus on the dynamics of many hard spheres with elastic collisions.

\subsection{Evolution equations of finitely many hard spheres}
The description of many-particle systems is based on the concepts of an observable and a state.
The mean value functional (expectation values) of observables defines the duality between
observables and a state, and as a result, there are two approaches to describing evolution.
The evolution of the system of finitely many colliding particles considered below is governed
by such fundamental evolution equations as the Liouville equation for observables or its dual
equation for a state.

Within the framework of a non-fixed, i.e., arbitrary but finite average number of identical
particles (non-equilibrium grand canonical ensemble), the observables and the state of a hard
sphere system are described by the sequences of functions
$A(t)=(A_0,A_{1}(t,x_1),\ldots,A_{n}(t,x_1,\ldots,x_n),\ldots)$ at instant $t\in\mathbb{R}$
and by the sequence $D(0)=(D_0,D_{1}^0(x_1),\ldots,D_{n}^0(x_1,\ldots,$ $x_n),\ldots)$ of the
probability distribution functions at the initial moment, respectively. These functions are
defined on the phase spaces of the corresponding number of particles, i.e.,
$x_i\equiv(q_i,p_i)\in\mathbb{R}^{3}\times\mathbb{R}^{3}$ is phase coordinates that characterize
a center of the $i$ hard sphere with a diameter of $\sigma>0$ in the space $\mathbb{R}^{3}$
and its momentum and are symmetrical with respect to arbitrary permutations of their arguments.
For configurations of a system of identical particles of a unit mass interacting as hard spheres
the following inequalities are satisfied: $|q_i-q_j|\geq\sigma,\,i\neq j\geq1$, i.e., the set
$\mathbb{W}_n\equiv\big\{(q_1,\ldots,q_n)\in\mathbb{R}^{3n}\big||q_i-q_j|<\sigma$ for at
least one pair $(i,j):\,i\neq j\in(1,\ldots,n)\big\}$, $n>1$, is the set of forbidden configurations.

A mean value functional of the observable of a hard sphere system is represented by the series
expansion \cite{PG90}
\begin{eqnarray}\label{averageD}
   &&\hskip-8mm \langle A\rangle(t)=(I,D(0))^{-1}(A(t),D(0)),
\end{eqnarray}
where the following abbreviated notation
\begin{eqnarray*}
    &&\hskip-10mm (A(t),D(0))\equiv\sum_{n=0}^\infty\,\frac{1}{n!}
         \int\limits_{(\mathbb{R}^{3}\times\mathbb{R}^{3})^n}
         A_{n}(t,x_1,\ldots,x_n)D_{n}^0(x_1,\ldots,x_n)dx_1\ldots dx_n
\end{eqnarray*}
was used and the coefficient $(I,D(0))$ is a normalizing factor (grand canonical partition function).

We remark that in the particular case of a system of $N < \infty$ hard spheres the observables
and a state are described by the one-component sequences: $A^{(N)}(t)=(0,\ldots,0,A_N(t),0, \ldots)$
and $D^{(N)}(0)=(0,\ldots,0,D^0_N,0, \ldots)$, respectively, and therefore, functional \eqref{averageD}
has the following representation
\begin{eqnarray*}
    &&\hskip-7mm \langle A^{(N)}\rangle(t)=(I,D^{(N)}(0))^{-1}(A^{(N)}(t),D^{(N)}(0))\equiv\\
    &&\hskip-5mm (I,D^{(N)}(0))^{-1}\frac{1}{N!}\int\limits_{(\mathbb{R}^{3}\times\mathbb{R}^{3})^N}
         A_{N}(t,x_1,\ldots,x_N)D_{N}^0(x_1,\ldots,x_N)dx_1\ldots dx_N,
\end{eqnarray*}
where $(I,D^{(N)}(0))=\frac{1}{N!}\int_{(\mathbb{R}^{3}\times\mathbb{R}^{3})^N}
D_{N}^0(x_1,\ldots,x_N)dx_1\ldots dx_N$ is the normalizing factor (canonical partition function),
and it is usually assumed that the normalization condition $(I,D^{(N)}(0))=1$ holds.

Let $\mathcal{C}_{\gamma}$ be the space of sequences $b=(b_0,b_1,\ldots,b_n,\ldots)$ of bounded
continuous functions $b_n=b_n(x_{1},\ldots,x_{n})$ that are symmetric with respect to permutations
of the arguments $x_1,\ldots,x_n$, equal to zero on the set of forbidden configurations $\mathbb{W}_n$
and equipped with the norm:
\begin{eqnarray*}
   &&\|b\|_{\mathcal{C}_{\gamma}}=\max_{n\geq 0}\,\frac{\gamma^{n}}{n!}\,\|b\|_{\mathcal{C}_{n}}=
            \max_{n\geq 0}\,\frac{\gamma^{n}}{n!}\,\sup_{x_{1},\ldots,x_{n}}|b_n(x_{1},\ldots,x_{n})|,
\end{eqnarray*}
where $0<\gamma<1$. We also introduce the space $L^{1}_{\alpha}=\oplus^{\infty}_{n=0}\alpha^n L^{1}_{n}$
of sequences $f=(f_0,f_1,\ldots,f_n,\ldots)$ of integrable functions $f_n=f_n(x_1,\ldots,x_n)$ that are
symmetric with respect to permutations of the arguments $x_1,\ldots,x_n$, equal to zero on the set
$\mathbb{W}_n$ and equipped with the norm:
\begin{eqnarray*}
   &&\|f\|_{L^{1}_{\alpha}}=\sum_{n=0}^\infty \alpha^n\int\limits_{(\mathbb{R}^{3}\times\mathbb{R}^{3})^n}
|f_n(x_1,\ldots,x_n)|dx_1\ldots dx_n,
\end{eqnarray*}
where $\alpha>1$ is a real number. If $A(t)\in\mathcal{C}_{\gamma}$ and $D(0)\in L^{1}_{\alpha}$ mean
value functional \eqref{averageD} exists and determines the duality between observables and states.

The evolution of the observables $A(t)=(A_0,A_{1}(t,x_1),\ldots,A_{n}(t,x_1,\ldots,$ $x_n),\ldots)$ is
described by the Cauchy problem for the sequence of the weak formulation of the Liouville equations
for hard spheres with elastic collisions \cite{CGP97}. On the space $\mathcal{C}_{\gamma}$ a
non-perturbative solution $A(t)=S(t)A(0)$ of the Liouville equation of many hard spheres is determined
by the following group of operators: \cite{PG90}
\begin{eqnarray} \label{Sspher}
  &&\hskip-12mm(S(t)b)_{n}(x_{1},\ldots,x_{n})=S_{n}(t,1,\ldots,n)b_{n}(x_{1},\ldots,x_{n})\doteq\\
  &&\begin{cases}
         b_{n}(X_{1}(t,x_{1},\ldots,x_{n}),\ldots,X_{n}(t,x_{1},\ldots,x_{n})),\\
         \hskip+49mm\mathrm{if}\,(x_{1},\ldots,x_{n})\in(\mathbb{R}^{3n}\times(\mathbb{R}^{3n}\setminus\mathbb{W}_{n})),\\
         0, \hskip+45mm \mathrm{if}\,(q_{1},\ldots,q_{n})\in\mathbb{W}_{n},
                    \end{cases}\nonumber
\end{eqnarray}
where for arbitrary $t\in\mathbb{R}$ the function $X_{i}(t)$ is a phase trajectory of $ith$ particle
constructed in book \cite{CGP97} and the set $\mathbb{M}_{n}^0$ of the zero Lebesgue measure, which
consists of the phase space points that are specified such initial data that during the evolution
generate multiple collisions, i.e., collisions of more than two particles, more than one two-particle
collision at the same instant, and an infinite number of collisions within a finite time interval.

On the space $\mathcal{C}_{\gamma}$ one-parameter mapping \eqref{Sspher} is an isometric $\ast$-weak
continuous group of operators, i.e., it is a $C_{0}^{\ast}$-group \cite{DL}. The infinitesimal generator
$\mathcal{L}=\oplus_{n=0}^\infty\mathcal{L}_{n}$ of the group of operators \eqref{Sspher} has the structure:
$\mathcal{L}_{n}=\sum_{j=1}^{n}\mathcal{L}(j)+\sum_{j_{1}<j_{2}=1}^{n}\mathcal{L}_{\mathrm{int}}(j_{1},j_{2}),$
where the operator $\mathcal{L}(j)$ defined on the set $C_{n,0}$ of continuously differentiable functions
with compact supports is the Liouville operator of free evolution of the $j$ hard sphere and for $t\geq0$ the
operators $\mathcal{L}(j)$ and $\mathcal{L}_{\mathrm{int}}(j_{1},j_{2})$ are defined by the formulas \cite{DE},\cite{PG90}:
\begin{eqnarray}\label{int}
   &&\hskip-12mm\mathcal{L}(j)\doteq\langle p_j,\frac{\partial}{\partial q_j}\rangle,\\
   &&\hskip-12mm\mathcal{L}_{\mathrm{int}}(j_1,j_{2})b_n\doteq
     \sigma^{2}\int_{\mathbb{S}_+^2}d\eta\langle\eta,(p_{j_1}-p_{j_2})\rangle
     \big(b_n(x_1,\ldots,q_{j_1},p_{j_1}^*,\ldots,q_{j_2},p_{j_2}^*,\ldots,x_n)-\nonumber\\
   &&b_n(x_1,\ldots,x_n)\big)\delta(q_{j_1}-q_{j_2}+\sigma\eta),\nonumber
\end{eqnarray}
respectively. In formulas (\ref{int}) the symbol $\langle \cdot,\cdot \rangle$ denotes
a scalar product, $\delta$ is the Dirac measure,
$\mathbb{S}_{+}^{2}\doteq\{\eta\in\mathbb{R}^{3}\big|\left|\eta\right|=1,\langle\eta,(p_{j_{1}}-p_{j_{2}})\rangle>0\}$
and the post-collision momenta: $p_{j_{1}}^*,p_{j_{2}}^*$ are defined by the equalities
\begin{eqnarray}\label{momenta}
     &&p_{j_{1}}^*\doteq p_{j_{1}}-\eta\left\langle\eta,\left(p_{j_{1}}-p_{j_{2}}\right)\right\rangle, \\
     &&p_{j_{2}}^*\doteq p_{j_{2}}+\eta\left\langle\eta,\left(p_{j_{1}}-p_{j_{2}}\right)\right\rangle \nonumber.
\end{eqnarray}
For $t<0$ operator (\ref{int}) is defined by the corresponding expression \cite{CGP97}.

It should be noted that the structure of generator (\ref{int}) is determined, on the one hand,
by the singular interaction potential of hard spheres and, on the other, by the fact that the
group of operators \eqref{Sspher} is defined for pairwise collisions outside the set
$\mathbb{M}_{n}^0$ of the zero Lebesgue measure defined above in \eqref{Sspher}.

As mentioned above, the evolution of observables for many hard spheres, i.e., the sequences
of functions $A_n(t)=S_n(t)A_n^0,\,n\geq1$, is governed by the Cauchy problem for the sequence
of the weak formulation of the Liouville equations with these generators \cite{GG18}:
\begin{eqnarray}\label{Le}
   &&\hskip-12mm \frac{\partial}{\partial t}A_n(t)=\big(\sum_{j=1}^{n}\mathcal{L}(j)+
        \sum_{j_{1}<j_{2}=1}^{n}\mathcal{L}_{\mathrm{int}}(j_{1},j_{2})\big)A_n(t),\\
        \nonumber\\
   &&\hskip-12mm  A_n(t)\big|_{t=0}=A_n^0,\quad n\geq1.\nonumber
\end{eqnarray}

For mean value functional \eqref{averageD} the following representation holds
\begin{eqnarray*}
   &&\hskip-12mm (A(t),D(0))=(A(0),D(t)),
\end{eqnarray*}
where the sequence $D(t)=(1,D_{1}(t,x_1),\ldots,D_{n}(t,x_1,\ldots,x_n),\ldots)$ of distribution functions
is defined as follows:
\begin{eqnarray*}\label{df}
   &&\hskip-12mm D(t)=S^{\ast}(t)D(0),
\end{eqnarray*}
and the mapping $S^{\ast}(t)$ is an adjoint operator to operator \eqref{Sspher} in the sense of mean
value functional \eqref{averageD}. We emphasize that this equality is a consequence of a fundamental
property of Hamiltonian systems, namely, the validity of the Liouville theorem for phase trajectories
\cite{CGP97}, i.e., isometry of the groups of operators  \eqref{Sspher}.

On the space $L^{1}_{\alpha}=\oplus^{\infty}_{n=0}\alpha^n L^{1}_{n}$ of sequences of integrable functions,
the group of operators  $S^\ast(t)=\oplus_{n=0}^\infty S^\ast_n(t)$ is an adjoint to the group of operators
\eqref{Sspher} in the sense of functional \eqref{averageD} and is defined as follows \cite{PG90}:
\begin{eqnarray}\label{S*}
      &&\hskip-12mm S^{\ast}(t)=S(-t).
\end{eqnarray}
On the space $L^{1}_{\alpha}$, one-parameter mapping (\ref{S*}) is an isometric strong
continuous group of operators, i.e., it is a $C_{0}$-group \cite{DL}. The infinitesimal generator
$\mathcal{L}^\ast=\oplus_{n=0}^\infty\mathcal{L}^\ast_{n}$ of this group of operators has the structure:
$\mathcal{L}_{n}^\ast\doteq\sum_{j=1}^{n}\mathcal{L}^\ast(j)+
\sum_{j_{1}<j_{2}=1}^{n}\mathcal{L}_{\mathrm{int}}^\ast(j_{1},j_{2}),$ where the operator $\mathcal{L}^\ast(j)$
defined on the set $L^{1}_{n,0}$ of continuously differentiable integrable functions
with compact supports is the free motion Liouville operator of the $j$ hard sphere, and for $t>0$ the operators
$\mathcal{L}^{\ast}(j)$ and $\mathcal{L}_{\mathrm{int}}^\ast(j_{1},j_{2})$ are defined by the formulas:
\begin{eqnarray}\label{L}
     &&\hskip-12mm \mathcal{L}^{\ast}(j) f_{n}\doteq-\langle p_j,\frac{\partial}{\partial q_j}\rangle f_{n},\\
     &&\hskip-12mm \mathcal{L}_{\mathrm{int}}^\ast(j_{1},j_{2})f_{n}
        \doteq\sigma^2\int_{\mathbb{S}_{+}^2}d\eta\langle\eta,(p_{j_{1}}-p_{j_{2}})\rangle
        \big(f_n(x_1,\ldots,p_{j_{1}}^*,q_{j_{1}},\ldots,\nonumber\\
     && p_{j_{2}}^*,q_{j_{2}},\ldots,x_n)\delta(q_{j_{1}}-q_{j_{2}}+\sigma\eta)-
        f_n(x_1,\ldots,x_n)\delta(q_{j_{1}}-q_{j_{2}}-\sigma\eta)\big),\nonumber
\end{eqnarray}
respectively, the pre-collision momenta $p^*_{j_{1}},p^*_{j_{2}}$ are defined by relations \eqref{momenta}
and notations accepted in formula (\ref{int}) are used. For $t<0$ these operators on the set $L^{1}_{n,0}$
are defined by the corresponding expressions \cite{GPM}.

We note that the evolution of the state, i.e., the sequence of probability distribution functions
$D_n(t)=S^{\ast}_n(t)D_n^0,\,n\geq 1$, describes by the Cauchy problem for the sequence of the weak
formulation of the Liouville equations for many hard spheres with these generators \cite{PG90}:
\begin{eqnarray}\label{dLe}
   &&\hskip-12mm \frac{\partial}{\partial t}D_n(t)=\big(\sum_{j=1}^{n}\mathcal{L}^\ast(j)+
         \sum_{j_{1}<j_{2}=1}^{n}\mathcal{L}_{\mathrm{int}}^\ast(j_{1},j_{2})\big)D_n(t),\\
            \nonumber\\
   &&\hskip-12mm D_n(t)\big|_{t=0}=D_n^0,\quad n\geq 1.\nonumber
\end{eqnarray}

Thus, the evolution of finitely many colliding particles is governed by the fundamental evolution
equations, such as the Liouville equation for observables \eqref{Le} or its dual equation for a
state \eqref{dLe}.

To formulate another representation of the mean value functional \eqref{averageD} in terms of sequences
of so-called reduced observables and reduced distribution functions, on sequences of bounded continuous
functions we introduce an analog of the creation operator
\begin{eqnarray}\label{a+}
   &&(\mathfrak{a}^{+}b)_{s}(x_{1},\ldots,x_{s})\doteq\sum_{j=1}^s\,b_{s-1}((x_{1},\ldots,x_{s})\setminus (x_j)),
\end{eqnarray}
and on sequences of integrable functions, we introduce an adjoint operator to operator \eqref{a+} in
the sense of mean value functional \eqref{averageD} which is an analogue of the annihilation operator
\begin{eqnarray}\label{a}
    &&(\mathfrak{a}f)_{n}(x_1,\ldots,x_n)=
      \int\limits_{\mathbb{R}^3\times\mathbb{R}^3}f_{n+1}(x_1,\ldots,x_n,x_{n+1})dx_{n+1}.
\end{eqnarray}
Then as a consequence of the validity of equalities:
$$(b,f)=(e^{\mathfrak{a^{+}}}e^{-\mathfrak{a^{+}}}b,f)=(e^{-\mathfrak{a^{+}}}b,e^{\mathfrak{a}}f),$$
for mean value functional \eqref{averageD} the following representation holds:
\begin{eqnarray}\label{B(t)}
   &&\hskip-12mm \langle A\rangle(t)=(I,D(0))^{-1}(A(t),D(0))=(B(t),F(0)),
\end{eqnarray}
where a sequence of the reduced observables is defined by the formula
\begin{eqnarray}\label{ro}
   &&\hskip-12mm B(t)=e^{-\mathfrak{a^{+}}}S(t)A(0),
\end{eqnarray}
and a sequence of so-called reduced distribution functions is defined as follows
(known as the non-equilibrium grand canonical ensemble \cite{PG83})
\begin{eqnarray*}\label{rdf}
   &&\hskip-12mm F(0)=(I,D(0))^{-1}e^{\mathfrak{a}}D(0),
\end{eqnarray*}
respectively.

Thus, according to the definition of the operator $e^{-\mathfrak{a^{+}}}$, the sequence of reduced
observables \eqref{ro} in component-wise form is represented by the expansions:
\begin{eqnarray*}\label{moo}
  &&\hskip-12mm B_s(t,x_1,\ldots,x_s)=\\
  &&\hskip-5mm \sum_{n=0}^s\,\frac{(-1)^n}{n!}\sum_{j_1\neq\ldots\neq j_{n}=1}^s
        (S(t)A(0))_{s-n}\big((x_1,\ldots,x_s)\setminus (x_{j_1},\ldots,x_{j_{n}})),\quad s\geq1,\nonumber
\end{eqnarray*}


The mean value functional \eqref{B(t)} also has the following representation:
\begin{eqnarray}\label{F(t)}
   &&\hskip-12mm (B(t),F(0))=(B(0),F(t)).
\end{eqnarray}
The sequence $F(t)=(1,F_{1}(t,x_1),\ldots,F_{n}(t,x_1,\ldots,x_n),\ldots)$ of reduced distribution functions
is defined as follows (known as the non-equilibrium grand canonical ensemble \cite{PG83})
\begin{eqnarray}\label{rdf}
   &&\hskip-12mm F(t)=(I,D(0))^{-1}e^{\mathfrak{a}}S^{\ast}(t)D(0),
\end{eqnarray}
where the mapping $S^{\ast}(t)$ is an adjoint operator \eqref{S*} to operator (\ref{Sspher}).
According to the definition of the operator $e^{\mathfrak{a}}$, the sequence of reduced
distribution functions \eqref{rdf} in component-wise form is represented by the series:
\begin{eqnarray*}
  &&\hskip-12mm F_s(t,x_1,\ldots,x_s)=\\
  &&\hskip-5mm(I,D(0))^{-1}\sum_{n=0}^\infty \frac{1}{n!}\int\limits_{(\mathbb{R}^{3}\times\mathbb{R}^{3})^n}
        (S^\ast(t)D(0))_{s+n}(x_1,\ldots,x_{s+n})dx_{s+1}\ldots dx_{s+n}, \quad s\geq1,\nonumber
\end{eqnarray*}
where the coefficient $(I,D(0))$ is the normalizing factor as above.

We emphasize that a widely used approach to the description of the evolution of many hard spheres
is based on the evolution of a state determined by the BBGKY hierarchy for reduced distribution
functions \cite{CGP97}. An equivalent approach to describing evolution is based on reduced
observables \eqref{ro} governed by the dual hierarchy of evolution equations \cite{GG21}.


\textcolor{blue!55!black}{\section{Hierarchies of evolution equations for colliding particles}}

As is well known, hierarchies of evolution equations for sequences of reduced functions
of observables and, accordingly, of a state for a finitely many hard spheres are equivalent
to the Liouville equations. Their advantages consist in the possibility of rigorously describing
the evolution of infinitely many hard spheres whose collective behavior exhibits thermodynamic
(statistical) features, namely, the existence of an equilibrium state in such a system as well
as the kinetic or hydrodynamic behavior in corresponding scaling approximations \cite{CGP97},\cite{DSS},\cite{GPM}.

An alternative approach to the description of the evolution of the state of a hard-sphere
system is based on functions determined by the cluster expansions of the probability
distribution functions. The cumulants of probability distribution functions are interpreted
as correlation functions and are governed by the Liouville hierarchy. The following outlines
the approach to the description of the evolution of a state by means of both reduced
distribution functions and reduced correlation functions, which is based on the dynamics
of correlations \cite{GG22}. It should be emphasized that on a microscopic scale, the
macroscopic characteristics of fluctuations of observables are directly determined by
the reduced correlation functions.

\bigskip
\subsection{Hierarchy of evolution equations for reduced observables}
The motivation for describing the evolution of many-particle systems in terms of reduced
observables is related to possible equivalent representations of the mean value functional
(mathematical expectation) of observables, namely as \eqref{B(t)} compared to the traditionally
used form \eqref{F(t)}.

The evolution of sequence \eqref{ro} of reduced observables of many hard spheres is determined by
the Cauchy problem of the following abstract hierarchy of evolution equations \cite{BGer},\cite{GG18}:
\begin{eqnarray}\label{dh}
   &&\hskip-12mm \frac{d}{dt}B(t)=\mathcal{L}B(t)+\big[\mathcal{L},\mathfrak{a}^+\big]B(t),\\
   \nonumber\\ \label{dhi}
   &&\hskip-12mm  B(t)\big|_{t=0}=B(0),
\end{eqnarray}
where the operator $\mathcal{L}$ is generator \eqref{int} of the group of operators \eqref{Sspher} for
hard spheres, the symbol $\big[ \cdot,\cdot \big]$ denotes the commutator of operators, which in equation
\eqref{dh} has the following component-wise form:
\begin{eqnarray*}
   &&\hskip-12mm(\big[\mathcal{L},\mathfrak{a}^+\big]b)_{s}(x_1,\ldots,x_s)=
      \sum_{j_1\neq j_{2}=1}^s\mathcal{L}_{\mathrm{int}}(j_1,j_{2})b_{s-1}(t,(x_1,\ldots,x_s)\setminus x_{j_1}),\quad s\geq1.
\end{eqnarray*}
In a component-wise form the hierarchy of evolution equations \eqref{dh} for hard-sphere fluids, in fact,
is a sequence of recurrence evolution equations (in literature it is known as the dual BBGKY hierarchy
\cite{GG21}). We adduce the simplest examples of recurrent evolution equations \eqref{dh}:
\begin{eqnarray*}
   &&\hskip-7mm\frac{\partial}{\partial t}B_{1}(t,x_1)=\mathcal{L}(1)B_{1}(t,x_1),\\
   &&\hskip-7mm\frac{\partial}{\partial t}B_{2}(t,x_1,x_2)=
      \big(\sum_{i=1}^{2}\mathcal{L}(j)+\mathcal{L}_{\mathrm{int}}(1,2)\big)B_{2}(t,x_1,x_2)+\\
  && \hskip+8mm\mathcal{L}_{\mathrm{int}}(1,2)\big(B_{1}(t,x_1)+B_{1}(t,x_2)\big),
\end{eqnarray*}
where the generators of these equations are defined by formula \eqref{int}.


The non-perturbative solution of the Cauchy problem of the dual BBGKY hierarchy
\eqref{dh},\eqref{dhi} for hard spheres is a sequence of reduced observables represented
by the following expansions \cite{GR02},\cite{GR03}:
\begin{eqnarray}\label{sed}
   &&\hskip-12mm B_{s}(t,x_1,\ldots,x_s)=(e^{\mathfrak{a}^{+}}\mathfrak{A}(t)B(0))_s(x_1,\ldots,x_s)=\\
   &&\hskip-5mm \sum_{n=0}^s\,\frac{1}{n!}\sum_{j_1\neq\ldots\neq j_{n}=1}^s
      \mathfrak{A}_{1+n}(t,\{(1,\ldots,s)\setminus(j_1,\ldots,j_{n})\},j_1,\ldots,j_{n}\big)\,B_{s-n}^{0}(x_1,\ldots,x_{j_1-1},\nonumber\\
   &&\hskip-5mm x_{j_1+1},\ldots,x_{j_n-1},x_{j_n+1},\ldots,x_s),\quad s\geq1,\nonumber
\end{eqnarray}
where the mappings $\mathfrak{A}_{1+n}(t),\,n\geq0,$ are the generating operators which are represented
as cumulant expansions with respect of groups of operators \eqref{Sspher}.
The simplest examples of reduced observables \eqref{sed} are given by the following expansions:
\begin{eqnarray*}
   &&\hskip-5mm B_{1}(t,x_1)=\mathfrak{A}_{1}(t,1)B_{1}^{0}(x_1),\\
   &&\hskip-5mm B_{2}(t,x_1,x_2)=\mathfrak{A}_{1}(t,\{1,2\})B_{2}^{0}(x_1,x_2)+
      \mathfrak{A}_{2}(t,1,2)(B_{1}^{0}(x_1)+B_{1}^{0}(x_2)).
\end{eqnarray*}

To determine the generating operators of expansions of reduced observables \eqref{sed},
we will introduce the notion of dual cluster expansions of groups of operators \eqref{Sspher}
in terms of operators interpreted as their cumulants. For this end on sequences of one-parametric
mappings $\mathfrak{u}(t)=(0,\mathfrak{u}_1(t),\ldots,$ $\mathfrak{u}_n(t),\ldots)$ we define the
following $\star$-product \cite{R69}
\begin{eqnarray}\label{Product}
    &&\hskip-12mm (\mathfrak{u}(t)\star\widetilde{\mathfrak{u}}(t))_{s}(1,\ldots,s)=
        \sum\limits_{Y\subset (1,\ldots,s)}\,\mathfrak{u}_{|Y|}(t,Y)
        \,\widetilde{\mathfrak{u}}_{s-|Y|}(t,(1,\ldots,s)\setminus Y),
\end{eqnarray}
where $\sum_{Y\subset (1,\ldots,s)}$ is the sum over all subsets $Y$ of the set $(1,\ldots,s)$.

Using the definition of the $\star$-product \eqref{Product}, the dual cluster expansions of groups of
operators \eqref{Sspher} are represented by the mapping ${\mathbb E}\mathrm{xp}_{\star}$ in the form
\begin{eqnarray*}\label{DtoGcircledStar}
    &&\hskip-7mm S(t)={\mathbb E}\mathrm{xp}_{\star}\,\mathfrak{A}(t)=\mathbb{I}+
      \sum\limits_{n=1}^{\infty}\frac{1}{n!}\mathfrak{A}(t)^{\star n},
\end{eqnarray*}
where $S(t)=(0,S_1(t,1),\ldots,S_n(t,1,\ldots,n),\ldots)$ and $\mathbb{I}=(1,0,\ldots,0,\ldots)$.
In component-wise form the dual cluster expansions are represented by the following
recursive relations:
\begin{eqnarray}\label{cexd}
   &&\hskip-12mm S_{s}(t,(1,\ldots,s)\setminus(j_1,\ldots,j_{n}),j_1,\ldots,j_{n})=\\
   &&\sum\limits_{\mathrm{P}:\,(\{(1,\ldots,s)\setminus(j_1,\ldots,j_{n})\},\,j_1,\ldots,j_{n})=
       \bigcup_i X_i}\,\prod\limits_{X_i\subset\mathrm{P}}\mathfrak{A}_{|X_i|}(t,X_i),\quad n\geq 0,\nonumber
\end{eqnarray}
where the set consisting of one element of indices $(1,\ldots,s)\setminus(j_1,\ldots,j_{n})$ we denoted by
the symbol $\{(1,\ldots,s)\setminus(j_1,\ldots,j_{n})\}$ and the symbol ${\sum}_\mathrm{P}$ means the sum
over all possible partitions $\mathrm{P}$ of the set $(\{(1,\ldots,s)\setminus(j_1,\ldots,j_{n})\},$ $j_1,\ldots,j_{n})$
into $|\mathrm{P}|$ nonempty mutually disjoint subsets $X_i\subset(1,\ldots,s)$.

The solution of recursive relations \eqref{cexd} are represented by the inverse mapping
${\mathbb L}\mathrm{n}_{\ast}$ in the form of the cumulant expansion
\begin{eqnarray*}
    &&\hskip-7mm \mathfrak{A}(t)={\mathbb L}\mathrm{n}_{\star}(\mathbb{I}+S(t))=
         \sum\limits_{n=1}^{\infty}\frac{(-1)^{n-1}}{n}S(t)^{\star n}.
\end{eqnarray*}
Then the $(1+n)th$-order dual cumulant of groups of operators \eqref{Sspher} is defined by
the following expansion:
\begin{eqnarray}\label{dcumulant}
    &&\hskip-12mm \mathfrak{A}_{1+n}(t,\{(1,\ldots,s)\setminus(j_1,\ldots,j_{n})\},j_1,\ldots,j_{n})\doteq\\
    &&\sum\limits_{\mathrm{P}:\,(\{(1,\ldots,s)\setminus(j_1,\ldots,j_{n})\},j_1,\ldots,j_{n})={\bigcup}_i X_i}
       (-1)^{\mathrm{|P|}-1}({\mathrm{|P|}-1})!\prod_{X_i\subset\mathrm{P}}S_{|\theta(X_i)|}(t,\theta(X_i)),\nonumber
\end{eqnarray}
where the above notation is used and the declusterization mapping $\theta$ is defined by
the formula: $\theta(\{(1,\ldots,$ $s)\setminus(j_1,\ldots,j_{n})\})=(1,\ldots,s)\setminus(j_1,\ldots,j_{n})$.
The dual cumulants \eqref{dcumulant} of the first two orders have the form:
\begin{eqnarray*}
    &&\hskip-12mm \mathfrak{A}_{1}(t,\{1,\ldots,s\})=S_{s}(t,1,\ldots,s),\\
    &&\hskip-12mm \mathfrak{A}_{1+1}(t,\{(1,\ldots,s)\setminus(j)\},j)=
       S_{s}(t,1,\ldots,s)-S_{s-1}(t,(1,\ldots,s)\setminus(j))S_{1}(t,j).
\end{eqnarray*}

If $b_{s}\in\mathcal{C}_{s}$, then for $(1+n)th$-order cumulant \eqref{dcumulant} of groups
of operators \eqref{Sspher} the estimate is valid
\begin{eqnarray}\label{estd}
   &&\hskip-12mm \big\|\mathfrak{A}_{1+n}(t)b_{s}\big\|_{\mathcal{C}_{s}}
      \leq \sum_{\mathrm{P}:\,(\{(1,\ldots,s)\setminus(j_1,\ldots,j_{n})\},j_1,\ldots,j_{n})={\bigcup}_i X_i}
      (|\mathrm{P}|-1)!\big\|b_{s}\big\|_{\mathcal{C}_{s}}\leq \\
   &&\leq \sum\limits_{k=1}^{n+1}\mathrm{s}(n+1,k)(k-1)!\big\|b_{s}\big\|_{\mathcal{C}_{s}}
      \leq n!e^{n+2}\big\|b_{s}\big\|_{\mathcal{C}_{s}},\nonumber
\end{eqnarray}
where $\mathrm{s}(n+1,k)$ are the Stirling numbers of the second kind. Then according to this estimate
\eqref{estd} for the generating operators of expansions \eqref{sed} provided that $\gamma<e^{-1}$
the inequality valid
\begin{eqnarray}\label{esd}
    &&\hskip-8mm \big\|B(t)\big\|_{C_{\gamma}}\leq e^2(1-\gamma e)^{-1}\big\|B(0)\big\|_{C_{\gamma}}.
\end{eqnarray}

In fact, the following criterion holds.

\begin{criterion*}
\emph{A solution of the Cauchy problem of the dual BBGKY hierarchy \eqref{dh},\eqref{dhi} is represented by
expansions \eqref{sed} if and only if the generating operators of expansions \eqref{sed} are solutions
of cluster expansions \eqref{cexd} of the groups of operators \eqref{Sspher} of the Liouville equations
for hard spheres.}
\end{criterion*}

The necessity condition means that cluster expansions \eqref{cexd} take place for groups of operators
\eqref{Sspher}. These recurrence relations are derived from definition \eqref{moo} of reduced observables,
provided that they are represented as expansions \eqref{sed} for the solution of the Cauchy problem
of the dual BBGKY hierarchy \eqref{dh},\eqref{dhi}.

The sufficient condition means that the infinitesimal generator of one-parameter mapping \eqref{sed}
coincides with the generator of the sequence of recurrence evolution equations \eqref{dh}.
Indeed, in the space $C_{\gamma}$ the following existence theorem is true \cite{GG18}.
\begin{theorem*}
\emph{A non-perturbative solution of the Cauchy problem \eqref{dh},\eqref{dhi} is represented by
expansions \eqref{sed} in which the generating operators are cumulants of the corresponding
order \eqref{dcumulant} of groups of operators \eqref{Sspher}:}
\begin{eqnarray}\label{sedc}
   &&\hskip-12mm B_{s}(t,x_1,\ldots,x_s)= \sum_{n=0}^s\,\frac{1}{n!}\sum_{j_1\neq\ldots\neq j_{n}=1}^s\,
       \sum_{\substack{\mathrm{P}:\,(\{(1,\ldots,s)\setminus(j_1,\ldots,j_{n})\},\\j_1,\ldots,j_{n})={\bigcup}_i X_i}}
       (-1)^{\mathrm{|P|}-1}({\mathrm{|P|}-1})!\times\\
   &&\hskip-5mm \prod_{X_i\subset\mathrm{P}}
       S_{|\theta(X_i)|}(t,\theta(X_i))B_{s-n}^{0}(x_1,\ldots,x_{j_1-1},x_{j_1+1},\ldots,x_{j_n-1},x_{j_n+1},\ldots,x_s),
       \quad s\geq1.\nonumber
\end{eqnarray}
\emph{Under the condition $\gamma<e^{-1}$ for initial data $B(0)\in C_{\gamma}^0$ of finite sequences
of infinitely differentiable functions with compact supports sequence \eqref{sedc} is a unique
global classical solution, and for arbitrary initial data $B(0)\in C_{\gamma}$ is a unique global
generalized solution.}
\end{theorem*}

We note that the one component sequences $B^{(1)}(0)=(0,b_{1}(x_1),0,\ldots)$ of reduced observables
correspond to the additive-type observable, and the sequences $B^{(k)}(0)=(0,\ldots,b_{k}(x_1,\ldots,x_k),0,\ldots)$
of reduced observables correspond to the $k$-ary-type observables \cite{BGer}.

If initial data \eqref{dhi} is specified by the additive-type reduced observable, then the structure
of solution expansion \eqref{sedc} is simplified and attains the form
\begin{eqnarray}\label{af}
     &&\hskip-8mm B_{s}^{(1)}(t,x_1,\ldots,x_s)=\mathfrak{A}_{s}(t,1,\ldots,s)\sum_{j=1}^s b_{1}(x_j), \quad s\geq 1,
\end{eqnarray}
where the generating operator $\mathfrak{A}_{s}(t)$ is the $sth$-order cumulant \eqref{dcumulant} of
the groups of operators \eqref{Sspher}.

An example of the additive-type observables is a number of particles, i.e., the sequence
$N^{(1)}(0)=(0,1,0,\ldots)$, then
\begin{eqnarray*}
    &&\hskip-8mm N^{(1)}_{s}(t)=\mathfrak{A}_{s}(t,1,\ldots,s)s=\sum\limits_{\mathrm{P}:\,(1,\ldots,s)={\bigcup}_i X_i}
        (-1)^{\mathrm{|P|}-1}({\mathrm{|P|}-1})!\sum_{j=1}^s 1=\\
    &&\sum\limits_{k=1}^s(-1)^{k-1}\mathrm{s}(s,k)(k-1)! s=s\delta_{s,1}=N^{(1)}_{s}(0),
\end{eqnarray*}
where $\mathrm{s}(s,k)$ are the Stirling numbers of the second kind and $\delta_{s,1}$ is a Kronecker symbol.
Consequently, the observable of a number of hard spheres is an integral of motion and, in particular, the
average number of particles is preserving in time.


In the case of initial $k$-ary-type, $k\geq2$, reduced observables solution expansion \eqref{sedc}
takes the form
\begin{eqnarray}\label{af-k}
     &&\hskip-12mm B_{s}^{(k)}(t)=0, \quad 1\leq s<k,\\ \nonumber\\
     &&\hskip-12mm B_{s}^{(k)}(t,x_1,\ldots,x_s)=\frac{1}{(s-k)!}\sum_{j_1\neq\ldots\neq j_{s-k}=1}^s
      \mathfrak{A}_{1+s-k}\big(t,\{(1,\ldots,s)\setminus (j_1,\ldots,j_{s-k})\},\nonumber\\
     &&j_1,\ldots,j_{s-k}\big)\,
      b_{k}(x_1,\ldots,x_{j_1-1},x_{j_1+1},\ldots,x_{j_s-k-1},x_{j_s-k+1},\ldots,x_s),\quad
    s\geq k,\nonumber
\end{eqnarray}
where the generating operator $\mathfrak{A}_{1+s-k}(t)$ is the $(1+s-k)th$-order cumulant \eqref{dcumulant}
of the groups of operators \eqref{Sspher}.

We emphasize that cluster expansions \eqref{cexd} of the groups of operators \eqref{Sspher} underlie
of the classification of possible solution representations of the Cauchy problem \eqref{dh},\eqref{dhi}
of the dual BBGKY hierarchy. Indeed, using cluster expansions \eqref{cexd} of the groups of operators
\eqref{Sspher}, other solution representations can be constructed.


For example, let us express the cumulants $\mathfrak{A}_{1+n}(t),\,n\geq2,$ of groups of
operators \eqref{Sspher} with respect to the $1st$-order and $2nd$-order cumulants. The
equalities are true:
\begin{eqnarray*}
   &&\hskip-12mm \mathfrak{A}_{1+n}(t,\{(1,\ldots,s)\setminus(j_1,\ldots,j_{n})\},j_1,\ldots,j_{n})=
\end{eqnarray*}
\begin{eqnarray*}
   &&\sum_{\substack{Y\subset(j_1,\ldots,j_{n}),\,Y\neq \emptyset}}
       \mathfrak{A}_{2}(t,\{(1,\ldots,s)\setminus(j_1,\ldots,j_{n})\},\{Y\})\times\\
   &&\sum_{\mathrm{P}:\,(j_1,\ldots,j_{n})\setminus Y ={\bigcup\limits}_i X_i}
       (-1)^{|\mathrm{P}|}\,|\mathrm{P}|!\,\prod_{i=1}^{|\mathrm{P}|}\mathfrak{A}_{1}(t,\{X_{i}\}), \quad n\geq2,
\end{eqnarray*}
where ${\sum\limits}_{\substack{Y\subset(j_1,\ldots,j_{n}),\\Y\neq\emptyset}}$ is a sum over all
nonempty subsets $Y\subset (j_1,\ldots,j_{n})$. Then, taking into account the identity
\begin{eqnarray}\label{id}
  &&\hskip-12mm \sum_{\mathrm{P}:\,(j_1,\ldots,j_{n})\setminus Y ={\bigcup\limits}_i X_i}
     (-1)^{|\mathrm{P}|}\,|\mathrm{P}|!\,\prod_{i=1}^{|\mathrm{P}|}
     \mathfrak{A}_{1}(t,\{X_{i}\})B^0_{s-n}((x_{1},\ldots,x_{s})\setminus(x_{j_1},\ldots,x_{j_{n}}))=\nonumber\\
   &&\sum_{\mathrm{P}:\,(j_1,\ldots,j_{n})\setminus Y ={\bigcup\limits}_i X_i}
     (-1)^{|\mathrm{P}|}\,|\mathrm{P}|!\,B^0_{s-n}((x_{1},\ldots,x_{s})\setminus(x_{j_1},\ldots,x_{j_{n}})),\nonumber
\end{eqnarray}
and the equalities
\begin{eqnarray}\label{aq}
  &&\hskip-12mm {\sum_{\mathrm{P}:\,(j_1,\ldots,j_{n})\setminus Y =
      {\bigcup\limits}_i X_i}}(-1)^{|\mathrm{P}|}\,|\mathrm{P}|!=(-1)^{|(j_1,\ldots,j_{n})\setminus Y|},
      \quad Y\subset\,(j_1,\ldots,j_{n}),
\end{eqnarray}
for solution expansions \eqref{sed} of the dual BBGKY hierarchy we derive the following representation:
\begin{eqnarray*}
    &&\hskip-12mm B _{s}(t,x_1,\ldots,x_s)=\mathfrak{A}_{1}(t,\{1,\ldots,s\})B_{s}^0(x_1,\ldots,x_s)+\\
    &&\hskip+5mm \sum_{n=1}^s\,\frac{1}{n!}\,\sum_{j_1\neq\ldots\neq j_{n}=1}^s\,\,
       \sum\limits_{\substack{Y\subset (1,\ldots,s)\setminus(j_1,\ldots,j_{n}),\,Y\neq \emptyset}}\,
       (-1)^{|(j_1,\ldots,j_{n})\setminus Y|}\,\times\\
    &&\hskip+5mm \mathfrak{A}_{2}(t,\{j_1,\ldots,j_{n}\},\{Y\})\,
       B^0_{s-n}((x_{1},\ldots,x_{s})\setminus(x_{j_1},\ldots,x_{j_{n}})),\\
    && \hskip-12mm  s\geq 1,
\end{eqnarray*}
where notations accepted above are used.


Taking into account that initial reduced observables depend only from the certain phase space arguments,
we deduce the reduced representation of expansions \eqref{sedc}:
\begin{eqnarray}\label{rsedc}
   &&\hskip-12mm B(t)=\sum\limits_{n=0}^{\infty}\frac{1}{n!}\,\sum\limits_{k=0}^{n}\,(-1)^{n-k}\,
      \frac{n!}{k!(n-k)!}\,(\mathfrak{a}^{+})^{n-k}S(t)(\mathfrak{a}^{+})^{k}B(0)=\\
   &&S(t)B(0)+\sum\limits_{n=1}^{\infty}\frac{1}{n!}
      \big[\ldots\big[S(t),\underbrace{\mathfrak{a}^{+}\big],\ldots,\mathfrak{a}^{+}}_{\hbox{n-times}}\big]B(0)=\nonumber\\
   && e^{-\mathfrak{a}^{+}}S(t)e^{\mathfrak{a}^{+}}B(0).\nonumber
\end{eqnarray}
Therefore, in component-wise form the generating operators of these expansions represented as expansions
\eqref{sed} are the following reduced cumulants of groups of operators \eqref{Sspher}:
\begin{eqnarray}\label{rcc}
  &&\hskip-12mm U_{1+n}(t,\{1,\dots,s-n\},s-n+1,\dots,s)=\sum^n_{k=0}(-1)^{k}\frac{n!}{k!(n-k)!}S_{s-k}(t,1,\dots,s-k).
\end{eqnarray}

Indeed, solutions of the recursive relations \eqref{cexd} with respect to first-order cumulants
can be represented as expansions in terms of cumulants acting on variables on which the initial
reduced observables depend, and in terms of cumulants not acting on these variables
\begin{eqnarray*}
   &&\hskip-12mm \mathfrak{A}_{1+n}(t,\{(1,\ldots,s)\setminus(j_1,\ldots,j_{n})\},j_1,\ldots,j_{n})=\\
   &&\sum_{\substack{Y\subset (j_1,\ldots,j_{n})}}
       \mathfrak{A}_{1}(t,\{(1,\ldots,s)\setminus((j_1,\ldots,j_{n})\cup Y)\})\times \nonumber\\
   &&\sum_{\mathrm{P}:\,(j_1,\ldots,j_{n})\setminus Y ={\bigcup\limits}_i X_i}
       (-1)^{|\mathrm{P}|}\,|\mathrm{P}|!\,\prod_{i=1}^{|\mathrm{P}|}\mathfrak{A}_{1}(t,\{X_{i}\}),
\end{eqnarray*}
where ${\sum\limits}_{\substack{Y\subset(j_1,\ldots,j_{n})}}$ is the sum over all possible subsets
$Y\subset(j_1,\ldots,j_{n})$. Then taking into account the identity \eqref{aq} and the equalities \eqref{id}
we derive expansions \eqref{rsedc} over reduced cumulants \eqref{rcc}.

We note that traditionally the solution of the BBGKY hierarchy for states of many hard spheres
is represented by perturbation series \cite{CGP97},\cite{GS-RT},\cite{PG90}. The expansions
\eqref{rsedc} can also be represented as expansions (iterations) of perturbation theory \cite{BGer}:
\begin{eqnarray*}
   &&\hskip-12mm B(t)=\sum\limits_{n=0}^{\infty}\,\int\limits_{0}^{t} dt_{1}\ldots
       \int\limits_{0}^{t_{n-1}}dt_{n}S(t-t_{1})\big[\mathcal{L},\mathfrak{a}^+\big]
        S(t_1-t_2)\ldots S(t_{n-1}-t_n)\big[\mathcal{L},\mathfrak{a}^+\big]S(t_{n})B(0).\nonumber
\end{eqnarray*}
Indeed, as a result of applying of analogs of the Duhamel equation to generating operators
\eqref{dcumulant} of expansions \eqref{sed} we derive in component-wise form, for examples,
\begin{eqnarray*}
    &&\hskip-12mm U_{1}(t,\{1,\ldots,s\})=S_s(t,1,\ldots,s),\\
    &&\hskip-12mm U_{2}(t,\{(1,\ldots,s)\setminus(j_1)\},j_1)=\nonumber\\
   &&\int\limits_{0}^{t}dt_{1}S_s(t-t_{1},1,\ldots,s)
       \sum_{j_{2}=1,\,j_2\neq j_{1}}^s\mathcal{L}_{\mathrm{int}}(j_1,j_{2})
       S_{s-1}(t_1,(1,\ldots,s)\setminus j_1).
\end{eqnarray*}

Recall that the mean value functional \eqref{B(t)} exists if $B(0)\in C_{\gamma}$ and $F(0)\in L^{1}_{\alpha}$.
In the case of the observable of a number of hard spheres $N^{(1)}(t)=(0,1,0,\ldots)$, this means that
\begin{eqnarray}\label{N}
    &&\big|(N^{(1)}(t),F(0))\big|=\big|\int\limits_{\mathbb{R}^{3}\times\mathbb{R}^{3}}\,F_{1}^0(x_{1})dx_{1}\big|\leq
       \big\|F_{1}^0\big\|_{L^{1}_{1}}<\infty.
\end{eqnarray}
Consequently, the states of a finite number of hard spheres are described by sequences of functions
from the space $L^{1}_{\alpha}$. To describe an infinite number of hard spheres, it is necessary to
consider reduced distribution functions from appropriate function spaces, for example, from the space
of sequences of bounded functions with respect to the configuration variables \cite{GP85},\cite{L75},\cite{PG90}.

\bigskip
\subsection{The BBGKY hierarchy for reduced distribution functions}
As mentioned already, the evolution of systems of many particles is traditionally described
as the evolution of the state of a system based on the representation of the mean value functional
for observables \eqref{F(t)}. In this case, the sequence of reduced distribution functions is
determined by the hierarchy of evolution equations, known as the BBGKY hierarchy, whose generator
is the operator adjoint to the generator of the hierarchy of evolution equations for reduced
observables \eqref{dh} in the sense of mean value functional \eqref{B(t)}.


The evolution of sequence \eqref{rdf} of reduced distribution functions from the space
$L^{1}_{\alpha}$ is governed by the Cauchy problem of the BBGKY hierarchy for many hard
spheres \cite{B1},\cite{CGP97},\cite{PG90}:
\begin{eqnarray}\label{h}
   &&\hskip-12mm \frac{d}{dt}F(t)=\mathcal{L}^\ast F(t)+\big[\mathfrak{a},\mathcal{L}^\ast\big]F(t),\\
   \nonumber\\ \label{hi}
   &&\hskip-12mm  F(t)\big|_{t=0}=F(0),
\end{eqnarray}
where the symbol $\big[ \cdot,\cdot \big]$ denotes the commutator of operator \eqref{a} and of the
Liouville operator (\ref{L}), which is the generator of the isometric group of operators \eqref{S*}.
Thus, in evolution equation \eqref{h}, the second term has the following component-wise form:
\begin{eqnarray*}
   &&\hskip-12mm (\big[\mathfrak{a},\mathcal{L}^*\big]f)_{s}(x_1,\ldots,x_s)=
       \sum_{i=1}^s\,\int\limits_{\mathbb{R}^{3}\times\mathbb{R}^{3}} dx_{s+1}\,
       \mathcal{L}^*_{\mathrm{int}}(i,s+1)f_{s+1}(t,x_1,\ldots,x_{s+1}),\quad s\geq1.
\end{eqnarray*}
For $t>0$ in a one-dimensional space, i.e., for gas of hard rods, this term of a generator
has the form \cite{PG83}:
\begin{eqnarray}\label{aLint1}
    &&\hskip-12mm\sum_{i=1}^{s}\,\int\limits_{\mathbb{R}\times\mathbb{R}}dx_{s+1}
       \mathcal{L}^*_{\mathrm{int}}(i,s+1)f_{s+1}(t)=\\
    &&\hskip-5mm\sum\limits_{i=1}^s\int\limits_{0}^{\infty}d P P
            \big(f_{s+1}(t,x_1,\ldots,q_i,p_i-P,\ldots,x_s,q_i-\sigma,p_i)-\nonumber\\
    &&\hskip-5mm f_{s+1}(t,x_1,\ldots,x_s,q_i-\sigma,p_i+P)+ \nonumber \\
    &&\hskip-5mm f_{s+1}(t,x_1,\ldots,q_i,p_i+P,\ldots,x_s,q_i+\sigma,p_i)- f_{s+1}(t,x_1,\ldots,x_s,q_i+\sigma,p_i-P)\big),\nonumber
\end{eqnarray}
and for $t<0$ this collision integral has the corresponding form \cite{PG90}.

It should be noted that for the system of a fixed, finite number of hard spheres, the BBGKY hierarchy
is an equation system for a finite sequence of reduced distribution functions. Such an equation system
is equivalent to the Liouville equation for the distribution function, which describes all possible states
of finitely many hard spheres. For a system of an infinite number of hard spheres, the BBGKY hierarchy is
an infinite chain of evolution equations, which can be derived as the thermodynamic limit of the BBGKY
hierarchy of a fixed finite number of hard spheres \cite{CGP97}. We note that since a sequence of functions
can be determined based on a generating functional, the corresponding hierarchy of evolution equations can
also be formulated as the evolution equation for such a generating functional \cite{GF12}.


A non-perturbative solution of the Cauchy problem of the BBGKY hierarchy \eqref{h},\eqref{hi}
is a sequence of reduced distribution functions represented by the following expansions \cite{GG21},\cite{GerRS}:
\begin{eqnarray}\label{se}
    &&\hskip-12mm F_s(t,x_1,\ldots,x_s)=(e^{\mathfrak{a}}\mathfrak{A}^\ast(t)F(0))_s(x_1,\ldots,x_s)= \\
    &&\hskip-7mm \sum_{n=0}^\infty\frac{1}{n!}\int\limits_{(\mathbb{R}^{3}\times\mathbb{R}^{3})^{n}}
     \mathfrak{A}_{1+n}^\ast(t,\{1,\ldots,s\},s+1,\ldots,s+n)F_{s+n}^0(x_{1},\ldots,x_{s+n})dx_{s+1}\ldots dx_{s+n},\nonumber\\
   &&\hskip-12mm  s\geq1,\nonumber
\end{eqnarray}
where the mappings $\mathfrak{A}^\ast_{1+n}(t),\,n\geq0,$ are the generating operators which are represented
by the cumulant expansions with respect to the group $S^\ast(t)=\oplus_{n=0}^\infty S^\ast_n(t)$ of operators
\eqref{S*}.

Using the definition of the $\star$-product \eqref{Product}, the cluster expansions of the groups of operators
\eqref{S*} are represented by the mapping ${\mathbb E}\mathrm{xp}_{\star}$ in the
form
\begin{eqnarray*}\label{ce}
    &&\hskip-7mm S^\ast(t)={\mathbb E}\mathrm{xp}_{\star}\,\mathfrak{A}^\ast(t).
\end{eqnarray*}
In component-wise form cluster expansions are represented by the following recursive relations:
\begin{eqnarray}\label{cex}
   &&\hskip-12mm S_{s+n}^\ast(t,1,\ldots,s,s+1,\ldots,s+n)=\\
   &&\sum\limits_{\mathrm{P}:\,(\{1,\ldots,s\},s+1,\ldots,s+n)=
       \bigcup_i X_i}\,\prod\limits_{X_i\subset\mathrm{P}}\mathfrak{A}^\ast_{|X_i|}(t,X_i),\quad n\geq 0,\nonumber
\end{eqnarray}
where the set consisting of one element of indices $(1,\ldots,s)$ we denoted by
the symbol $\{(1,\ldots,s)\}$ and the symbol ${\sum}_\mathrm{P}$ means the sum
over all possible partitions $\mathrm{P}$ of the set $(\{1,\ldots,s\},s+1,\ldots,s+n)$
into $|\mathrm{P}|$ nonempty mutually disjoint subsets $X_i$.

The solution of recursive relations \eqref{cex} are represented by the inverse mapping
${\mathbb L}\mathrm{n}_{\star}$ in the form of the cumulant expansion
\begin{eqnarray*}
    &&\hskip-7mm \mathfrak{A}^\ast(t)={\mathbb L}\mathrm{n}_{\star}(\mathbb{I}+S^\ast(t)).
\end{eqnarray*}
Then the $(1+n)th$-order cumulant of the groups of operators $S^\ast(t)=\oplus_{n=0}^\infty S^\ast_n(t)$
is defined by the following expansion:
\begin{eqnarray}\label{cumulant}
    &&\hskip-12mm \mathfrak{A}_{1+n}^\ast(t,\{1,\ldots,s\},s+1,\ldots,s+n)\doteq\\
    && \sum\limits_{\mathrm{P}:\,(\{1,\ldots,s\},s+1,\ldots,s+n)={\bigcup}_i X_i}
       (-1)^{\mathrm{|P|}-1}({\mathrm{|P|}-1})!\prod_{X_i\subset\mathrm{P}}
       S^\ast_{|\theta(X_i)|}(t,\theta(X_i)),\nonumber
\end{eqnarray}
where the declusterization mapping $\theta$ is defined by the formula:
$\theta(\{1,\ldots,$ $s\})=(1,\ldots,s)$ and the above notation is used.
The simplest examples of cumulants \eqref{cumulant} of the groups of operators \eqref{S*} have the form:
\begin{eqnarray*}
    &&\hskip-12mm \mathfrak{A}_{1}^{\ast}(t,\{1,\ldots,s\})\doteq S_{s}^{\ast}(t,1,\ldots,s),\\
    &&\hskip-12mm \mathfrak{A}_{1+1}^{\ast}(t,\{1,\ldots,s\},s+1)\doteq S_{s+1}^{\ast}(t,1,\ldots,s+1) -
                  S_{s}^{\ast}(t,1,\ldots,s)S_{1}^{\ast}(t,s+1),\\
    &&\hskip-12mm \mathfrak{A}_{1+2}^{\ast}(t,\{1,\ldots,s\},s+1,s+2)\doteq S_{s+2}^{\ast}(t,1,\ldots,s+2) -\\
    && S_{s+1}^{\ast}(t,1,\ldots,s+1)S_{1}^{\ast}(t,s+2)-S_{s+1}^{\ast}(t,1,\ldots,s,s+2)S_{1}^{\ast}(t,s+1) - \\
    &&S_{s}^{\ast}(t,1,\ldots,s)S_{2}^{\ast}(t,s+1,s+2)+2!S_{s}^{\ast}(t,1,\ldots,s)S_{1}^{\ast}(t,s+1)S_{1}^{\ast}(t,s+2).
\end{eqnarray*}

If $f_{s}\in L^{1}_{s}$, then taking into account that $\big\|S^\ast_{n}(t)\big\|_{L^{1}_{n}}=1$,
for the $(1+n)th$-order cumulant \eqref{cumulant} the following estimate is valid:
\begin{eqnarray}\label{est}
   &&\hskip-12mm \big\|\mathfrak{A}^\ast_{1+n}(t)f_{s+n}\big\|_{L^{1}_{s+n}}
      \leq \sum\limits_{\mathrm{P}:\,(\{1,\ldots,s\},s+1,\ldots,s+n)={\bigcup}_i X_i}
      (|\mathrm{P}|-1)!\big\|f_{s+n}\big\|_{L^{1}_{s+n}}\leq \\
   && \sum\limits_{k=1}^{n+1}\mathrm{s}(n+1,k)(k-1)!\big\|f_{s+n}\big\|_{L^{1}_{s+n}}\leq
       n!e^{n+2}\big\|f_{s+n}\big\|_{ L^{1}_{s+n}},\nonumber
\end{eqnarray}
where $\mathrm{s}(n+1,k)$ are the Stirling numbers of the second kind.

Then, according to this estimate \eqref{est} for the generating operators of expansions \eqref{se},
provided that $\alpha>e$ series \eqref{se} converges on the norm of the space $L^{1}_{\alpha}$, and
the inequality holds
\begin{eqnarray*}\label{Fes}
    &&\|F(t)\|_{L^{1}_{\alpha}}\leq c_{\alpha}\|F(0)\|_{L^{1}_{\alpha}},
\end{eqnarray*}
where $c_{\alpha}=e^{2}(1-\frac{e}{\alpha})^{-1}$. The parameter $\alpha$ can be interpreted as
the value inverse to the average number of hard spheres.

In fact, the following criterion holds.

\begin{criterion*}
\emph{A solution of the Cauchy problem of the BBGKY hierarchy \eqref{h},\eqref{hi} is represented by
expansions \eqref{se} if and only if the generating operators of expansions \eqref{se} are
solutions of cluster expansions \eqref{cex} of the groups of operators \eqref{S*}.}
\end{criterion*}

The necessity condition means that cluster expansions \eqref{cex} take place for groups of
operators \eqref{S*}. These recurrence relations are derived from definition \eqref{rdf} of reduced
distribution functions, provided that they are represented as expansions \eqref{se} for the solution
of the Cauchy problem of the BBGKY hierarchy \eqref{h},\eqref{hi}.

The sufficient condition means that the infinitesimal generator of one-parameter mapping \eqref{se}
coincides with the generator of the BBGKY hierarchy \eqref{h}.
Indeed, in the space $L^{1}_{\alpha}$ the following existence theorem is true \cite{GR02}.
\begin{theorem*}
\emph{If $\alpha>e$, a non-perturbative solution of the Cauchy problem of the BBGKY hierarchy
\eqref{h},\eqref{hi} is represented by series expansions \eqref{se} in which the generating
operators are cumulants of the corresponding order \eqref{cumulant} of groups of operators \eqref{S*}:}
\begin{eqnarray}\label{sec}
    &&\hskip-12mm F_s(t,x_1,\ldots,x_s)=
        \sum_{n=0}^\infty\frac{1}{n!}\int\limits_{(\mathbb{R}^{3}\times\mathbb{R}^{3})^{n}}
         \,\sum\limits_{\mathrm{P}:\,(\{1,\ldots,s\},s+1,\ldots,s+n)={\bigcup}_i X_i}
         (-1)^{\mathrm{|P|}-1}({\mathrm{|P|}-1})!\times\\
    &&\prod_{X_i\subset\mathrm{P}}S^\ast_{|\theta(X_i)|}(t,\theta(X_i))
         F_{s+n}^0(x_{1},\ldots,x_{s+n})dx_{s+1}\ldots dx_{s+n}, \quad s\geq1.\nonumber
\end{eqnarray}
\emph{For initial data $F(0)\in L^{1}_{0}$ of finite sequences of infinitely differentiable functions
with compact supports sequence \eqref{sec} is a unique global classical solution and for arbitrary
initial data $F(0)\in L^{1}_{\alpha}$ is a unique global generalized solution.}
\end{theorem*}


We observe that cluster expansions \eqref{cex} of the groups of operators \eqref{S*} underlie the
classification of possible solution representations \eqref{se} of the Cauchy problem of the BBGKY
hierarchy \eqref{h},\eqref{hi}. In a particular case, non-perturbative solution \eqref{sec}
of the BBGKY hierarchy for many hard spheres can be represented in the form of the perturbation
(iteration) series as a result of applying analogs of the Duhamel equation to cumulants
\eqref{cumulant} of groups of operators.

Indeed, let us put groups of operators in the expression of cumulant \eqref{cumulant} into a new
order with respect to the groups of operators which act on the variables $(x_{1},\ldots,x_{s})$
\begin{eqnarray}\label{peregr}
    &&\hskip-12mm \mathfrak{A}^\ast_{1+n}(t,\{1,\ldots,s\},s+1,\ldots,s+n)=\\
    &&\hskip-7mm\sum\limits_{Y\subset\,(s+1,\ldots,s+n)}S^\ast_{s+|Y|}(t,(1,\ldots,s)\cup\,Y)
    \sum\limits_{\mathrm{P}\,:(s+1,\ldots,s+n)\setminus Y={\bigcup\limits}_i Y_i}
        (-1)^{|\mathrm{P}|}|\mathrm{P}|!\prod_{Y_i\subset\mathrm{P}}
        S^\ast_{|Y_{i}|}(t,Y_{i}).\nonumber
\end{eqnarray}
If $Y_{i}\subset(s+1,\ldots,s+n)$, then for the integrable functions $F_{s+n}^{0}$ and the unitary
group of operators $S^\ast(t)=\oplus_{n=0}^\infty S^\ast_n(t)$ the equality is valid
\begin{eqnarray*}
    &&\hskip-12mm\int\limits_{(\mathbb{R}^{3}\times\mathbb{R}^{3})^{n}}dx_{s+1}\ldots dx_{s+n}
      \prod_{Y_i\subset \mathrm{P}}S^\ast_{|Y_{i}|}(t;Y_{i})
        F_{s+n}^{0}(x_{1},\ldots,x_{s+n})=\\
    &&\int\limits_{(\mathbb{R}^{3}\times\mathbb{R}^{3})^{n}}dx_{s+1}\ldots dx_{s+n}F_{s+n}^{0}(x_{1},\ldots,x_{s+n}).
\end{eqnarray*}
Then, taking into account the validity for arbitrary $Y\subset\,(s+1,\ldots,s+n)$ the following equality:
\begin{eqnarray*}
    &&\hskip-7mm\sum\limits_{\mathrm{P}\,:(s+1,\ldots,s+n)\setminus Y={\bigcup\limits}_i Y_i}
       (-1)^{|\mathrm{P}|}|\mathrm{P}|!=(-1)^{|(s+1,\ldots,s+n)\setminus Y|},
\end{eqnarray*}
according to expression \eqref{peregr} for series expansions \eqref{sec} of the BBGKY hierarchy, we obtain
\begin{eqnarray}\label{cherez1}
    &&\hskip-12mm F_s(t,x_1,\ldots,x_s)=\sum_{n=0}^\infty\frac{1}{n!}
       \int\limits_{(\mathbb{R}^{3}\times\mathbb{R}^{3})^{n}}U_{1+n}^\ast(t,\{1,\ldots,s\},\\
    && s+1,\ldots,s+n)F_{s+n}^0(x_{1},\ldots,x_{s+n})dx_{s+1}\ldots dx_{s+n},\quad s\geq1,\nonumber
\end{eqnarray}
where $U_{1+n}^\ast(t)$ is the $(1+n)th$-order reduced cumulant of the groups of operators \eqref{S*}
\begin{eqnarray*}\label{rc}
    &&\hskip-12mm U_{1+n}^\ast(t,\{1,\ldots,s\},s+1,\ldots,s+n)=\\
    &&\sum\limits_{Y\subset (s+1,\ldots,s+n)}(-1)^{|(s+1,\ldots,s+n)\setminus Y|}
       S^\ast_{|(1,\ldots,s)\cup Y|}(t,(1,\ldots,s)\cup Y).\nonumber
\end{eqnarray*}
Using the symmetry property of initial reduced distribution functions, for integrand functions
in every term of series \eqref{cherez1} the following equalities are valid
\begin{eqnarray*}
    &&\hskip-12mm \sum\limits_{Y\subset (s+1,\ldots,s+n)}
       (-1)^{|(s+1,\ldots,s+n)\setminus Y)|}S^\ast_{|(1,\ldots,s)\cup Y|}(t,(1,\ldots,s)\cup Y)
       F_{s+n}^0=\nonumber
\end{eqnarray*}
\begin{eqnarray*}
    && \sum\limits_{k=0}^{n}(-1)^{k}\sum\limits_{i_{1}<\ldots<i_{n-k}=s+1}^{s+n}
       S^\ast_{s+n-k}(t,1,\ldots,s,i_{1},\ldots,i_{n-k})F_{s+n}^0=\nonumber\\
    && \sum\limits_{k=0}^{n}(-1)^{k}\frac{n!}{k!(n-k)!}
       S^\ast_{s+n-k}(t,1,\ldots,s+n-k)F_{s+n}^0(x_{1},\ldots,x_{s+n}).\nonumber
\end{eqnarray*}
Thus, the $(1+n)th$-order reduced cumulant represents by the following expansion \cite{L61}:
\begin{eqnarray*}
    &&\hskip-12mm U_{1+n}^\ast(t,\{1,\ldots,s\},s+1,\ldots,s+n)=
        \sum^n_{k=0}(-1)^k \frac{n!}{k!(n-k)!}S_{s+n-k}^\ast(t,1,\ldots,s+n-k),
\end{eqnarray*}
and consequently, we derive the representation for series expansions of a solution of the BBGKY
hierarchy \cite{PG83} which is can be written down in terms of an analogue of the annihilation
operator \eqref{a}:
\begin{eqnarray}\label{rcexp}
   &&\hskip-12mm F(t)= \sum\limits_{n=0}^{\infty}\frac{1}{n!}\,\sum\limits_{k=0}^{n}\,(-1)^{k}\,
      \frac{n!}{k!(n-k)!}\,\mathfrak{a}^{n-k}S^\ast(t)\mathfrak{a}^{k}F(0)=\\
   &&S^\ast(t)F(0)+\sum\limits_{n=1}^{\infty}\frac{1}{n!}
      \big[\underbrace{\mathfrak{a},\ldots,\big[{\mathfrak{a}}}_{\hbox{n-times}},S^\ast(t)\big]\ldots\big]F(0)=\nonumber\\
   && e^{\mathfrak{a}}S^\ast(t)e^{-\mathfrak{a}}F(0).\nonumber
\end{eqnarray}
Finally, in view of the validity of the equality
\begin{eqnarray*}
   &&S^\ast(t-\tau)
       \big[\mathfrak{a},\mathcal{L}^\ast\big]S^\ast(\tau)F(0)=\frac{d}{d\tau}S^\ast(t-\tau)\mathfrak{a}S^\ast(\tau)F(0),
\end{eqnarray*}
expansion \eqref{rcexp} is represented in the form of perturbation (iteration) series
of the BBGKY hierarchy \eqref{h} for many hard spheres
\begin{eqnarray*}
   &&\hskip-12mm F(t)=\sum\limits_{n=0}^{\infty}\,\int\limits_{0}^{t} dt_{1}\ldots
        \int\limits_{0}^{t_{n-1}}dt_{n}S^\ast(t-t_{1})\big[\mathfrak{a},\mathcal{L}^\ast\big]
        S^\ast(t_1-t_2)\ldots S^\ast(t_{n-1}-t_n)\big[\mathfrak{a},\mathcal{L}^\ast\big]S^\ast(t_{n})F(0),\nonumber
\end{eqnarray*}
or in component-wise form \cite{B1},\cite{GP85},\cite{L75}:
\begin{eqnarray}\label{hiter}
   &&\hskip-12mm F_s(t,x_1,\ldots,x_s)=\\
   &&\hskip-7mm \sum\limits_{n=0}^{\infty}\,\int\limits_{0}^{t}dt_{1}\ldots
       \int\limits_{0}^{t_{n-1}}dt_{n}\int\limits_{(\mathbb{R}^{3}\times\mathbb{R}^{3})^{n}}
       dx_{s+1}\ldots dx_{s+n}\,S^\ast_s(t-t_{1})\sum\limits_{j_1=1}^{s}\mathcal{L}^{\ast}_{\mathrm{int}}(j_1,s+1)
       S^\ast_{s+1}(t_1-t_2)\ldots\nonumber\\
   &&\hskip-7mm S^\ast_{s+n-1}(t_{n-1}-t_n)\sum\limits_{j_n=1}^{s+n-1}\mathcal{L}^{\ast}_{\mathrm{int}}(j_n,s+n)
      S^\ast_{s+n}(t_{n})F_{s+n}^0(x_1,\ldots,x_{s+n}), \quad s\geq1.\nonumber
\end{eqnarray}

Let us make some comments concerning the existence of solutions to the Cauchy problem of the BBGKY
hierarchy for initial data from various function spaces.

In the spaces of sequences of integrable functions, the existence and uniqueness of a global in time
non-perturbative solution was proved in the papers \cite{GerRS},\cite{PG83} (see also book \cite{CGP97}).
It should be noted that the first few terms of the \eqref{sec} series were established in papers
\cite{C62},\cite{Ch},\cite{G56},\cite{GP63} as an analog of cluster expansions of the reduced
equilibrium distribution functions.

The BBGKY hierarchy describes both the non-equilibrium and equilibrium states. Non-equilibrium states
are described by the solution of the initial value problem for this hierarchy, and, correspondingly,
equilibrium states are solutions of the steady BBGKY hierarchy. The existence of equilibrium solutions
to the steady BBGKY hierarchy has been reviewed in books \cite{CGP97},\cite{GPM}.

As is known, to describe the evolution of a state of infinitely many particles, the suitable
functional space is the space of sequences of functions bounded with respect to the configuration
variables and decreasing with respect to the momentum ones; in particular, the equilibrium
distribution functions belong to this space. For the solution extension from the space of
sequences of integrable functions to this space, the method of the thermodynamic limit was
developed \cite{CGP97},\cite{GP85},\cite{GP85_D}.

For one-dimensional many-particle systems with short-range potential, using the method of
interaction region developed by Petrina \cite{P79} for solution representation \eqref{cherez1},
the existence theorem for the BBGKY hierarchy was proved for the first time in this functional
space. By a similar method for the initial reduced distribution functions from such a space,
the existence of a mean value functional for solution \eqref{sedc} of the dual BBGKY hierarchy
\eqref{dh} was established in the paper \cite{R10}.

As mentioned above, for a solution representation of the Cauchy problem of the BBGKY hierarchy
for hard spheres is widely used in the representation as a series of perturbation theory
\eqref{hiter} (an iteration series over the evolution of the state of selected groups of
particles) \cite{B1},\cite{CIP},\cite{GP85},\cite{L75},\cite{Sp91}. In this form, the solution
is applied to construct its Boltzmann--Grad asymptotics, which is governed by the Boltzmann
kinetic equation (see section 3.1).
The justification of a solution represented as an iteration series for hard spheres is based on
giving a rigorous mathematical meaning to every term of the iteration series and on the proof of
its convergence. The main difficulty in this problem is that the phase trajectories of particles
for a system with a singular interaction potential are defined almost everywhere in the phase space,
and initial distribution functions in the iteration series are concentrated on lower-dimensional
manifolds. It is necessary to ensure that the trajectories are defined on these manifolds. This
problem was completely solved in the papers \cite{GP90},\cite{PG90}.

In the case of infinitely many hard spheres a local in time solution \cite{GP85} of the Cauchy problem
of the BBGKY hierarchy is represented by iteration series for arbitrary initial data from the space of
sequences of functions bounded with respect to configuration variables and for initial data close to
the equilibrium state it is a global in time solution \cite{PG90}. For such initial data in a one-dimensional
space for hard sphere system the existence of global in time solution was proved in the paper \cite{Ger}.

In addition, we remark that the correlation decay property, known as the Bogolyubov correlation weakening
principle \cite{B1}, for solution \eqref{sec} of the BBGKY hierarchy for hard spheres was proved in the
article \cite{GSh08}.

\bigskip
\subsection{The Liouville hierarchy for correlation functions}
An alternative approach to the description of a state of finitely many hard spheres
consists in the employment of functions determined by the cluster expansions of the
probability distribution functions. The solutions to such cluster expansions are
cumulants (semi-invariants) of probability distribution functions and are interpreted,
from a physical point of view, as correlations of a state or correlation functions.
The evolution of correlation functions is governed by the so-called Liouville hierarchy
\cite{GG22R}. Historically, there have been several approaches to describing correlations
in many-particle systems. Among them, we mention the well-known approach to the dynamics
of correlations by I. Prigogine \cite{P62} and R. Balescu \cite{B71} and its applications
in plasma theory.

Further, it will be established that the constructed dynamics of correlation underlie
the description of the dynamics of infinitely many hard spheres governed by the BBGKY
hierarchy for reduced distribution functions or the hierarchy of nonlinear evolution
equations for reduced correlation functions, i.e., of the cumulants of reduced distribution
functions.


We introduce a sequence of correlation functions $g(t)=(1,g_{1}(t,x_1),\ldots,$ $g_{s}(t,x_1,\ldots,x_s),\ldots)$
by means of cluster expansions of the probability distribution functions
$D(t)=(1,D_{1}(t,x_1),\ldots,$ $D_{n}(t,x_1,\ldots,x_n),\ldots)$, defined on the set
of allowed configurations $\mathbb{R}^{3n}\setminus \mathbb{W}_n$ as follows:
\begin{eqnarray}\label{D_(g)N}
    &&\hskip-12mm D_{n}(t,x_1,\ldots,x_n)= g_{n}(t,x_1,\ldots,x_n)+
       \sum\limits_{\mbox{\scriptsize $\begin{array}{c}\mathrm{P}:
       (x_1,\ldots,x_n)=\bigcup_{i}X_{i},\\|\mathrm{P}|>1 \end{array}$}}
        \prod_{X_i\subset \mathrm{P}}g_{|X_i|}(t,X_i),\quad n\geq1,\nonumber
\end{eqnarray}
where ${\sum\limits}_{\mathrm{P}:(x_1,\ldots,x_n)=\bigcup_{i} X_{i},\,|\mathrm{P}|>1}$ is the sum over all
possible partitions $\mathrm{P}$ of the set of the arguments $(x_1,\ldots,x_n)$ into
$|\mathrm{P}|>1$ nonempty mutually disjoint subsets $X_i\subset(x_1,\ldots,x_n)$.

On the set $\mathbb{R}^{3n}\setminus \mathbb{W}_n$ solutions of recursion relations (\ref{D_(g)N}) are given
by the following expansions:
\begin{eqnarray}\label{gfromDFB}
   &&\hskip-12mm g_{s}(t,x_1,\ldots,x_s)=D_{s}(t,x_1,\ldots,x_s)+\\
   &&\hskip-8mm \sum\limits_{\mbox{\scriptsize $\begin{array}{c}\mathrm{P}:(x_1,\ldots,x_s)=
       \bigcup_{i}X_{i},\\|\mathrm{P}|>1\end{array}$}}(-1)^{|\mathrm{P}|-1}(|\mathrm{P}|-1)!\,
       \prod_{X_i\subset \mathrm{P}}D_{|X_i|}(t,X_i), \quad s\geq1.\nonumber
\end{eqnarray}
The structure of expansions (\ref{gfromDFB}) is such that the correlation functions can be treated as cumulants
(semi-invariants) of the probability distribution functions \cite{D19}. Such an interpretation of these functions
is due to the fact that the probability distribution function of statistically independent hard spheres on allowed
configurations is described by the product of one-particle correlation functions (probability distribution
functions of each hard sphere): $g_{s}(t,x_1,\ldots,x_s)=\prod_{i=1}^{s}g_{1}(t,x_i)
\mathcal{X}_{\mathbb{R}^{3s}\setminus \mathbb{W}_s}\delta_{s,1},$ $s\geq1 $.


The evolution of the sequence of correlation functions (\ref{gfromDFB}) of many hard spheres is determined
by the Cauchy problem of the weak formulation of the Liouville hierarchy of the following evolution equations \cite{GG22}:
\begin{eqnarray}\label{Lh}
   &&\hskip-12mm\frac{\partial}{\partial t}g_{s}(t,x_1,\ldots,x_s)=
      \mathcal{L}^{\ast}_{s}(1,\ldots,s)g_{s}(t,x_1,\ldots,x_s)+\\
   &&\sum\limits_{\mathrm{P}:\,(x_1,\ldots,x_s)=X_{1}\bigcup X_2}\,\sum\limits_{i_{1}\in \widehat{X}_{1}}
      \sum\limits_{i_{2}\in \widehat{X}_{2}}\mathcal{L}_{\mathrm{int}}^{\ast}(i_{1},i_{2})
      g_{|X_{1}|}(t,X_{1})g_{|X_{2}|}(t,X_{2}), \nonumber
\end{eqnarray}
\begin{eqnarray}
 \label{Lhi}
   &&\hskip-12mm g_{s}(t,x_1,\ldots,x_s)\big|_{t=0}=g_{s}^{0}(x_1,\ldots,x_s),\quad s\geq1,
\end{eqnarray}
where ${\sum\limits}_{\mathrm{P}:\,(x_1,\ldots,x_s)=X_{1}\bigcup X_2}$ is the sum over all
possible partitions $\mathrm{P}$ of the set $(x_1,\ldots,x_s)$ into two nonempty mutually
disjoint subsets $X_1$ and $X_2$, the symbol $\widehat{X}_i$ means the set of indexes of
the set $X_i$ of phase space coordinates and the operator $\mathcal{L}^{\ast}_{s}$ is
defined on the subspace $L^{1}_{0}\subset L^{1}$ by formulas (\ref{L}). It should be noted
that the Liouville hierarchy (\ref{Lh}) is the evolution recurrence equations set.

For $t\geq0$ we give a few examples of recurrence equations set (\ref{Lh}) for a system
of hard spheres:
\begin{eqnarray*}
   &&\hskip-12mm\frac{\partial}{\partial t}g_{1}(t,x_1)=
     -\langle p_1,\frac{\partial}{\partial q_1}\rangle g_{1}(t,x_1),\\
   &&\hskip-12mm\frac{\partial}{\partial t}g_{2}(t,x_1,x_2)=
     -\sum\limits_{j=1}^{2}\langle p_j,\frac{\partial}{\partial q_j}\rangle g_{2}(t,x_1,x_2)+ \\
   && \sigma^2\int\limits_{\mathbb{S}_{+}^2}d\eta\langle\eta,(p_{1}-p_{2})\rangle
        \big(g_2(t,q_1,p_{1}^\ast,q_{2},p_{2}^\ast)\delta(q_{1}-q_{2}+\sigma\eta)-\\
   &&g_2(t,x_1,x_2)\delta(q_{1}-q_{2}-\sigma\eta)\big)+\\
   &&\sigma^2\int\limits_{\mathbb{S}_{+}^2}d\eta\langle\eta,(p_{1}-p_{2})\rangle
        \big(g_1(t,q_1,p_{1}^\ast)g_1(t,q_{2},p_{2}^\ast)\delta(q_{1}-q_{2}+\sigma\eta)-\\
   &&g_1(t,x_1)g_1(t,x_2)\delta(q_{1}-q_{2}-\sigma\eta)\big),
\end{eqnarray*}
where it was used notations accepted above in definition (\ref{L}).

Thus, in terms of correlation functions (\ref{gfromDFB}), the evolution of the states of a finite
number of hard spheres is described by an equivalent method compared to probability distribution
functions, namely, within the framework of the dynamics of correlations.

We note that because the Liouville hierarchy (\ref{Lh}) is the recurrence evolution equations
set, we can construct a solution of the Cauchy problem (\ref{Lh}),(\ref{Lhi}), integrating
each equation of the hierarchy as the inhomogeneous Liouville equation. For example, as a result
of the integration of the first two equations of the Liouville hierarchy (\ref{Lh}), we obtain
the following equalities:
\begin{eqnarray*}
    &&\hskip-12mm g_{1}(t,x_1)=S_{1}(-t,1)g_{1}^{0}(x_1),\\
    &&\hskip-12mm g_{2}(t,1,2)=S_{2}(-t,1,2)g_{2}^0(x_1,x_2)+\\
    && \int\limits_{0}^{t}dt_{1}S_{2}(t_{1}-t,1,2)\mathcal{L}^\ast_{\mathrm{int}}(1,2)
       S_{1}(-t_{1},1)S_{1}(-t_{1},2)g_{1}^0(x_1)g_{1}^0(x_2).
\end{eqnarray*}
Then for the corresponding term on the right-hand side of the second equality, an analog of the
Duhamel equation holds
\begin{eqnarray*}
    &&\hskip-12mm\int\limits_{0}^{t}dt_{1}S_{2}(t_{1}-t,1,2)\mathcal{L}^\ast_{\mathrm{int}}(1,2)
       S_{1}(-t_{1},1)S_{1}(-t_{1},2)=
\end{eqnarray*}
\begin{eqnarray*}
    &&-\int\limits_{0}^{t}dt_{1}\frac{d}{dt_{1}}\big(S_{2}(t_{1}-t,1,2)
       S_{1}(-t_{1},1)S_{1}(-t_{1},2)\big)=\nonumber\\
    &&S_{2}(-t,1,2)-S_{1}(-t,1)S_{1}(-t,2)=\mathfrak{A}^\ast_{2}(t,1,2),\nonumber
\end{eqnarray*}
where $\mathfrak{A}^\ast_{2}(t)$ is the second-order cumulant (\ref{cumulant}) of groups of operators
(\ref{S*}). As a result of similar transformations for $s>2$, the solution of the Cauchy problem
(\ref{Lh}),(\ref{Lhi}), constructed using an iterative procedure, can be represented as expansions
in cumulants of groups of operators (\ref{S*}).

If the initial state is specified by the sequence $g(0)=(1,g_{1}^{0}(x_1),\ldots,$ $g_{n}^{0}(x_1,\ldots,x_n),\ldots)$,
of correlation functions $g_{n}^{0}\in L^{1}_{n},\,n\geq1,$ then the evolution of all possible states, i.e., the sequence
$g(t)=(1,$ $g_{1}(t,x_1),\ldots,$ $g_{s}(t,x_1,\ldots,x_s),\ldots)$ of the correlation functions $g_{s}(t),\,s\geq1$,
is represented by the following expansions \cite{GG22}:
\begin{eqnarray}\label{gLhs}
   &&\hskip-12mm g_{s}(t,x_1,\ldots,x_s)=\sum\limits_{\mathrm{P}:\,(x_1,\ldots,x_s)=\bigcup_j X_j}
      \mathfrak{A}^\ast_{|\mathrm{P}|}(t,\{\widehat{X}_1\},\ldots,\{\widehat{X}_{|\mathrm{P}|}\})
      \prod_{X_j\subset \mathrm{P}}g_{|X_j|}^{0}(X_j),\quad s\geq1,
\end{eqnarray}
where the symbol $\sum_{\mathrm{P}:\,(x_1,\ldots,x_s)=\bigcup_j X_j}$ denotes the sum over all possible partitions
$\mathrm{P}$ of the set $(x_1,\ldots,$ $x_s)$ into $|\mathrm{P}|$ nonempty mutually disjoint subsets $X_j$, the
symbol $\widehat{X}$ means the set of indexes of the set $X$ of phase space coordinates and the set
$(\{\widehat{X}_1\},\ldots,\{\widehat{X}_{|\mathrm{P}|}\})$ consists of elements that are subsets
$\widehat{X}_j\subset (1,\ldots,s)$, i.e., $|(\{\widehat{X}_1\},\ldots,\{\widehat{X}_{|\mathrm{P}|}\})|=|\mathrm{P}|$.
The generating operator $\mathfrak{A}^\ast_{|\mathrm{P}|}(t)$ of expansions (\ref{gLhs}) is the $|\mathrm{P}|th$-order
cumulant of the groups of operators (\ref{S*}) which is defined by the expansion
\begin{eqnarray}\label{cumulantP}
   &&\hskip-12mm \mathfrak{A}^\ast_{|\mathrm{P}|}(t,\{\widehat{X}_1\},\ldots,\{\widehat{X}_{|\mathrm{P}|}\})\doteq\\
   && \sum\limits_{\mathrm{P}^{'}:\,(\{\widehat{X}_1\},\ldots,\{\widehat{X}_{|\mathrm{P}|}\})=
      \bigcup_k Z_k}(-1)^{|\mathrm{P}^{'}|-1}({|\mathrm{P}^{'}|-1})!
      \prod\limits_{Z_k\subset\mathrm{P}^{'}}S^\ast_{|\theta(Z_{k})|}(t,\theta(Z_{k})),\nonumber
\end{eqnarray}
where the symbol $\theta$ is the declusterization mapping: $\theta(\{\widehat{X}_i\})\doteq(\widehat{X}_i)$.
The simplest examples of correlation functions (\ref{gLhs}) are given as follows:
\begin{eqnarray*}
   &&g_{1}(t,x_1)=\mathfrak{A}^\ast_{1}(t,1)g_{1}^{0}(x_1),\\
   &&g_{2}(t,x_1,x_2)=\mathfrak{A}^\ast_{1}(t,\{1,2\})g_{2}^{0}(x_1,x_2)+
     \mathfrak{A}^\ast_{2}(t,1,2)g_{1}^{0}(x_1)g_{1}^{0}(x_2).
\end{eqnarray*}

The structure of expansions (\ref{gLhs}) is established as a result of the permutation of the terms
of cumulant expansions (\ref{gfromDFB}) for correlation functions and cluster expansions (\ref{D_(g)N})
for initial probability distribution functions. Thus, the cumulant origin of correlation
functions induces the cumulant structure of their dynamics (\ref{gLhs}).

In particular, in the absence of correlations between hard spheres at the initial moment (initial state
satisfying the chaos condition \cite{CGP97},\cite{Sp91}) the sequence of the initial correlation functions
on allowed configurations has the form $g^{(c)}(0)=(0,g_{1}^{0}(x_1),0,\ldots,0,\ldots)$. In terms of a
sequence of the probability distribution functions, the chaos condition means that initial data is specified
in the form
$D^{(c)}(0)=(1,D_{1}^0(x_1),D_{1}^0(x_1)D_{1}^0(x_2)\mathcal{X}_{\mathbb{R}^{6}\setminus \mathbb{W}_2},\ldots,$
$\prod^n_{i=1}D_{1}^0(x_i)\mathcal{X}_{\mathbb{R}^{3n}\setminus \mathbb{W}_n},\ldots)$,
where the function $\mathcal{X}_{\mathbb{R}^{3n}\setminus \mathbb{W}_{n}}$ is the Heaviside step function of
allowed configurations of $n$ hard spheres.
In this case for $(x_1,\ldots,x_s)\in\mathbb{R}^{3s}\times(\mathbb{R}^{3s}\setminus \mathbb{W}_s)$
expansions (\ref{gLhs}) are represented as follows:
\begin{eqnarray}\label{gth}
   &&\hskip-12mm g_{s}(t,x_1,\ldots,x_s)=\mathfrak{A}^\ast_{s}(t,1,\ldots,s)\,\prod\limits_{i=1}^{s}g_{1}^{0}(x_i)
         \mathcal{X}_{\mathbb{R}^{3s}\setminus \mathbb{W}_s},\quad s\geq1,
\end{eqnarray}
where the generating operator $\mathfrak{A}^\ast_{s}(t)$ of this expansion is the $sth$-order cumulant
of groups of operators (\ref{S*}) defined by the expansion
\begin{eqnarray}\label{cumcp}
   &&\hskip-14mm\mathfrak{A}^\ast_{s}(t,1,\ldots,s)=\sum\limits_{\mathrm{P}:\,(1,\ldots,s)=
       \bigcup_i X_i}(-1)^{|\mathrm{P}|-1}({|\mathrm{P}|-1})!
      \prod\limits_{X_i\subset\mathrm{P}}S^\ast_{|X_i|}(t,X_i),
\end{eqnarray}
with notations accepted in formula (\ref{gLhs}). From the structure of series (\ref{gth}) it is clear
that in case of the absence of correlations at the initial instant the correlations generated by the
dynamics of a system of hard spheres are completely determined by cumulants (\ref{cumcp}) of the groups
of operators (\ref{S*}).


We note that in the case of initial data $g^{(c)}(0)$ expansions (\ref{gth}) can be rewritten in another
representation that explains their physical meaning. Indeed, for $n=1$ we have
\begin{eqnarray*}
     &&g_{1}(t,x_1)=\mathfrak{A}^\ast_{1}(t,1)g_{1}^0(x_1)=g_{1}^0(p_1,q_1-p_1t).
 \end{eqnarray*}
Then, according to formula (\ref{gth}) and the definition of the first-order cumulant
$\mathfrak{A}^\ast_{1}(t)=S_1(-t)$, and its inverse group of operators $S_1^{-1}(-t)=S_1(t)$, we express the
correlation functions $g_{s}(t),$ $s\geq 2$, in terms of the one-particle correlation function $g_{1}(t)$.
Therefore, for $s\geq2$ expansions (\ref{gth}) are represented in the following form:
\begin{eqnarray*}
     &&g_{s}(t,x_1,\ldots,x_s)=
         \widehat{\mathfrak{A}}^\ast_{s}(t,1,\ldots,s)\,\prod_{i=1}^{s}\,g_{1}(t,x_i),\quad s\geq 2,
\end{eqnarray*}
where $\widehat{\mathfrak{A}}^\ast_{s}(t,1,\ldots,s)$ is the $s$-order cumulant (\ref{cumcp}) of the scattering
operators
\begin{eqnarray*}
     &&\widehat{S}_n(t,1,\ldots,n)\doteq
        S_{n}(-t,1,\ldots,n)\mathcal{X}_{\mathbb{R}^{3n}\setminus\mathbb{W}_n}\prod_{i=1}^{n}S_{1}(t,i),\quad n\geq1.
\end{eqnarray*}
On the subspace $L^{1}_{n,0}$ a generator of the scattering operator $\widehat{S}_n(t,1,\ldots,n)$ is determined
by the operator:
$\frac{d}{dt}\widehat{S}_n(t,1,\ldots,n)\mid_{t=0}=\sum_{j_{1}<j_{2}=1}^{n}\mathcal{L}_{\mathrm{int}}^\ast(j_{1},j_{2}),$
where for $t\geq0$ the operator $\mathcal{L}_{\mathrm{int}}^\ast(j_{1},j_{2})$ is defined by formula (\ref{L}).

If $g^0_{n}\in L^{1}_{n},\,n\geq1$, one-parameter mapping (\ref{gLhs}) generates strong continuous group
of nonlinear operators
\begin{eqnarray}\label{Sspherc}
   &&\mathcal{G}(t;1,\ldots,s\mid g(0))\doteq g_{s}(t,x_1,\ldots,x_s),
\end{eqnarray}
and it is bounded, and the following estimate holds:
$\big\|\mathcal{G}(t;1,\ldots,s\mid g)\big\|_{L^{1}_{s}}\leq s!c^{s},$
where $c\equiv \max(1,$ $\max_{\mathrm{P}:\,(1,\ldots,s)=\bigcup_i X_i}\|g_{|X_{i}|}\|_{L^{1}_{|X_{i}|}})$.
For $g_{n}\in L^{1}_{n,0},\,n\geq1$, the infinitesimal generator of this group of nonlinear operators has
the following structure
\begin{eqnarray}\label{LL}
   &&\hskip-12mm \mathcal{L}(1,\ldots,s\mid g)\doteq\mathcal{L}^{\ast}_{s}(1,\ldots,s)g_{s}(x_1,\ldots,x_s)+\\
   &&\sum\limits_{\mathrm{P}:\,(x_1,\ldots,x_s)=X_{1}\bigcup X_2}\,\sum\limits_{i_{1}\in \widehat{X}_{1}}
      \sum\limits_{i_{2}\in \widehat{X}_{2}}\mathcal{L}_{\mathrm{int}}^{\ast}(i_{1},i_{2})
      g_{|X_{1}|}(X_{1})g_{|X_{2}|}(X_{2}),\nonumber
\end{eqnarray}
where we used the notation adopted above in expansions (\ref{gLhs}).


The following statement is true.

\begin{theorem*}
\emph{If $t\in\mathbb{R}$, a unique solution of the Cauchy problem of the Liouville hierarchy
(\ref{Lh}),(\ref{Lhi}) is represented by a sequence of expansions (\ref{gLhs}).
For $g_{n}^0\in L^{1}_{n,0}\subset L^{1}_{n},\,n\geq1$, a sequence of functions (\ref{gLhs})
is a classical solution and for arbitrary initial data $g_{n}^0\in L^{1}_{n},\,n\geq1$, one has
a generalized solution.}
\end{theorem*}

The proof of the theorem is similar to the proof of the existence theorem for the BBGKY hierarchy
in the space of sequences of integrable functions \cite{CGP97},\cite{GerRS}. Indeed, if the initial
data is $g_{n}^0\in L^{1}_{n,0},\,n\geq1$, then the infinitesimal generator of the group of nonlinear
operators (\ref{Sspherc}) coincides with the operator (\ref{LL}) and hence the Cauchy problem
(\ref{Lh}),(\ref{Lhi}) has a classical (strong) solution (\ref{gLhs}).


We remark that a steady solution of the Liouville hierarchy (\ref{Lh}) is a sequence of
the Ursell functions on the allowed configurations of hard spheres, i.e., it is the sequence
$g^{(eq)}=(0,e^{-\beta \frac{p^2_1}{2}},0,\ldots,)$, where $\beta$ is a parameter
inversely proportional to temperature \cite{GPM}.

Finally, we emphasize that the dynamics of correlations, that is, the fundamental equations
(\ref{Lh}) describing the evolution of correlations of states of hard spheres, can be used
as a basis for describing the evolution of the state of both a finite and an infinite number
of hard spheres instead of the Liouville equations (\ref{dLe}).


In what follows, we outline an approach to describing the evolution of a state using reduced
distribution functions based on the dynamics of correlations in a system of many hard spheres
governed by the Liouville hierarchy for correlation functions (\ref{Lh}).

Remind that reduced distribution functions are defined by means of sequence \eqref{rdf} of the
probability distribution functions:
\begin{eqnarray}\label{ms}
     &&\hskip-12mm F_{s}(t,x_1,\ldots,x_s)\doteq(I,D(t))^{-1}\sum\limits_{n=0}^{\infty}\frac{1}{n!}
         \,\int\limits_{(\mathbb{R}^{3}\times\mathbb{R}^{3})^{n}}dx_{s+1}\ldots dx_{s+n}
         \,D_{s+n}(t,x_1,\ldots,x_{s+n}),\\
    && \hskip-12mm   s\geq1,\nonumber
\end{eqnarray}
where the normalizing factor $(I,D(t))\doteq\sum_{n=0}^{\infty}\frac{1}{n!}
\int_{(\mathbb{R}^{3}\times\mathbb{R}^{3})^{n}}dx_{1}\ldots dx_{n}D_{n}(t,x_1,\ldots,x_n)$
is a grand canonical partition function.
The possibility of redefining of the reduced distribution functions naturally arises as a result
of dividing the series in expression (\ref{ms}) by the series of the normalization factor.

A definition of reduced distribution functions equivalent to definition (\ref{ms})
is formulated on the basis of correlation functions (\ref{gLhs}) of a system of hard
spheres by means of the following series expansion \cite{GG22}:
\begin{eqnarray}\label{FClusters}
    &&\hskip-14mm  F_{s}(t,x_1,\ldots,x_s)\doteq\sum\limits_{n=0}^{\infty}\frac{1}{n!}\,
       \int\limits_{(\mathbb{R}^{3}\times\mathbb{R}^{3})^{n}}dx_{s+1}\ldots dx_{s+n}\,
       g_{1+n}(t,\{x_1,\ldots,x_s\},x_{s+1},\ldots,x_{s+n}), \\
    && \hskip-14mm   s\geq1,\nonumber
\end{eqnarray}
where on the set of allowed configurations $\mathbb{R}^{3(s+n)}\setminus \mathbb{W}_{s+n}$ the correlation
functions of clusters of hard spheres $g_{1+n}(t), n\geq0,$ are determined by the expansions:
\begin{eqnarray}\label{rozvL-Nclusters}
    &&\hskip-12mm g_{1+n}(t,\{x_1,\ldots,x_s\},x_{s+1},\ldots,x_{s+n})=\\
    &&\hskip-7mm \sum_{\mbox{\scriptsize$\begin{array}{c}\mathrm{P}:(\{x_1,\ldots,x_s\},\\
       x_{s+1},\ldots,x_{s+n})=\bigcup_i X_i\end{array}$}}
       \mathfrak{A}_{|\mathrm{P}|}^\ast\big(t,\{\theta(\widehat{X}_1)\},\ldots,\{\theta(\widehat{X}_{|\mathrm{P}|})\}\big)
       \prod_{X_i\subset \mathrm{P}}g_{|X_i|}^0(X_i),\quad n\geq0.\nonumber
\end{eqnarray}
We remind that in expansions (\ref{rozvL-Nclusters}) the generating operator $\mathfrak{A}_{|\mathrm{P}|}^\ast(t)$
is the $|\mathrm{P}|th$-order cumulant (\ref{cumulantP}) of the groups of operators (\ref{S*}),
and the symbol
$\sum_{\mathrm{P}:(\{x_1,\ldots,x_s\},x_{s+1},\ldots,x_{s+n})=\bigcup_i X_i}$
means the sum over all possible partitions $\mathrm{P}$ of the set $(\{x_1,\ldots,x_s\},x_{s+1},\ldots,x_{s+n})$
into nonempty mutually disjoint subsets $X_i$.

On allowed configurations the correlation functions of particle clusters in series (\ref{FClusters}),
i.e., the functions $g_{1+n}(t,\{x_{1},\ldots,x_{s}\},$ $x_{s+1},\ldots,x_{s+n}),\,n\geq0$, are defined as solutions
of generalized cluster expansions of a sequence of solutions of the Liouville equations:
\begin{eqnarray}\label{a8}
    &&\hskip-12mm D_{s+n}(t,x_{1},\ldots,x_{s+n})=\sum_{\mbox{\scriptsize$\begin{array}{c}\mathrm{P}:(\{x_1,\ldots,x_s\},\\
      x_{s+1},\ldots,x_{s+n})=\bigcup_i X_i\end{array}$}}
      \prod_{X_i\subset\mathrm{P}}g_{|X_i|}(t,X_i),\quad s\geq1,\,n\geq0,
\end{eqnarray}
namely,
\begin{eqnarray*}
    &&\hskip-12mm g_{1+n}(t,\{x_{1},\ldots,x_{s}\},x_{s+1},\ldots,x_{s+n})=\\
    &&\hskip-7mm \sum_{\mbox{\scriptsize$\begin{array}{c}\mathrm{P}:(\{x_1,\ldots,x_s\},\\
       x_{s+1},\ldots,x_{s+n})=\bigcup_i X_i\end{array}$}}
      (-1)^{|\mathrm{P}|-1}(|\mathrm{P}| -1)!\,\prod_{X_i\subset\mathrm{P}}D_{|\theta(X_{i})|}(t,\theta(X_{i})),
       \quad s\geq1,\,n\geq0,\nonumber
\end{eqnarray*}
where $\theta$ is the declusterization mapping defined in formula (\ref{cumulantP}), the probability
distribution function $D_{|\theta(X_{i})|}(t,\theta(X_{i}))$ is a solution of the Liouville equation.

The correlation functions of particle clusters satisfy the Liouville hierarchy of evolution equations
with the following generator
\begin{eqnarray}\label{LcL}
   &&\hskip-12mm \mathcal{L}(\{1,\ldots,s\},s+1,\ldots,s+n\mid \mathfrak{d}_{\{x_1,\ldots,x_s\}}g(t))\doteq\\
    &&\mathcal{L}^{\ast}_{s+n}(1,\ldots,s+n)g_{1+n}(t,\{x_1,\ldots,x_s\},x_{s+1},\ldots,x_{s+n})+\nonumber \\
   && \sum\limits_{\mathrm{P}:\,(\{x_1,\ldots,x_s\},x_{s+1},\ldots,x_{s+n})=X_{1}\bigcup X_2}\,
      \sum\limits_{i_{1}\in\theta(\widehat{X}_{1})}
      \sum\limits_{i_{2}\in\theta(\widehat{X}_{2})}\mathcal{L}_{\mathrm{int}}^{\ast}(i_{1},i_{2})
      g_{|X_{1}|}(t,X_{1})g_{|X_{2}|}(t,X_{2}),\nonumber\\
   &&\hskip-12mm n\geq0,\nonumber
\end{eqnarray}
where the sequence of solutions of generalized cluster expansions (\ref{a8}) is denoted
by means of the mapping
\begin{eqnarray*}
   &&\hskip-12mm(\mathfrak{d}_{\{x_1,\ldots,x_s\}}g)_{n}(x_1,\ldots,x_n)\doteq
        g_{1+n}(\{x_1,\ldots,x_s\},x_{s+1},\ldots,x_{s+n}),\quad n\geq0,
\end{eqnarray*}
and we also used the notations adopted above in expansion (\ref{gLhs}).

We note that on the allowed configurations correlation functions of hard-sphere clusters can be expressed
through correlation functions of hard spheres (\ref{gLhs}) by the following relations:
\begin{eqnarray}\label{gClusters}
    &&\hskip-12mm g_{1+n}(t,\{x_1,\ldots,x_s\},x_{s+1},\ldots,x_{s+n})=\\
    &&\sum_{\mbox{\scriptsize$\begin{array}{c}\mathrm{P}:(\{x_1,\ldots,x_s\},\\
      x_{s+1},\ldots,x_{s+n})=\bigcup_i X_i\end{array}$}}(-1)^{|\mathrm{P}|-1}(|\mathrm{P}|-1)!
      \prod_{X_i\subset\mathrm{P}}\,\sum\limits_{\mathrm{P'}:\,\theta(X_{i})=\bigcup_{j_i} Z_{j_i}}\,
      \prod_{Z_{j_i}\subset\mathrm{P'}}g_{|Z_{j_i}|}(t,Z_{j_i}),\nonumber\\
    &&\hskip-12mm n\geq0.\nonumber
\end{eqnarray}
In particular case $n=0$, i.e., the correlation function of a cluster of the $s$ hard spheres,
these relations take the form
\begin{eqnarray*}
  &&\hskip-12mm g_{1+0}(t,\{x_1,\ldots,x_s\})=\sum\limits_{\mathrm{P}:\,\theta(\{x_1,\ldots,x_s\})=\bigcup_{i} X_{i}}\,
      \prod_{X_{i}\subset \mathrm{P}}g_{|X_{i}|}(t,X_{i}).\nonumber
\end{eqnarray*}

As a consequence of these relations, for the initial state satisfying the chaos condition,
from (\ref{rozvL-Nclusters}) the following generalization of expansions (\ref{gth}) holds:
\begin{eqnarray}\label{gcph}
   &&\hskip-12mm g_{s+n}(t,\{x_1,\ldots,x_s\},x_{s+1},\ldots,x_{s+n})=\\
   &&\hskip-5mm \mathfrak{A}_{1+n}(t,\{1,\ldots,s\},s+1,\ldots,s+n)\,
      \prod\limits_{i=1}^{s+n}g_{1}^{0}(x_i)\mathcal{X}_{\mathbb{R}^{3(s+n)}\setminus\mathbb{W}_{s+n}},\quad
    s\geq1,\,n\geq0.\nonumber
\end{eqnarray}

As we noted above, the possibility of the description of the evolution of a state based on the dynamics
of correlations (\ref{FClusters}) occurs naturally in consequence of dividing the series of expressions
(\ref{ms}) by the series of the normalizing factor. To provide evidence of this statement, we will
introduce the necessary notions and prove the validity of some auxiliary equalities.

On sequences of functions $f,\widetilde{f}\in L^{1}\oplus_{n=0}^\infty L^{1}_n$ we define the following
$\ast$-product \cite{R69}
\begin{eqnarray}\label{Product}
    (f\ast\widetilde{f})_{s}(x_1,\ldots,x_s)=\sum\limits_{Z\subset (x_1,\ldots,x_s)}\,f_{|Z|}(Z)
        \,\widetilde{f}_{s-|Z|}((x_1,\ldots,x_s)\setminus Z),
\end{eqnarray}
where $\sum_{Z\subset(x_1,\ldots,x_s)}$ is the sum over all subsets $Z$ of the set $(x_1,\ldots,x_s)$.
Using the definition of the $\ast$-product (\ref{Product}), we introduce the mapping
${\mathbb E}\mathrm{xp}_{\ast}$ and the inverse mapping ${\mathbb L}\mathrm{n}_{\ast}$ on sequences
$h=(0,h_1(x_1),\ldots,h_n(x_1,\ldots,$ $x_n),\ldots)$ of functions $h_n\in L^{1}_n$ by the expansions:
\begin{eqnarray}\label{circledExp}
   &&\hskip-12mm({\mathbb E}\mathrm{xp}_{\ast}\,h)_{s}(x_1,\ldots,x_s)=\big(\mathbb{I}+
      \sum\limits_{n=1}^{\infty}\frac{h^{\ast n}}{n!}\big)_{s}(x_1,\ldots,x_s)=\\
   &&\hskip+12mm\delta_{s,0}+\sum\limits_{\mathrm{P}:\,(x_1,\ldots,x_s)=\bigcup_{i}X_{i}}\,
      \prod_{X_i\subset \mathrm{P}}h_{|X_i|}(X_i),\nonumber
\end{eqnarray}
where we used the notations accepted in formula (\ref{D_(g)N}), $\mathbb{I}=(1,0,\ldots,0,\ldots)$
and $\delta_{s,0}$ is the Kronecker symbol, and respectively,
\begin{eqnarray}\label{circledLn}
   &&\hskip-12mm({\mathbb L}\mathrm{n}_{\ast}(\mathbb{I}+h))_{s}(x_1,\ldots,x_s)=
       \big(\sum\limits_{n=1}^{\infty} (-1)^{n-1}\,\frac{h^{\ast n}}{n}\big)_{s}(x_1,\ldots,x_s)=\\
   && \sum\limits_{\mathrm{P}:\,(x_1,\ldots,x_s)=\bigcup_{i}X_{i}}(-1)^{|\mathrm{P}|-1}(|\mathrm{P}|-1)!\,
       \prod_{X_i\subset\mathrm{P}}h_{|X_i|}(X_i).\nonumber
\end{eqnarray}
Therefore in terms of sequences of functions recursion relations (\ref{D_(g)N}) are rewritten
in the form
\begin{eqnarray*}\label{DtoGcircledStar}
    &&\hskip-8mm D(t)={\mathbb E}\mathrm{xp}_{\ast}\,\,g(t),
\end{eqnarray*}
where $D(t)=\mathbb{I}+(0,D_1(t,x_1),\ldots,D_n(t,x_1,\ldots,x_n),\ldots)$. As a result, we get
\begin{eqnarray*}
    &&\hskip-8mm g(t)={\mathbb L}\mathrm{n}_{\ast}\,\,D(t).
\end{eqnarray*}

Thus, according to definition (\ref{Product}) of the $\ast$-product and mapping (\ref{circledLn}),
in the component-wise form solutions of recursion relations (\ref{D_(g)N}) are represented by
expansions (\ref{gfromDFB}).

For arbitrary $f=(f_{0},f_{1},\ldots,f_{n},\ldots)\in L^{1}$ and the set $Y\equiv(x_1,\ldots,x_s)$
we define the linear mapping $\mathfrak{d}_{Y}:f\rightarrow \mathfrak{d}_{Y}f$, by the formula
\begin{eqnarray}\label{oper_d}
   &&\hskip-12mm (\mathfrak{d}_{Y}f)_{n}(x_1,\ldots,x_n)\doteq f_{s+n}(x_1,\ldots,x_s,x_{s+1},\ldots,x_{s+n}),
       \quad n\geq0.
\end{eqnarray}
For the set $\{Y\}$ consisting of the one element $Y=(x_1,\ldots,x_s)$, we have, respectively
\begin{eqnarray}\label{oper_c}
   &&\hskip-14mm(\mathfrak{d}_{\{Y\}}f)_{n}(x_1,\ldots,x_n)\doteq
        f_{1+n}(\{x_1,\ldots,x_s\},x_{s+1},\ldots,x_{s+n}),\,\, n\geq0.
\end{eqnarray}
On sequences $\mathfrak{d}_{Y}f$ and $\mathfrak{d}_{Y'}\widetilde{f}$ we introduce the $\ast$-product
\begin{eqnarray*}
    &&\hskip-9mm (\mathfrak{d}_{Y}f\ast\mathfrak{d}_{Y'}\widetilde{f})_{|X|}(X)\doteq
       \sum\limits_{Z\subset X}f_{|Z|+|Y|}(Y,Z)\,\widetilde{f}_{|X\backslash Z|+|Y'|}(Y',X\setminus Z),
\end{eqnarray*}
where $X,Y,Y'$ are the sets, which characterize clusters of hard spheres, and $\sum_{Z\subset X}$
is the sum over all subsets $Z$ of the set $X$. In particular case $Y=\emptyset,\,Y'=\emptyset$,
this definition reduces to the definition of $\ast$-product (\ref{Product}).

Let us establish some properties of introduced mappings (\ref{circledExp}) and (\ref{oper_c}).

If $f_{n}\in L^{1}_n,\,n\geq 1$ for the sequences $f=(0,f_{1},\ldots,f_{n},\ldots)$, according to
definitions of mappings (\ref{circledExp}) and (\ref{oper_c}), the following equality holds
\begin{eqnarray}\label{d_gamma}
    &&\mathfrak{d}_{\{Y\}}\mathbb{E}\mathrm{xp}_{\ast}f=
      \mathbb{E}\mathrm{xp}_{\ast}f\ast\mathfrak{d}_{\{Y\}}f,
\end{eqnarray}
and for mapping (\ref{oper_d}), respectively
\begin{eqnarray*}
    &&\mathfrak{d}_{Y}\mathbb{E}\mathrm{xp}_{\ast}f=
       \mathbb{E}\mathrm{xp}_{\ast} f\ast\sum\limits_{\mathrm{P}:\,Y=\bigcup_i X_{i}}
       \mathfrak{d}_{X_1}f\ast\ldots\ast \mathfrak{d}_{X_{|\mathrm{P}|}}f,
\end{eqnarray*}
where ${\sum\limits}_{\mathrm{P}:\,Y=\bigcup_i X_{i}}$ is the sum over all possible partitions
$\mathrm{P}$ of the set $Y\equiv(x_1,\ldots,x_s)$ into $|\mathrm{P}|$ nonempty mutually disjoint
subsets $X_i\subset Y$.

Hence in terms of mappings (\ref{oper_d}) and (\ref{oper_c}) generalized cluster expansions
(\ref{a8}) take the form
\begin{eqnarray}\label{gcea}
    &&\mathfrak{d}_{Y}D(t)=\mathfrak{d}_{\{Y\}}{\mathbb E}\mathrm{xp}_{\ast}\,\,g(t).
\end{eqnarray}

On sequences of functions $f\in L^{1}=\oplus_{n=0}^\infty L^{1}_n$ we also define the analogue
of the annihilation operator
\begin{eqnarray}\label{a1}
    &&(\mathfrak{a}f)_{n}(x_1,\ldots,x_n)=
      \int\limits_{\mathbb{R}^3\times\mathbb{R}^3}dx_{n+1}f_{n+1}(x_1,\ldots,x_n,x_{n+1}).
\end{eqnarray}
Then for sequences $f,\widetilde{f}\in L^{1}$, the following equality holds
\begin{eqnarray}\label{efg}
    &&(e^\mathfrak{a}f\ast\widetilde{f})_0=(e^\mathfrak{a}f)_0(e^\mathfrak{a}\widetilde{f})_0,
\end{eqnarray}
where such a notation was used
\begin{eqnarray}\label{series}
    &&(e^\mathfrak{a}f)_0=\sum\limits_{n=0}^{\infty}\frac{1}{n!}
       \int\limits_{(\mathbb{R}^{3}\times\mathbb{R}^{3})^{n}}dx_{1}\ldots dx_{n}\,f_{n}(x_1,\ldots,x_n).
\end{eqnarray}


Now let us prove the equivalence of definition (\ref{ms}) of the reduced distribution
functions and their definition (\ref{FClusters}) within the framework of the dynamics
of correlations.

In terms of mapping (\ref{oper_d}) and notation (\ref{series}) the definition of reduced distribution
functions (\ref{ms}) is written as follows
\begin{eqnarray*}
    &&F_{s}(t,x_1,\ldots,x_s)=(e^\mathfrak{a}D(t))^{-1}_0(e^\mathfrak{a}\mathfrak{d}_{Y}D(t))_0.
\end{eqnarray*}
Using generalized cluster expansions (\ref{gcea}), and as a consequence of equalities
(\ref{d_gamma}) and (\ref{efg}), we find
\begin{eqnarray*}
    &&(e^\mathfrak{a}\mathfrak{d}_{Y}D(t))_0=
       (e^\mathfrak{a}\mathfrak{d}_{\{Y\}}{\mathbb E}\mathrm{xp}_{\ast}\,g(t))_0=\\
    &&(e^\mathfrak{a}\mathbb{E}\mathrm{xp}_{\ast}g(t)\ast\mathfrak{d}_{\{Y\}}g(t))_0
       =(e^\mathfrak{a}\mathbb{E}\mathrm{xp}_{\ast}g(t))_0(e^\mathfrak{a}\mathfrak{d}_{\{Y\}}g(t))_0.
\end{eqnarray*}
Taking into account that, according to the particular case $Y=\emptyset,$ of cluster
expansions (\ref{a8}), the equality holds
\begin{eqnarray*}
    &&(e^\mathfrak{a}\mathbb{E}\mathrm{xp}_{\ast}g(t))_0=(e^\mathfrak{a}D(t))_0,
\end{eqnarray*}
as a result, we establish the following representation for the reduced distribution functions
\begin{eqnarray*}
    &&F_{s}(t,x_1,\ldots,x_s)=(e^\mathfrak{a}\mathfrak{d}_{\{Y\}}g(t))_0.
\end{eqnarray*}
Therefore, in componentwise-form we obtain relation (\ref{FClusters}).

Since the correlation functions $g_{1+n}(t),\,n\geq0,$ are governed by the corresponding
Liouville hierarchy for the cluster of hard spheres and hard spheres, the reduced distribution
functions (\ref{FClusters}) are governed by the BBGKY hierarchy for hard spheres
\begin{eqnarray}\label{BBG}
    &&\frac{\partial}{\partial t}F(t)=e^{\mathfrak{a}}\mathcal{L}(\{\cdot\},\cdot\mid e^{-\mathfrak{a}}F(t)),
\end{eqnarray}
where the operator $\mathcal{L}(\{\cdot\},\cdot\mid f)$ is generator (\ref{LcL}) of the Liouville hierarchy
for a cluster of hard spheres and hard spheres. For a generator of this hierarchy of evolution equations
takes place the following representation:
\begin{eqnarray*}
    &&e^{\mathfrak{a}}\mathcal{L}(\{\cdot\},\cdot\mid e^{-\mathfrak{a}}F(t))=
      e^\mathfrak{a}\mathcal{L}^\ast e^{-\mathfrak{a}}F(t),
\end{eqnarray*}
where the operator $\mathcal{L}^\ast=\oplus_{n=0}^\infty\mathcal{L}_n^\ast$ is a direct sum of the Liouville
operators and the operator $\mathfrak{a}$ is defined by formula (\ref{a1}). Due to the fact that
pairwise collisions occur during the evolution, a generator of this hierarchy is reduced to the operator
of such a structure \cite{CGP97}
\begin{eqnarray*}
    &&e^\mathfrak{a}\mathcal{L}^\ast e^{-\mathfrak{a}}=\mathcal{L}^\ast + [\mathfrak{a},\mathcal{L}^\ast],
\end{eqnarray*}
where, as above, the bracket $[\cdot,\cdot]$ is the commutator of operators.

We note that  for the first time the BBGKY hierarchy for many hard spheres (\ref{BBG}) was mathematically
justified in paper \cite{PG90} (see also \cite{CGP97}).

In consequence of definition (\ref{FClusters}) and the cumulant structure of representation of a
solution (\ref{gLhs}) for the Liouville hierarchy (\ref{Lh}), if the initial state specified by the sequence
of reduced distribution functions $F(0)=(1,F_{1}^{0}(x_1),\ldots,F_{n}^{0}(x_1,\ldots,x_n),\ldots)$,
then the evolution of all possible states, i.e., a sequence of the reduced distribution functions
$F_{s}(t),\,s\geq1$, is determined by the series expansions (\ref{sec}).

We remark that the representation (\ref{sec}) is directly established for the initial
states satisfying the chaos condition due to the validity in this case of the representation
(\ref{gcph}) for the correlation functions of the hard-sphere cluster and of the hard spheres.

Consequently, as follows from the above, the cumulant structure of generating operators of
expansions for correlation functions (\ref{gLhs}) or (\ref{rozvL-Nclusters}) induces the cumulant
structure (\ref{cumulant}) of generating operators of series expansions for reduced distribution
functions (\ref{sec}) or in other words, the evolution of the state of a system of an infinite
number of hard spheres is governed by the dynamics of correlations on a microscopic scale.

Thus, we have established relation (\ref{FClusters}) between the reduced distribution functions
and correlation functions governed by the Liouville hierarchy.

\bigskip
\subsection{The hierarchy of nonlinear evolution equations \\ for reduced correlation functions}
As is known, on a microscopic scale, the macroscopic characteristics of fluctuations of observables
are directly determined by means of the reduced correlation functions. Assuming as a basis an
alternative approach to the description of the evolution of states of a hard-sphere system within
the framework of correlation functions (\ref{gLhs}), then the reduced correlation functions are
defined by means of a solution of the Cauchy problem of the Liouville hierarchy (\ref{Lh}),(\ref{Lhi})
as follows \cite{GG22}:
\begin{eqnarray}\label{Gexpg}
   &&\hskip-12mm G_{s}(t,x_1,\ldots,x_s)\doteq \sum\limits_{n=0}^{\infty}\frac{1}{n!}\,
      \int\limits_{(\mathbb{R}^{3}\times\mathbb{R}^{3})^{n}}dx_{s+1}\ldots dx_{s+n}
      \,g_{s+n}(t,x_1,\ldots,x_{s+n}),\quad s\geq1,
\end{eqnarray}
where the generating function $g_{s+n}(t,x_1,\ldots,x_{s+n})$ is defined by expansion (\ref{gLhs}),
or in terms of mapping (\ref{oper_d}) and notation (\ref{series}) this definition takes the form
\begin{eqnarray*}
   &&G_{s}(t,x_1,\ldots,x_s)=(e^\mathfrak{a}\mathfrak{d}_{Y}g(t))_0,
\end{eqnarray*}
or in terms of sequences of functions this expression has the form
\begin{eqnarray*}
   &&G(t)=e^\mathfrak{a}g(t).
\end{eqnarray*}
We emphasize that $nth$ term of expansions (\ref{Gexpg}) of the reduced correlation functions are
determined by the $(s+n)th$-particle correlation function (\ref{gLhs}) in contrast with the expansions
of reduced distribution functions (\ref{FClusters}) which are determined by the $(1+n)th$-particle
correlation function of clusters of hard spheres (\ref{rozvL-Nclusters}).

Such a representation for reduced correlation functions (\ref{Gexpg}) can be derived as a result
of the fact that the reduced correlation functions are cumulants of reduced distribution functions
(\ref{FClusters}). Indeed, traditionally reduced correlation functions are introduced by means of
the cluster expansions of the reduced distribution functions similar to the cluster expansions of
the probability distribution functions (\ref{D_(g)N}) and on the set of allowed configurations
$\mathbb{R}^{3n}\setminus \mathbb{W}_n$ they have the form:
\begin{eqnarray}\label{FG}
   &&\hskip-12mm F_{s}(t,x_1,\ldots,x_s)=
      \sum\limits_{\mbox{\scriptsize$\begin{array}{c}\mathrm{P}:(x_1,\ldots,x_s)=\bigcup_{i}X_{i}\end{array}$}}
      \prod_{X_i\subset\mathrm{P}}G_{|X_i|}(t,X_i), \, s\geq1,
\end{eqnarray}
where, as above, the symbol ${\sum\limits}_{\mathrm{P}:(x_1,\ldots,x_s)=\bigcup_{i} X_{i}}$ is the sum
over all possible partitions $\mathrm{P}$ of the set $(x_1,\ldots,x_s)$ into $|\mathrm{P}|$ nonempty
mutually disjoint subsets $X_i\subset(x_1,\ldots,x_s)$. As a consequence of this, the solution of
recurrence relations (\ref{FG}) are represented through reduced distribution functions as follows:
\begin{eqnarray}\label{gBigfromDFB}
   &&\hskip-12mm G_{s}(t,x_1,\ldots,x_s)=
      \sum\limits_{\mbox{\scriptsize $\begin{array}{c}\mathrm{P}:(x_1,\ldots,x_s)=\bigcup_{i}X_{i}\end{array}$}}
     (-1)^{|\mathrm{P}|-1}(|\mathrm{P}|-1)!\prod_{X_i\subset\mathrm{P}}F_{|X_i|}(t,X_i), \quad s\geq1.
\end{eqnarray}
Functions (\ref{gBigfromDFB}) are interpreted as the functions which describe the correlations of
hard-sphere states. The structure of expansions (\ref{gBigfromDFB}) is such that the reduced correlation
functions are cumulants (semi-invariants) of the reduced distribution functions (\ref{sec}).

Thus, taking into account representation (\ref{FClusters}) of the reduced distribution functions,
in consequence of the validity of relations (\ref{gClusters}) we derive representation (\ref{Gexpg})
of the reduced correlation functions through correlation functions
\begin{eqnarray*}
    &&\hskip-12mm G_{s}(t,x_1,\ldots,x_s)=
      \sum\limits_{\mathrm{P}:(x_1,\ldots,x_s)=\bigcup_{i}X_{i}}(-1)^{|\mathrm{P}|-1}(|\mathrm{P}|-1)!\,
      \prod_{X_i\subset\mathrm{P}}(e^\mathfrak{a}\mathfrak{d}_{\{X_i\}}g(t))=\\
   &&\hskip-5mm (e^\mathfrak{a}\mathfrak{d}_{(x_1,\ldots,x_s)}g(t))_0.
\end{eqnarray*}

Since the correlation functions $g_{s+n}(t),\,n\geq0,$ are governed by the Liouville hierarchy
for hard spheres (\ref{Lh}), the reduced correlation functions defined as (\ref{Gexpg}) are
governed by the hierarchy of nonlinear equations for hard spheres (the nonlinear BBGKY hierarchy) \cite{GG22}:
\begin{eqnarray}\label{gBigfromDFBa}
   &&\hskip-12mm\frac{\partial}{\partial t}G_s(t,x_1,\ldots,x_s)=\mathcal{L}^{\ast}_{s}G_{s}(t,x_1,\ldots,x_s)+\\
   &&\hskip-5mm \sum\limits_{\mathrm{P}:\,(x_1,\ldots,x_s)=X_{1}\bigcup X_2}\,\sum\limits_{i_{1}\in\widehat{X}_{1}}
      \sum\limits_{i_{2}\in \widehat{X}_{2}}\mathcal{L}_{\mathrm{int}}^{\ast}(i_{1},i_{2})
      G_{|X_{1}|}(t,X_{1})G_{|X_{2}|}(t,X_{2}))+\nonumber\\
   &&\hskip-5mm \int\limits_{\mathbb{R}^{3}\times\mathbb{R}^{3}}dx_{s+1}
      \big(\sum_{i=1}^{s}\mathcal{L}^{\ast}_{\mathrm{int}}(i,s+1)G_{s+1}(t,x_1,\ldots,x_{s+1})+\nonumber\\
   &&\hskip-5mm  \sum\limits_{\mathrm{P}:\,(x_1,\ldots,x_{s+1})=X_{1}\bigcup X_2}
      \sum_{i\in\widehat{X}_1;s+1\in \widehat{X}_2}
      \mathcal{L}^{\ast}_{\mathrm{int}}(i,s+1)G_{|X_1|}(t,X_1)G_{|X_2|}(t,X_2)\big), \nonumber\\ \nonumber\\
 \label{gBigfromDFBai}
   &&\hskip-12mmG_{s}(t,x_1,\ldots,x_s)\big|_{t=0}=G_{s}^{0}(x_1,\ldots,x_s), \quad s\geq1,
\end{eqnarray}
where the symbol
$\sum_{\mbox{\scriptsize $\begin{array}{c}\mathrm{P}:(x_1,\ldots,x_{s+1})=X_1\bigcup X_2,\end{array}$}}$
means the sum over all possible partitions of the set $(x_1,\ldots,x_{s+1})$ into two mutually disjoint
subsets $X_1$ and $X_2$, the sum over the index $i$ which takes values from the subset $\widehat{X}_1$
provided that the index $s+1$ belongs to the subset $\widehat{X}_2$ is denoted by
$\sum_{\mbox{\scriptsize $\begin{array}{c}i\in\widehat{X}_1;s+1\in\widehat{X}_2\end{array}$}}$ and
notations accepted in the Liouville hierarchy (\ref{Lh}) are used.

A generator of this hierarchy of nonlinear evolution equations has the following structure:
\begin{eqnarray*}\label{B}
    &&\frac{\partial}{\partial t}G(t)=e^{\mathfrak{a}}\mathcal{L}(\cdot\mid e^{-\mathfrak{a}}G(t)),
\end{eqnarray*}
where the operator $\mathcal{L}(\cdot\mid f)=\oplus_{n=0}^\infty\mathcal{L}(1,\ldots,n\mid f)$
is a direct sum of generators (\ref{LL}) of the Liouville hierarchy (\ref{Lh}).
Here are some component-wise examples of hierarchy (\ref{gBigfromDFBa}):
\begin{eqnarray*}
   &&\hskip-12mm\frac{\partial}{\partial t}G_1(t,x_1)=\mathcal{L}^{\ast}_{1}(1)G_{1}(t,x_1)+\\
   && \int\limits_{\mathbb{R}^{3}\times\mathbb{R}^{3}}dx_{2}
      \mathcal{L}^{\ast}_{\mathrm{int}}(1,2)\big(G_{2}(t,x_1,x_{2})+
      G_{1}(t,x_1)G_{1}(t,x_2)\big),\nonumber\\
   &&\hskip-12mm\frac{\partial}{\partial t}G_2(t,x_1,x_2)=\mathcal{L}^{\ast}_{2}(1,2)G_{2}(t,x_1,x_2)+
        \mathcal{L}_{\mathrm{int}}^{\ast}(1,2)G_{1}(t,x_{1})G_{1}(t,x_{2})+\nonumber\\
   && \int\limits_{\mathbb{R}^{3}\times\mathbb{R}^{3}}dx_{3}
     \Big(\sum_{i=1}^2\mathcal{L}^{\ast}_{\mathrm{int}}(i,3)\big(G_{3}(t,x_1,x_2,x_{3})+
     G_{2}(t,x_1,x_2)G_{1}(t,x_3)\big)+\nonumber\\
   && \mathcal{L}^{\ast}_{\mathrm{int}}(2,3)G_{2}(t,x_1,x_3)G_{1}(t,x_2)+
      \mathcal{L}^{\ast}_{\mathrm{int}}(1,3)G_{2}(t,x_2,x_3)G_{1}(t,x_1)\Big),\nonumber
\end{eqnarray*}
where it was used notations accepted above in definition (\ref{L}).

If $G(0)=(1,G_1^{0}(x_1),\ldots,G_s^{0}(x_1,\ldots,x_s),\ldots)$ is a sequence of reduced correlation
functions at the initial instant, then by means of mappings (\ref{Sspherc}) the evolution of all possible
states, i.e., the sequence of the reduced correlation functions $G_{s}(t),\,s\geq1$, is determined by
the following series expansions:
\begin{eqnarray}\label{sss}
    &&\hskip-12mm G_{s}(t,x_1,\ldots,x_s)=\\
    &&\hskip-5mm \sum\limits_{n=0}^{\infty}\frac{1}{n!}
        \,\int\limits_{(\mathbb{R}^{3}\times\mathbb{R}^{3})^{n}}dx_{s+1}\ldots dx_{s+n}\,
        \,\mathfrak{A}_{1+n}(t;\{1,\ldots,s\}, s+1,\ldots,s+n\mid G(0)), \quad s\geq1,\nonumber
\end{eqnarray}
where the generating operator $\mathfrak{A}_{1+n}(t;\{1,\ldots,s\},s+1,\ldots,s+n\mid G(0))$ of this
series is the $(1+n)th$-order cumulant of groups of nonlinear operators (\ref{gLhs}):
\begin{eqnarray}\label{cc}
   &&\hskip-12mm\mathfrak{A}_{1+n}(t;\{1,\ldots,s\},s+1,\ldots,s+n\mid G(0))\doteq\\
   &&\hskip-5mm \sum\limits_{\mathrm{P}:\,(\{1,\ldots,s\},s+1,\ldots,s+n)=
      \bigcup_k X_k}(-1)^{|\mathrm{P}|-1}({|\mathrm{P}|-1})!
      \mathcal{G}(t;\theta(X_1)\mid\ldots \mathcal{G}(t;\theta(X_{|\mathrm{P}|})\mid G(0))\ldots), \nonumber\\
   &&\hskip-12mm   n\geq0,\nonumber
\end{eqnarray}
and where the composition of mappings (\ref{gLhs}) of the corresponding noninteracting groups of
particles was denoted by
$\mathcal{G}(t;\theta(X_1)\mid \ldots\mathcal{G}(t;\theta(X_{|\mathrm{P}|})\mid G(0))\ldots)$,
for example,
\begin{eqnarray*}
    &&\hskip-12mm\mathcal{G}\big(t;1\mid\mathcal{G}(t;2\mid G(0))\big)=
        \mathfrak{A}_{1}(t,1)\mathfrak{A}_{1}(t,2)G^{0}_{2}(x_1,x_2),\\
    &&\hskip-12mm\mathcal{G}\big(t;1,2\mid\mathcal{G}(t;3\mid G(0))\big)=
        \mathfrak{A}_{1}(t,\{1,2\})\mathfrak{A}_{1}(t,3)G^{0}_{3}(x_1,x_2,x_3)+\\
    &&\hskip+3mm\mathfrak{A}_{2}(t,1,2)\mathfrak{A}_{1}(t,3)
       \big(G^{0}_{1}(x_1)G^{0}_{2}(x_2,x_3)+G^{0}_{1}(x_2)G^{0}_{2}(x_1,x_3)\big).
\end{eqnarray*}

We will adduce examples of expansions (\ref{cc}). The first order cumulant of the groups
of nonlinear operators (\ref{gLhs}) is the group of these nonlinear operators
\begin{eqnarray*}
     &&\hskip-8mm \mathfrak{A}_{1}(t;\{1,\ldots,s\}\mid G(0))=\mathcal{G}(t;1,\ldots,s\mid G(0)).
\end{eqnarray*}
In case of $s=2$ the second order cumulant of nonlinear operators (\ref{gLhs}) has the structure
\begin{eqnarray*}
     &&\hskip-12mm \mathfrak{A}_{1+1}(t;\{1,2\},3\mid G(0))=\mathcal{G}(t;1,2,3\mid G(0))-
       \mathcal{G}\big(t;1,2\mid\mathcal{G}(t;3\mid G(0))\big)=\\
     &&\mathfrak{A}^\ast_{1+1}(t,\{1,2\},3)G^{0}_{3}(1,2,3)+\\
     && \big(\mathfrak{A}^\ast_{1+1}(t,\{1,2\},3)-
        \mathfrak{A}_{2}(t,2,3)\mathfrak{A}^\ast_{1}(t,1)\big)G^{0}_{1}(x_1)G^{0}_{2}(x_2,x_3)+\\
     &&\big(\mathfrak{A}_{1+1}(t,\{1,2\},3)-
        \mathfrak{A}^\ast_{2}(t,1,3)\mathfrak{A}^\ast_{1}(t,2)\big)G^{0}_{1}(x_2)G^{0}_{2}(x_1,x_3)+\\
     && \mathfrak{A}^\ast_{1+1}(t,\{1,2\},3)G^{0}_{1}(x_3)G^{0}_{2}(x_1,x_2)+
        \mathfrak{A}^\ast_{3}(t,1,2,3)G^{0}_{1}(x_1)G^{0}_{1}(x_2)G^{0}_{1}(x_3),
\end{eqnarray*}
where the operator $ \mathfrak{A}^\ast_{3}(t,1,2,3)=\mathfrak{A}^\ast_{1+1}(t,\{1,2\},3)-
       \mathfrak{A}^\ast_{2}(t,2,3)\mathfrak{A}^\ast_{1}(t,1)-
       \mathfrak{A}^\ast_{2}(t,1,3)\mathfrak{A}^\ast_{1}(t,2)$
is the third-order cumulant (\ref{cumcp}) of groups of operators (\ref{S*}) of a system
of hard spheres.

The following statement is true \cite{GG22}.

\begin{theorem*}
\emph{Let $G(0)\in\oplus_{n=0}^{\infty}L^{1}_{n}$, then for arbitrary $t\in\mathbb{R}$ provided
that $\max_{n\geq1}\big\|G_n^{0}\big\|_{L^{1}_{n}}<(2e^{3})^{-1},$ the sequence of
reduced correlation functions (\ref{sss}) is a unique solution of the Cauchy problem
of nonlinear hierarchy (\ref{gBigfromDFBa}),(\ref{gBigfromDFBai}) for hard spheres.}
\end{theorem*}

In the particular case of the initial state specified by the sequence of reduced correlation
functions $G^{(c)}=(0,G_1^{0},0,\ldots,0,\ldots)$ on the allowed configurations, that is, in
the absence of correlations between hard spheres at the initial moment of time \cite{CGP97},
according to definition (\ref{cc}) of the generating operators, reduced correlation functions
(\ref{sss}) are represented by the following series expansions:
\begin{eqnarray}\label{mcc}
   &&\hskip-12mm G_{s}(t,x_1,\ldots,x_s)=\\
    &&\hskip-8mm\sum\limits_{n=0}^{\infty}\frac{1}{n!}
     \,\int\limits_{(\mathbb{R}^{3}\times\mathbb{R}^{3})^{n}}dx_{s+1}\ldots dx_{s+n}\,
     \mathfrak{A}^\ast_{s+n}(t;1,\ldots,s+n)\prod_{i=1}^{s+n}G_1^{0}(x_i)
      \mathcal{X}_{\mathbb{R}^{3(s+n)}\setminus \mathbb{W}_{s+n}}, \quad s\geq1, \nonumber
\end{eqnarray}
where the generating operator $\mathfrak{A}^\ast_{s+n}(t)$ is the $(s+n)th$-order cumulant (\ref{cumcp})
of the groups of operators (\ref{S*}).

We emphasize that in the absence of correlations of states of hard spheres on allowed configurations
at the initial moment of time, the generators of expansions into a series of reduced correlation
functions (\ref{mcc}) and reduced distribution functions (\ref{sec}) differ only in the order
of cumulants of groups of operators of hard spheres. Therefore, by means of such reduced distribution
functions or reduced correlation functions, the process of creating correlations in a system of hard
spheres is described.

We note that the reduced correlation functions give an equivalent approach to the description of the
evolution of states of many hard spheres, along with the reduced distribution functions. Indeed, the
macroscopic characteristics of fluctuations of observables are directly determined by the reduced
correlation functions on the microscopic scale \cite{B1} for example, the functional of the dispersion
of an additive-type observable, i.e., the sequence
$A^{(1)}=(0,a_{1}(x_1),\ldots,\sum_{i_{1}=1}^{n}a_1(x_{i_{1}}),\ldots)$, is represented by the formula
\begin{eqnarray*}
    &&\hskip-12mm \langle(A^{(1)}-\langle A^{(1)}\rangle)^2\rangle(t)=\\
    &&\int\limits_{\mathbb{R}^{3}\times\mathbb{R}^{3}}dx_{1}\,(a_1^2(x_1)-
      \langle A^{(1)}\rangle^2(t))G_{1}(t,x_1)+
      \int\limits_{(\mathbb{R}^{3}\times\mathbb{R}^{3})^2}dx_{1}dx_{2}\,a_{1}(x_1)a_{1}(x_2)G_{2}(t,x_1,x_2),
\end{eqnarray*}
where
\begin{eqnarray*}
    &&\hskip-12mm \langle A^{(1)}\rangle(t)=\int_{\mathbb{R}^{3}\times\mathbb{R}^{3}}dx_{1}\,a_{1}(x_1)G_{1}(t,x_1)
\end{eqnarray*}
is the mean value functional of an additive-type observable.


\textcolor{blue!55!black}{\section{Nonlinear kinetic equations for many hard spheres}}

The conventional philosophy of the description of kinetic evolution is that if the initial
state is specified by a one-particle (reduced) distribution function, then at an arbitrary
time the evolution of the state in an appropriate scaling limit can be effectively described
by means of a one-particle distribution function that is governed by the nonlinear kinetic
equation. Below, we give an answer to the question about the description of the kinetic
evolution of colliding particles, not on the basis of a common interpretation but within the
framework of the evolution of the observables of many hard spheres.

The problem of a rigorous description of the kinetic evolution by means of hard sphere
observables will be considered by giving the example of the Boltzmann--Grad asymptotics
of a non-perturbative solution of the Cauchy problem of the dual BBGKY hierarchy \cite{GG18}.

\bigskip
\subsection{On the Boltzmann--Grad scaling approximation}
The present notion of the Boltzmann--Grad approximation was first introduced in Grad's paper
\cite{G49}. From a physical point of view, this approximation means that we deal with a
low-density gas in a situation where the diameter of a hard sphere, or, in other words, the
radius of the short-range interaction potential, is sufficiently less in comparison with the
average length of a free path of hard spheres.

In a dimensionless form, the generator of the BBGKY hierarchy for hard spheres contains a
scaling parameter: the ratio of the diameter of hard spheres to their mean free path \cite{G13}.
The finite value of the mean free path of hard spheres means that in this approximation the
average number of particles tends to infinity; in other words, according to the definition
of \eqref{N}, the state must be described by functions from the appropriate function spaces,
for example, from the space to which the sequences of reduced equilibrium distribution
functions belong \cite{GPM},\cite{R69}. In this case, the initial state is described by the
reduced distribution functions from the space $L^{\infty}_\xi$ of sequences of functions
bounded with respect to the  configuration variables and decreasing with respect to the
momentum ones, equipped with the norm
\begin{eqnarray*}
    &&\hskip-8mm \|f\|_{L^{\infty}_\xi}=
       \sup_{n\geq0}\xi^{-n}\sup_{x_1,\ldots,x_n}|f_n(x_1,\ldots,x_n)|\exp\big(\beta\sum_{i=1}^n\frac{p_i^2}{2}\big),
\end{eqnarray*}
where $\xi>0$ and $\beta>0$ are parameters.

For such initial data, the Boltzmann--Grad asymptotics of a solution of the Cauchy problem
of the BBGKY hierarchy for hard spheres are described by the so-called Boltzmann hierarchy
\cite{L75}. As a consequence, for factorized initial data, i.e., for  the initial state
without correlations, which describes molecular chaos \cite{CGP97}, the equation determining
the evolution of an initial state is a closed equation for a one-particle distribution function,
that is to say Boltzmann's kinetic equation \cite{CC}.

The detailed analysis of the problem of the construction of such asymptotics for a solution of
the Cauchy problem of the BBGKY hierarchy shows that the basic difficulty consists in proving
the term-by-term convergence of the iteration series that represents this solution to the
corresponding limit, that is, to the series representing the solution of the Cauchy problem of
the Boltzmann hierarchy.
This difficulty is related to the fact that the integrands in each term of the iteration series
do not converge to the limit uniformly across the whole domain of integration. We note that in
early works on the justification of the Boltzmann--Grad limit, attention was not properly paid
to this property, and a precise mathematical meaning was not given to the individual terms of
the iteration series representing a solution of the BBGKY hierarchy.
In the papers \cite{GP90},\cite{PG90} a complete discussion of these problems was presented.

From a mathematical point of view, the existence of the Boltzmann--Grad asymptotics of a perturbative
solution of the BBGKY hierarchy for hard spheres was discussed in Cercignani's paper \cite{C72} and
later in Lanford's work \cite{L75}. A rigorous mathematical proof of the Boltzmann--Grad limit theorem
has been given in a series of papers
\cite{GP85_D},\cite{GP85},\cite{GP87},\cite{GP90},\cite{PG90} by D.~Ya.~Petrina and V.~I.~Gerasimenko.
The Boltzmann--Grad limit theorem for equilibrium states was proved in the paper \cite{GP88}.

Recently, there has been unflagging interest in the problem of deriving kinetic equations from
the dynamics of many colliding particles as an asymptotic behavior of the BBGKY hierarchy in
the scaling limits. In particular, progress in the rigorous solution of this problem on the
basis of perturbation theory was achieved in the Boltzmann--Grad limit in the works
\cite{Bo2020},\cite{Bo2022},\cite{BGS-RS23},\cite{D16},\cite{DS},\cite{G19},\cite{PS16},\cite{PS17};
also see links therein.

\bigskip
\subsection{The Boltzmann--Grad limit of reduced observables}
To determine the scaling parameter, we rewrite the dual BBGKY hierarchy in dimensionless form.
Then generator (\ref{int}) of the hierarchy takes the form:
\begin{eqnarray}\label{com}
   &&\hskip-12mm \mathcal{L}(j)b_n\doteq \langle p_{j},\frac{\partial}{\partial q_{j}}\rangle b_n,\\
   &&\hskip-12mm \mathcal{L}_{\mathrm{int}}(j_1,j_{2})b_n\doteq \epsilon^{2}\int_{\mathbb{S}_+^2}d\eta\langle\eta,(p_{j_1}-p_{j_2})\rangle
     \big(b_n(x_1,\ldots,q_{j_1},p_{j_1}^*,\ldots,\nonumber\\
   &&\hskip+5mmq_{j_2},p_{j_2}^*,\ldots,x_n)-b_n(x_1,\ldots,x_n)\big)\delta(q_{j_1}-q_{j_2}+\epsilon\eta),\nonumber
\end{eqnarray}
where the coefficient $\epsilon>0$ is a scaling parameter, which is the ratio of the diameter
$\sigma>0$ to the mean free path of hard spheres. For $t\leq0$, a generator of the dimensionless dual BBGKY
hierarchy is determined by the corresponding expression \cite{GG18}.

Then the Boltzmann--Grad asymptotic behavior of dimensionless reduced observables (\ref{sed})
is described by the following statement \cite{GG18}.

\begin{theorem*}
{\it Assume that for the initial data $B_{n}^{\epsilon,0}\in\mathcal{C}_n,\, n\geq1,$ there is a limit
$b_{n}^0\in\mathcal{C }_n $ in the sense of $\ast$-weak convergence of space $\mathcal{C}_n$}
\begin{eqnarray}\label{asumdin}
    &&\mathrm{w^{\ast}-}\lim\limits_{\epsilon\rightarrow 0}\big(\epsilon^{-2n}B_{n}^{\epsilon,0}-b_{n}^0\big)=0.
\end{eqnarray}
{\it Then, for an arbitrary finite time interval, the Boltzmann--Grad limit of dimensionless reduced
observables (\ref{sed}) exists in the same sense}
\begin{eqnarray}\label{asymt}
    &&\mathrm{w^{\ast}-}\lim\limits_{\epsilon\rightarrow 0}\big(\epsilon^{-2s} B_{s}(t)-b_{s}(t)\big)=0,
\end{eqnarray}
{\it and it is determined by the expansions:}
\begin{eqnarray}\label{Iterd}
   &&\hskip-12mm b_{s}(t,x_1,\ldots,x_s)=
      \sum\limits_{n=0}^{s-1}\,\int\limits_0^tdt_{1}\ldots\int\limits_0^{t_{n-1}}dt_{n}
      \prod\limits_{j\in(1,\ldots,s)}S_{1}(t-t_{1},j)\sum\limits_{i_{1}\neq j_{1}=1}^{s}
      \mathcal{L}_{\mathrm{int}}^0(i_{1},j_{1})\times\\
   &&\hskip-5mm \prod\limits_{j\in(1,\ldots,s)\setminus (j_{1})}S_{1}(t_{1}-t_{2},j)
      \ldots\prod\limits_{j\in(1,\ldots,s)\setminus(j_{1},\ldots,j_{n-1})}S_{1}(t_{n-1}-t_{n},j)\times\nonumber \\
   &&\hskip-7mm \sum\limits^{s}_{\mbox{\scriptsize $\begin{array}{c}i_{n}\neq j_{n}=1,\\
      i_{n},j_{n}\neq(j_{1},\ldots,j_{n-1})\end{array}$}}
      \mathcal{L}_{\mathrm{int}}^0(i_{n},j_{n})\prod\limits_{j\in(1,\ldots,s)\setminus(j_{1},\ldots,j_{n})}S_{1}(t_{n},j) b_{s-n}^0((x_1,\ldots,x_s)\setminus(x_{j_{1}},\ldots,x_{j_{n}})), \nonumber\\
   &&\hskip-12mm s\geq1,\nonumber
\end{eqnarray}
{\it where for the collision operator of point particles, the notation $\mathcal{L}_{\mathrm{int}}^{0}(j_1,j_{2})$
is used}
\begin{eqnarray}\label{int0}
   &&\hskip-12mm \mathcal{L}_{\mathrm{int}}^{0}(j_1,j_{2})b_n\doteq
     \int_{\mathbb{S}_+^2}d\eta\langle\eta,(p_{j_1}-p_{j_2})\rangle
     \big(b_n(x_1,\ldots,q_{j_1},p_{j_1}^*,\ldots,q_{j_2},p_{j_2}^*,\ldots,x_n)-\\
   &&\hskip+9mmb_n(x_1,\ldots,x_n)\big)\delta(q_{j_1}-q_{j_2}).\nonumber
\end{eqnarray}
\end{theorem*}

Let us make several comments on this theorem.

Consider the existence of the Boltzmann--Grad limit for a special case of reduced observables,
namely additive-type reduced observables. Let's say that for the initial additive-type dimensionless
reduced observable $B^{(1)}(0)=(0,b_{1}^{\epsilon},0,\ldots)$ the following condition is satisfied:
\begin{eqnarray*}
    &&\hskip-5mm \mathrm{w^{\ast}-}\lim\limits_{\epsilon\rightarrow 0}\big(\epsilon^{-2}
        b_{1}^{\epsilon}-b_{1}^{0}\big)=0,
\end{eqnarray*}
then, according to statement (\ref{asymt}), for additive-type reduced observables (\ref{af})
we derive
\begin{eqnarray*}
    &&\hskip-5mm \mathrm{w^{\ast}-}\lim\limits_{\epsilon\rightarrow 0}
                 \big(\epsilon^{-2s}B_{s}^{(1)}(t)-b_{s}^{(1)}(t)\big)=0,
\end{eqnarray*}
where the limit reduced observable $b_{s}^{(1)}(t)$ is determined as a special case
of expansion (\ref{Iterd}):
\begin{eqnarray}\label{itvad}
   &&\hskip-12mm b_{s}^{(1)}(t,x_1,\ldots,x_s)=\int\limits_0^t dt_{1}\ldots\int\limits_0^{t_{s-2}}dt_{s-1}\,
       \prod\limits_{j\in(1,\ldots,s)}S_{1}(t-t_{1},j)\sum\limits_{i_{1}\neq j_{1}=1}^{s}
       \mathcal{L}_{\mathrm{int}}^0(i_{1},j_{1})\times \\
   &&\hskip-5mm \prod\limits_{j\in(1,\ldots,s)\setminus (j_{1})}S_{1}(t_{1}-t_{2},j)\ldots
       \prod\limits_{j\in(1,\ldots,s)\setminus(j_{1},\ldots,j_{s-2})}S_{1}(t_{s-2}-t_{s-1},j)\times \nonumber\\
   &&\hskip-5mm \sum\limits^{s}_{\mbox{\scriptsize $\begin{array}{c}i_{s-1}\neq j_{s-1}=1,\\
       i_{s-1},j_{s-1}\neq (j_{1},\ldots,j_{s-2})\end{array}$}}\hskip-2mm\mathcal{L}_{\mathrm{int}}^0(i_{s-1},j_{s-1})
       \prod\limits_{j\in(1,\ldots,s)\setminus(j_{1},\ldots,j_{s-1})}S_{1}(t_{s-1},j)\times\nonumber \\
   &&\hskip-5mm b_{1}^{0}\big((x_1,\ldots,x_s)\setminus(x_{j_{1}},\ldots,x_{j_{s-1}})\big),\quad s\geq1.\nonumber
\end{eqnarray}
We make several examples of expansions (\ref{itvad}) of the limit additive-type reduced
observables:
\begin{eqnarray*}
   &&\hskip-8mm b_{1}^{(1)}(t,x_1)=S_{1}(t,1)\,b_{1}^{0}(x_1),\\
   &&\hskip-8mm b_{2}^{(1)}(t,x_1,x_2)=\int_0^t dt_{1}\prod\limits_{i=1}^{2}S_{1}(t-t_{1},i)\,
      \mathcal{L}_{\mathrm{int}}^0(1,2)\sum\limits_{j=1}^{2}S_{1}(t_{1},j)\,b_{1}^{0}(x_j).
\end{eqnarray*}

Also suppose that the following condition is valid for the initial $k$-ary-type reduced
observable $B^{(k)}(0)=(0,\ldots,b_{k}^{\epsilon},0,\ldots)$:
\begin{eqnarray*}
    &&\hskip-5mm \mathrm{w^{\ast}-}\lim\limits_{\epsilon\rightarrow 0}\big(\epsilon^{-2}
        b_{k}^{\epsilon}-b_{k}^{0}\big)=0,
\end{eqnarray*}
then, according to statement (\ref{asymt}), for $k$-ary-type dimensionless reduced observables (\ref{af-k}),
we derive
\begin{eqnarray*}
    &&\hskip-5mm \mathrm{w^{\ast}-}\lim\limits_{\epsilon\rightarrow 0}
                 \big(\epsilon^{-2s}B_{s}^{(k)}(t)-b_{s}^{(k)}(t)\big)=0,
\end{eqnarray*}
where the limit reduced observable $b_{s}^{(k)}(t)$ is determined as a special case
of expansion (\ref{Iterd}):
\begin{eqnarray}\label{kIterd}
   &&\hskip-12mm b_{s}^{(k)}(t,x_1,\ldots,x_s)=\int\limits_0^tdt_{1}\ldots\int\limits_0^{t_{s-k-1}}dt_{s-k}
      \prod\limits_{j\in(1,\ldots,s)}S_{1}(t-t_{1},j)\sum\limits_{i_{1}\neq j_{1}=1}^{s}
      \mathcal{L}_{\mathrm{int}}^0(i_{1},j_{1})\times\\
   &&\hskip-5mm \prod\limits_{j\in(1,\ldots,s)\setminus (j_{1})}S_{1}(t_{1}-t_{2},j)
      \ldots\prod\limits_{j\in(1,\ldots,s)\setminus(j_{1},\ldots,j_{s-k-1})}S_{1}(t_{s-k-1}-t_{s-k},j)\times\nonumber \\
   &&\hskip-5mm \hskip-2mm\sum\limits^{s}_{\mbox{\scriptsize $\begin{array}{c}i_{s-k}\neq j_{s-k}=1,\\
      i_{s-k},j_{s-k}\neq (j_{1},\ldots,j_{s-k-1})\end{array}$}}\hskip-2mm\mathcal{L}_{\mathrm{int}}^0(i_{s-k},j_{s-k})
      \prod\limits_{j\in(1,\ldots,s)\setminus(j_{1},\ldots,j_{s-k})}S_{1}(t_{s-k},j)\times\nonumber \\
   &&\hskip-5mm b_{k}^0((x_1,\ldots,x_s)\setminus(x_{j_{1}},\ldots,x_{j_{s-k}})),\quad 1\leq s\leq k.\nonumber
\end{eqnarray}

If $b^0\in\mathcal{C}_{\gamma}$, then the sequence $b(t)=(b_0,b_1(t),\ldots,b_{s}(t),\ldots)$ of
limit reduced observables (\ref{Iterd}) is a generalized global solution of the Cauchy problem
of the dual Boltzmann hierarchy with hard sphere collisions \cite{GG18}:
\begin{eqnarray}\label{vdh}
   &&\hskip-12mm \frac{\partial}{\partial t}b_{s}(t)=
     \sum\limits_{j=1}^{s}\mathcal{L}(j)\,b_{s}(t)+
     \sum_{j_1\neq j_{2}=1}^s\mathcal{L}_{\mathrm{int}}^{0}(j_1,j_{2})\,b_{s-1}(t,(x_1,\ldots,x_s)\setminus(x_{j_1})), \\
  \nonumber \\
  \label{vdhi}
   &&\hskip-12mm  b_{s}(t,x_1,\ldots,x_s)\mid_{t=0}=b_{s}^0(x_1,\ldots,x_s),\quad s\geq1,
\end{eqnarray}
where it was used notations accepted in (\ref{Iterd}).

This fact is proved similar to the case of an iteration series of the dual BBGKY hierarchy \cite{BGer}.

It should be noted that equations set (\ref{vdh}) has the structure of recurrence evolution equations.
We make a few examples of the dual Boltzmann hierarchy with hard sphere collisions (\ref{vdh}):
\begin{eqnarray*}
    &&\hskip-12mm  \frac{\partial}{\partial t}b_{1}(t,x_1)=
      \langle p_1,\frac{\partial}{\partial q_{1}}\rangle\,b_{1}(t,x_1),\\
    &&\hskip-12mm  \frac{\partial}{\partial t}b_{2}(t,x_1,x_2)=
      \sum\limits_{j=1}^{2}\langle p_j,\frac{\partial}{\partial q_{j}}\rangle\,b_{2}(t,x_1,x_2)+\\
    &&\hskip-5mm  \int_{\mathbb{S}_+^2}d\eta\langle\eta,(p_{1}-p_{2})\rangle
     \big(b_1(q_{1},p_{1}^*)-b_1(x_1)+b_1(q_{2},p_{2}^*)-b_1(x_2)\big)\delta(q_{1}-q_{2}).
\end{eqnarray*}

Thus, in the Boltzmann--Grad scaling asymptotics, the kinetic evolution of hard sphere observables is described
in terms of limit reduced observables (\ref{Iterd}) governed by the dual Boltzmann hierarchy (\ref{vdh})
with hard sphere collisions.

\bigskip
\subsection{The Boltzmann kinetic equation}
We now establish the relationship between the constructed Boltzmann--Grad asymptotics of the reduced
observables and the description of the kinetic evolution of states in terms of the one-particle reduced
distribution function described by the Boltzmann kinetic equation.

In the case of the absence of correlations between particles at initial time, i.e., initial
states satisfying a chaos condition \cite{CGP97}, the sequence of initial reduced distribution
functions for a system of hard spheres has the form
\begin{eqnarray}\label{h2}
    &&F^{(c)}\equiv\big(1,F_1^{\epsilon,0}(x_1),\ldots,\prod_{i=1}^s F_1^{\epsilon,0}(x_i)
        \mathcal{X}_{\mathbb{R}^{3s}\setminus \mathbb{W}_s},\ldots\big),
\end{eqnarray}
where $\mathcal{X}_{\mathbb{R}^{3s}\setminus \mathbb{W}_s}$ is the Heaviside step function of the
allowed configurations. This assumption about initial state is intrinsic for the kinetic theory,
because in this case all possible states of gases are described by means of a one-particle distribution
function.

Let $F_1^{0,\epsilon}\in L^{\infty}_\xi(\mathbb{R}^3\times\mathbb{R}^3)$, i.e., the following inequality holds:
$|F_1^{0,\epsilon}(x_i)|\leq \xi\exp(-\beta\frac{p^2_i}{2})$, where $\xi>0,\beta\geq0$ are parameters.
We assume that the Boltzmann--Grad limit of the initial one-particle (reduced) distribution function
$F_{1}^{0,\epsilon}\in L^{\infty}_\xi(\mathbb{R}^3\times\mathbb{R}^3)$ exists in the sense of a weak
convergence of the space $L^{\infty}_\xi(\mathbb{R}^3\times\mathbb{R}^3)$, namely,
\begin{eqnarray}\label{lh2}
 &&\mathrm{w-}\lim_{\epsilon\rightarrow 0}(\epsilon^2\,F_{1}^{0,\epsilon}-f_{1}^0)=0,
\end{eqnarray}
then the Boltzmann--Grad limit of the initial state (\ref{h2}) satisfies a chaos property too, i.e.,
$f^{(c)}\equiv\big(1,f_1^0(x_1),\ldots,\prod_{i=1}^{s}f_{1}^0(x_i),\ldots\big)$.

We note that assumption (\ref{lh2}) with respect to the Boltzmann--Grad limit of initial states
holds true for the equilibrium state \cite{GP88}.

If $b(t)\in\mathcal{C}_{\gamma}$ and $|f_1^{0}(x_i)|\leq \xi\exp(-\beta\frac{p^2_i}{2})$, then
the Boltzmann--Grad limit of mean value functional $\big(B(t),F^{(c)}\big)$ exists under the
condition that \cite{PG90}:
$t<t_{0}\equiv\big(\text{const}(\xi,\beta)\|f_1^0\|_{L^{\infty}_\xi(\mathbb{R}^3\times\mathbb{R}^3)}\big)^{-1}$,
and it is determined by the following series expansion:
\begin{eqnarray*}
   &&\hskip-8mm \big(b(t),f^{(c)}\big)=\sum\limits_{s=0}^{\infty}\,\frac{1}{s!}\,
       \int\limits_{(\mathbb{R}^{3}\times\mathbb{R}^{3})^{s}}
      dx_{1}\ldots dx_{s}\,b_{s}(t,x_1,\ldots,x_s)\prod\limits_{i=1}^{s} f_1^0(x_i).
\end{eqnarray*}

For the limit of additive-type reduced observables (\ref{itvad}) the following equality holds \cite{GG18}:
\begin{eqnarray}\label{avmar-2}
    &&\hskip-12mm\big(b^{(1)}(t),f^{(c)}\big)=\sum\limits_{s=0}^{\infty}\,\frac{1}{s!}\,
       \int\limits_{(\mathbb{R}^{3}\times\mathbb{R}^{3})^{s}}
      dx_{1}\ldots dx_{s}\,b_{s}^{(1)}(t,x_1,\ldots,x_s)\prod \limits_{i=1}^{s}f_{1}^0(x_i)=\nonumber\\
  &&\int\limits_{\mathbb{R}^{3}\times\mathbb{R}^{3}}dx_{1}\,b_{1}^{0}(x_1)f_{1}(t,x_1),\nonumber
\end{eqnarray}
where the function $b_{s}^{(1)}(t)$ is given by expansion (\ref{itvad}) and the distribution function
$f_{1}(t,x_1)$ is represented by the series
\begin{eqnarray}\label{viter}
   &&\hskip-12mm f_{1}(t,x_1)=\sum\limits_{n=0}^{\infty}\int\limits_0^tdt_{1}\ldots\int\limits_0^{t_{n-1}}dt_{n}\,
        \int\limits_{(\mathbb{R}^{3}\times\mathbb{R}^{3})^{n}}dx_{2}\ldots dx_{n+1}\,
        S_{1}^{\ast}(t-t_{1},1)\mathcal{L}_{\mathrm{int}}^{0,\ast}(1,2)\times
\end{eqnarray}
\begin{eqnarray}
   &&\hskip-5mm \prod\limits_{j_1=1}^{2}S_{1}^{\ast}(t_{1}-t_{2},j_1)\ldots
        \prod\limits_{i_{n}=1}^{n}S_{1}^{\ast}(t_{n-1}-t_{n},i_{n})
        \sum\limits_{k_{n}=1}^{n}\mathcal{L}_{\mathrm{int}}^{0,\ast}(k_{n},n+1)\times\nonumber\\
   &&\hskip-5mm \prod\limits_{j_n=1}^{n+1}S_{1}^{\ast}(t_{n},j_n)\prod\limits_{i=1}^{n+1}f_1^0(x_i),\nonumber
\end{eqnarray}
and the following operator was introduced:
\begin{eqnarray}\label{aLint}
   &&\hskip-12mm\int\limits_{\mathbb{R}^{3}\times\mathbb{R}^{3}}dx_{n+1}
     \mathcal{L}_{\mathrm{int}}^{0,\ast}(i,n+1)f_{n+1}(x_1,\ldots,x_{n+1})\equiv\\
   &&\int\limits_{\mathbb{R}^3\times\mathbb{S}_{+}^{2}}d p_{n+1}d\eta\,\langle\eta,(p_i-p_{n+1})
      \rangle\big(f_{n+1}(x_1,\ldots,q_i,p_i^{*},\ldots,x_s,q_i,p_{n+1}^{*})-\nonumber\\
   && f_{n+1}(x_1,\ldots,x_s,q_i,p_{n+1})\big).\nonumber
\end{eqnarray}

A one-particle distribution function represented as a series (\ref{viter}) is a solution of the Cauchy
problem of the Boltzmann kinetic equation:
\begin{eqnarray}
  \label{Bolz}
    &&\hskip-12mm\frac{\partial}{\partial t}f_{1}(t,x_1)=
       -\langle p_1,\frac{\partial}{\partial q_1}\rangle f_{1}(t,x_1)+\\
    &&\int\limits_{\mathbb{R}^3\times\mathbb{S}^2_+}d p_2\, d\eta
       \,\langle\eta,(p_1-p_2)\rangle\big(f_1(t,q_1,p_1^{*})f_1(t,q_1,p_2^{*})-
       f_1(t,x_1)f_1(t,q_1,p_2)\big), \nonumber\\ \nonumber\\
  \label{Bolzi}
    &&\hskip-12mm f_1(t,x_1)_{\mid t=0}=f_{1}^0(x_1).
\end{eqnarray}

Thus, we establish that the dual Boltzmann hierarchy (\ref{vdh}) for additive-type reduced observables
and initial state (\ref{lh2}) describe the evolution of hard sphere systems just as the Boltzmann
kinetic equation (\ref{Bolz}).

We remark that in a one-dimensional space, the collision integral of the Boltzmann equation with elastic
hard sphere collisions identically equals zero. In a one-dimensional space, the Boltzmann--Grad limit is
not trivial in the case of hard sphere dynamics with inelastic collisions. In the paper \cite{GB14} for
one-dimensional granular gas, the process of the creation and propagation of correlations in the Boltzmann--Grad
scaling limit was also described (see also section 5.1).

Correspondingly, if the initial state of hard spheres is given by a sequence of reduced distribution functions
(\ref{h2}), then in the Boltzmann--Grad limit, the property of the propagation of initial chaos holds \cite{GG18}.
It is a result of the validity of the following equality for the limit $k$-ary
reduced observables, i.e., for the sequences $b^{(k)}(0)=(0,\ldots,b_{k}^{0}(x_1,\ldots,x_k),0,\ldots)$,
\begin{eqnarray}\label{pchaos}
    &&\hskip-12mm\big(b^{(k)}(t),f^{(c)}\big)=\sum\limits_{s=0}^{\infty}\,\frac{1}{s!}\,
       \int\limits_{(\mathbb{R}^{3}\times\mathbb{R}^{3})^{s}}
      dx_{1}\ldots dx_{s}\,b_{s}^{(k)}(t,x_1,\ldots,x_s) \prod \limits_{i=1}^{s} f_1^0(x_i)=\\
    &&\hskip-5mm\frac{1}{k!}\int\limits_{(\mathbb{R}^{3}\times\mathbb{R}^{3})^{k}}
      dx_{1}\ldots dx_{k}\,b_{k}^{0}(x_1,\ldots,x_k)
       \prod\limits_{i=1}^{k}f_{1}(t,x_i),\quad k\geq2,\nonumber
\end{eqnarray}
where the limit one-particle reduced distribution function $f_{1}(t)$ is defined by expansion
(\ref{viter}) and therefore it is governed by the Cauchy problem of the Boltzmann kinetic equation
(\ref{Bolz}),(\ref{Bolzi}).

Thus, in the Boltzmann--Grad scaling limit, an equivalent approach to the description of the kinetic
evolution of hard spheres in terms of the Cauchy problem of the Boltzmann kinetic equation
(\ref{Bolz}),(\ref{Bolzi}) is given by the Cauchy problem of the dual Boltzmann hierarchy with hard
sphere collisions (\ref{vdh}),(\ref{vdhi}) for the additive-type reduced observables. In the case of
non-additive-type reduced observables, a solution of the dual Boltzmann hierarchy with hard sphere
collisions (\ref{vdh}) is equivalent to the property of the propagation of initial chaos in the sense
of equality (\ref{pchaos}).

\bigskip
\subsection{The Boltzmann kinetic equation with initial correlations}
We now consider the case of the more general initial state of a hard sphere system specified
by the one-particle reduced distribution function $F_1^{0,\epsilon}\in L^{\infty}_\xi(\mathbb{R}^3\times\mathbb{R}^3)$
in the presence of correlations, i.e., the initial state that is specified by the following
sequence of reduced distribution functions:
\begin{eqnarray}\label{ins}
   &&\hskip-8mm F^{(cc)}=\big(1,F_1^{0,\epsilon}(x_1),g_{2}^{\epsilon}\prod_{i=1}^{2}F_1^{0,\epsilon}(x_i),\ldots,
        g_{n}^{\epsilon}\prod_{i=1}^{n}F_1^{0,\epsilon}(x_i),\ldots\big),
\end{eqnarray}
where the functions
$g_{n}^{\epsilon}(x_1,\ldots,x_n)\equiv g_{n}^{\epsilon}\in C_n(\mathbb{R}^{3n}\times(\mathbb{R}^{3n}\setminus\mathbb{W}_n)),\,n\geq2$,
specify the initial correlations. Since many-particle systems in condensed states are characterized
by correlations, sequence (\ref{ins}) describes the initial state of the kinetic evolution of hard
sphere fluids.

We assume that the Boltzmann--Grad limit of the initial one-particle reduced distribution function
$F_{1}^{0,\epsilon}\in L^{\infty}_\xi(\mathbb{R}^3\times\mathbb{R}^3)$ exists in the sense as above, i.e.,
in the sense of a weak convergence the equality holds:
$\lim_{\epsilon\rightarrow 0}(\epsilon^2\,F_{1}^{0,\epsilon}-f_{1}^0)=0$,
and in the case of correlation functions, let be:
$\lim_{\epsilon\rightarrow 0}(g_{n}^{\epsilon}-g_{n})=0,\,n\geq2$,
then in the Boltzmann--Grad limit, initial state (\ref{ins}) is specified by the following sequence
of the limit reduced distribution functions:
\begin{eqnarray}\label{lins}
   &&\hskip-8mm f^{(cc)}=\big(1,f_1^{0}(x_1),g_{2}\prod_{i=1}^{2}f_1^{0}(x_i),\ldots,
        g_{n}\prod_{i=1}^{n}f_1^{0}(x_i),\ldots\big).
\end{eqnarray}

We now consider relationships between the constructed Boltzmann--Grad asymptotic behavior of
reduced observables and the nonlinear Boltzmann-type kinetic equation in the case of initial
state specified by sequence (\ref{lins}).

For the limit additive-type reduced observables (\ref{itvad}) and initial state (\ref{lins}) the
following equality is true:
\begin{eqnarray*}
  &&\hskip-12mm \big(b^{(1)}(t),f^{(cc)}\big)=\sum\limits_{s=0}^{\infty}\,\frac{1}{s!}\,
      \int\limits_{(\mathbb{R}^{3}\times\mathbb{R}^{3})^{s}}dx_{1}\ldots dx_{s}
     \,b_{s}^{(1)}(t,x_{1},\ldots,x_{s})g_{s}(x_{1},\ldots,x_{s})\prod_{i=1}^{s}f_1^{0}(x_i)=\\
  &&\int\limits_{\mathbb{R}^{3}\times\mathbb{R}^{3}}dx_{1}\,b_{1}^{0}(x_1)f_{1}(t,x_1),\nonumber
\end{eqnarray*}
where the functions $b_{s}^{(1)}(t)$ are represented by expansions (\ref{itvad}) and the limit
reduced distribution function $f_{1}(t)$ is represented by the following series expansion
\begin{eqnarray}\label{viterc}
   &&\hskip-12mm f_{1}(t,x_1)=\sum\limits_{n=0}^{\infty}\,\int\limits_0^tdt_{1}\ldots\int\limits_0^{t_{n-1}}dt_{n}\,
        \int\limits_{(\mathbb{R}^{3}\times\mathbb{R}^{3})^{n}}dx_{2}\ldots dx_{n+1}
        S_{1}^{\ast}(t-t_{1},1)\times\\
   &&\hskip-5mm \mathcal{L}_{\mathrm{int}}^{0,\ast}(1,2)S_{1}^{\ast}(t_{1}-t_{2},j_1)\ldots
       \prod\limits_{i_{n}=1}^{n}S_{1}^{\ast}(t_{n}-t_{n},i_{n})
        \sum\limits_{k_{n}=1}^{n}\mathcal{L}_{\mathrm{int}}^{0,\ast}(k_{n},n+1)\times\nonumber\\
   &&\hskip-5mm \prod\limits_{j_n=1}^{n+1}S_{1}^{\ast}(t_{n},j_n)g_{1+n}(x_1,\ldots,x_{n+1})\prod\limits_{i=1}^{n+1}f_1^0(x_i).\nonumber
\end{eqnarray}
Series (\ref{viterc}) is uniformly convergent for a finite time interval under the condition as above.

The function $f_{1}(t)$ represented by series (\ref{viterc}) is a weak solution of the Cauchy
problem of the Boltzmann kinetic equation with initial correlations \cite{G14},\cite{GK15}:
\begin{eqnarray}\label{Bc}
   &&\hskip-12mm\frac{\partial}{\partial t}f_{1}(t,x_1)=-\langle p_1,\frac{\partial}{\partial q_1}\rangle f_{1}(t,x_1)+\\
   &&\hskip-5mm\int\limits_{\mathbb{R}^3\times\mathbb{S}^2_+}d p_2\,d\eta
       \,\langle\eta,(p_1-p_2)\rangle\Big(g_{2}(q_1-p_1^{*}t,p_1^{*},q_2-p_2^{*}t,p_2^{*})
       f_1(t,q_1,p_1^{*})f_1(t,q_1,p_2^{*})-\nonumber\\
   &&\hskip-5mm g_{2}(q_1-p_1t,p_1,q_2-p_2t,p_2)f_1(t,x_1)f_1(t,q_1,p_2)\Big), \nonumber\\ \nonumber\\
\label{Bci}
   &&\hskip-12mmf_{1}(t,x_1)\big|_{t=0}=f_{1}^0(x_1).
\end{eqnarray}
This fact is proved similarly to the case of a perturbative solution of the BBGKY hierarchy for hard
spheres represented by the iteration series \cite{CGP97},\cite{GP85}.

Thus, in the case of initial states specified by a one-particle reduced distribution function (\ref{lins})
we establish that the dual Boltzmann hierarchy with hard sphere collisions (\ref{vdh}) for additive-type
reduced observables describes the evolution of a hard sphere system just as the Boltzmann kinetic equation
with initial correlations (\ref{Bc}).

The property of the propagation of initial correlations is a consequence of the validity of the following
equality for the mean value functional of the limit $k$-ary reduced observables:
\begin{eqnarray}\label{dchaos}
    &&\hskip-12mm \big(b^{(k)}(t),f^{(cc)}\big)=\sum\limits_{s=0}^{\infty}\,\frac{1}{s!}\,
       \int\limits_{(\mathbb{R}^{3}\times\mathbb{R}^{3})^{s}}dx_{1}\ldots dx_{s}\,b_{s}^{(k)}(t,x_1,\ldots,x_s)
        g_{s}(x_1,\ldots,x_s)\prod \limits_{j=1}^{s} f_1^0(x_j)=\\
    &&\hskip-5mm \frac{1}{k!}\int\limits_{(\mathbb{R}^{3}\times\mathbb{R}^{3})^{k}}dx_{1}\ldots dx_{k}
       \,b_{k}^{0}(x_1,\ldots,x_k)\prod_{i_1=1}^{k}S_{1}^{\ast}(t,i_1)g_{k}(x_1,\ldots,x_k)\times\nonumber\\
   &&\hskip-5mm\prod_{i_2=1}^{k}(S_{1}^{\ast})^{-1}(t,i_2)\prod\limits_{j=1}^{k}f_{1}(t,x_j), \quad k\geq2,\nonumber
\end{eqnarray}
where the one-particle reduced distribution function $f_{1}(t,x_j)$ is solution (\ref{viterc}) of the
Cauchy problem of the Boltzmann kinetic equation with initial correlations (\ref{Bc}),(\ref{Bci}), and
the inverse group to the group of operators
$S_{1}^{\ast}(t)$ we denote by $(S_{1}^{\ast})^{-1}(t)=S_{1}^{\ast}(-t)=S_{1}(t)$.

This fact is proved similarly to the proof of a property on the propagation of initial chaos
(\ref{pchaos}).

We note that, according to equality (\ref{dchaos}), in the Boltzmann--Grad limit, the reduced correlation
functions are defined as cluster expansions of reduced distribution functions, namely,
\begin{eqnarray*}
   &&\hskip-12mm f_{s}(t,x_1,\ldots,x_s)=
      \sum\limits_{\mbox{\scriptsize $\begin{array}{c}\mathrm{P}:(x_1,\ldots,x_s)=\bigcup_{i}X_{i}\end{array}$}}
      {\prod\limits}_{X_i\subset\mathrm{P}}\,g_{|X_i|}(t,X_i),\quad s\geq1,
\end{eqnarray*}
and they have the explicit form:
\begin{eqnarray}\label{cfmf}
    &&\hskip-12mm g_{1}(t,x_1)=f_{1}(t,x_1),\\
    &&\hskip-12mm g_{s}(t,x_1,\ldots,x_s)=\tilde{g}_{s}(q_1-p_1t,p_1,\ldots,q_s-p_st,p_s)\prod\limits_{j=1}^{s}f_{1}(t,x_j),
                \quad s\geq2,\nonumber
\end{eqnarray}
where for initial correlation functions (\ref{lins}) it is used the following notations:
\begin{eqnarray*}\label{FG1}
   &&\hskip-12mm \tilde{g}_{s}(x_1,\ldots,x_s)=
      \sum\limits_{\mbox{\scriptsize$\begin{array}{c}\mathrm{P}:(x_1,\ldots,x_s)=\bigcup_{i}X_{i}\end{array}$}}
      \prod_{X_i\subset \mathrm{P}}g_{|X_i|}(X_i),
\end{eqnarray*}
the symbol $\sum_\mathrm{P}$ means the sum over possible partitions $\mathrm{P}$ of the set of arguments
$(x_1,\ldots,x_s)$ on $|\mathrm{P}|$ nonempty  subsets $X_i$, and the one-particle reduced distribution
function $f_{1}(t)$ is a solution of the Cauchy problem of the Boltzmann kinetic equation with initial
correlations (\ref{Bc}),(\ref{Bci}).

Thus, in the case of the limit $k$-ary reduced observables, a solution of the dual Boltzmann hierarchy
with hard sphere collisions (\ref{vdh}) is equivalent to a property of the propagation of initial
correlations for the $k$-particle reduced distribution function in the sense of equality (\ref{dchaos})
or in other words, the Boltzmann--Grad scaling dynamics does not create new correlations.

\bigskip
\textcolor{blue!55!black}{\section{Origin of kinetic equations}}

One of the challenges of kinetic theory, as mentioned above, is understanding the nature
of the possibility of describing the evolution of the state of a system of many hard
spheres by means of the state of a typical hard sphere. More precisely, we further focus
on the problem of the origin of the description of the evolution of the state of hard
spheres by the Enskog-type kinetic equation.

In the circumstances where the initial state is specified by a one-particle reduced
distribution function, for the mean value functional of observables at an arbitrary
instant, the representation is also valid in terms of a one-particle reduced distribution
function, the evolution of which is governed by a non-Markovian nonlinear evolution equation.
In other words, for such initial data, the Cauchy problem of the BBGKY hierarchy for hard
spheres is equivalent to the nonlinear Enskog-type kinetic equation and a sequence of reduced
functionals determined by the solution of this evolution equation.

\bigskip
\subsection{The generalized Enskog kinetic equation}
In the case of initial state (\ref{h2}) the dual picture of the evolution to the picture described
by employing observables  governed by the dual BBGKY hierarchy (\ref{dh}) for hard spheres is the
picture of the evolution of a state described by means of the non-Markovian Enskog kinetic equation
and by a sequence of explicitly defined functionals of the solution of such a kinetic equation that
describe the evolution of all possible correlations in a system of hard spheres \cite{GG11b},\cite{GG12}.

In view of the fact that the initial state is completely specified by a one-particle reduced distribution
function on allowed configurations (\ref{h2}), for mean value functional (\ref{B(t)}) the following
representation holds \cite{GG12}:
\begin{eqnarray}\label{w}
    &&\big(B(t),F^{(c)}\big)=\big(B(0),F(t\mid F_{1}(t))\big),
\end{eqnarray}
where $F^{(c)}$ is the sequence of initial reduced distribution functions (\ref{h2}), and the sequence
$F(t\mid F_{1}(t))=\big(1,F_1(t),F_2(t\mid F_{1}(t)),\ldots,F_s(t\mid F_{1}(t)),\ldots\big)$
is a sequence of the reduced functionals of the state $F_{s}(t,x_1,\ldots,x_s \mid F_{1}(t)),$ $s\geq2,$
represented by the series expansions over the products of the one-particle distribution function $F_{1}(t)$,
namely
\begin{eqnarray}\label{f}
   &&\hskip-12mm F_{s}(t,x_1,\ldots,x_s\mid F_{1}(t))\doteq \\
   &&\hskip-7mm \sum _{n=0}^{\infty }\frac{1}{n!}\,\int\limits_{(\mathbb{R}^{3}\times\mathbb{R}^{3})^{n}}
      dx_{s+1}\ldots dx_{s+n}\,\mathfrak{V}_{1+n}(t,\{1,\ldots,s\},s+1,\ldots,s+n)\prod_{i=1}^{s+n}F_{1}(t,x_i),\nonumber\\
   &&\hskip-12mm  s\geq 2.\nonumber
\end{eqnarray}
where the generating operators of series (\ref{f}) are the $(1+n)th$-order operators
$\mathfrak{V}_{1+n}(t),\,n\geq0$, determined by the following expansions \cite{GG12}:
\begin{eqnarray}\label{skrrc}
   &&\hskip-12mm \mathfrak{V}_{1+n}\bigl(t,\{1,\ldots,s\},s+1,\ldots,s+n\bigr)=
                 n!\,\sum_{k=0}^{n}\,(-1)^k\,\sum_{n_1=1}^{n}\ldots\\
   && \sum_{n_k=1}^{n-n_1-\ldots-n_{k-1}}\frac{1}{(n-n_1-\ldots-n_k)!}\,
               \widehat{\mathfrak{A}}_{1+n-n_1-\ldots-n_k}(t,\{1,\ldots,s\},\nonumber\\
   && s+1,\ldots, s+n-n_1-\ldots-n_k)\prod_{j=1}^k\,\sum\limits_{\mbox{\scriptsize$\begin{array}{c}
       \mathrm{D}_{j}:Z_j=\bigcup_{l_j}X_{l_j},\\
       |\mathrm{D}_{j}|\leq s+n-n_1-\dots-n_j\end{array}$}}\frac{1}{|\mathrm{D}_{j}|!}\times\nonumber\\
   &&  \sum_{i_1\neq\ldots\neq i_{|\mathrm{D}_{j}|}=1}^{s+n-n_1-\ldots-n_j}\,
       \prod_{X_{l_j}\subset \mathrm{D}_{j}}\,\frac{1}{|X_{l_j}|!}\,
       \widehat{\mathfrak{A}}_{1+|X_{l_j}|}(t,i_{l_j},X_{l_j}),\nonumber
\end{eqnarray}
In expansion \eqref{skrrc} the symbol $\sum_{\mathrm{D}_{j}:Z_j=\bigcup_{l_j} X_{l_j}}$ means the sum over
all possible dissections of the linearly ordered set $Z_j\equiv(s+n-n_1-\ldots-n_j+1,\ldots,s+n-n_1-\ldots-n_{j-1})$
on no more than $s+n-n_1-\ldots-n_j$ linearly ordered subsets, and
the $(1+n)th$-order scattering cumulant we denoted by the operator:
\begin{eqnarray}\label{scacu}
   &&\hskip-12mm\widehat{\mathfrak{A}}_{1+n}(t,\{1,\ldots,s\},s+1,\ldots,s+n)\doteq\\
   &&\mathfrak{A}_{1+n}^\ast(t,\{1,\ldots,s\},s+1,\ldots,s+n)\mathcal{X}_{\mathbb{R}^{3(s+n)}\setminus\mathbb{W}_{s+n}}
      \prod_{i=1}^{s+n}\mathfrak{A}_{1}^\ast(t,i)^{-1},\nonumber
\end{eqnarray}
where the operator $\mathfrak{A}_{1+n}^\ast(t)$ is the $(1+n)th$-order cumulant of the groups of operators
(\ref{S*}) of hard spheres.

We provide some examples of expressions for the generating operators of series (\ref{f}) for reduced
functionals of the state:
\begin{eqnarray*}
   &&\hskip-12mm\mathfrak{V}_{1}(t,\{1,\ldots,s\})=\widehat{\mathfrak{A}}_{1}(t,\{1,\ldots,s\})\doteq
      S_s^\ast(t,1,\ldots,s)\mathcal{X}_{\mathbb{R}^{3(s)}\setminus \mathbb{W}_{s}}\prod_{i=1}^{s}S_1^\ast(t,i)^{-1},\\
   &&\hskip-12mm\mathfrak{V}_{2}(t,\{1,\ldots,s\},s+1)=\widehat{\mathfrak{A}}_{2}(t,\{1,\ldots,s\},s+1)-
    \widehat{\mathfrak{A}}_{1}(t,\{1,\ldots,s\})\sum_{i_1=1}^s \widehat{\mathfrak{A}}_{2}(t,i_1,s+1).
\end{eqnarray*}

The method of constructing reduced state functionals \eqref{f} is based on the application of the variation
of cluster expansions, the so-called kinetic cluster expansions, to generating operators \eqref{cumulant} of
series representing solutions of hierarchies of evolution equations \cite{GG12}.

The one-particle distribution function $F_{1}(t)$, i.e., the first element of the sequence $F(t\mid F_{1}(t))$,
is determined by series (\ref{se}), namely
\begin{eqnarray}\label{F(t)1}
    &&\hskip-12mm F_{1}(t,x_1)=\sum\limits_{n=0}^{\infty}\frac{1}{n!}
      \int\limits_{(\mathbb{R}^3\times\mathbb{R}^3)^n}dx_2\ldots dx_{n+1}\,
      \mathfrak{A}_{1+n}^\ast(t,1,\ldots,n+1)\prod_{i=1}^{n+1}F_{1}^{\epsilon,0}(x_i)\mathcal{X}_{\mathbb{R}^{3(1+n)}\setminus\mathbb{W}_{1+n}},
\end{eqnarray}
where the generating operator $\mathfrak{A}_{1+n}^\ast(t)$ is the $(1+n)$th-order cumulant
of groups of operators (\ref{S*}).

Let us note that in particular case of initial data (\ref{dhi}) specified by the additive-type
reduced observables, according to solution expansion (\ref{af}), equality
(\ref{w}) takes the form
\begin{eqnarray}\label{avmar-11}
   &&\hskip-8mm\big( B^{(1)}(t),F(0)\big)=
      \int\limits_{\mathbb{R}^{3}\times\mathbb{R}^{3}}dx_{1}\,b_{1}^{\epsilon}(x_1)F_{1}(t,x_1),\nonumber
\end{eqnarray}
where the one-particle distribution function $F_{1}(t)$ is determined by series (\ref{F(t)1}).
In the case of initial data (\ref{dhi}) specified by the $s$-ary reduced observable $s\geq2$,
equality (\ref{w}) has the form
\begin{eqnarray*}\label{avmar-12}
   &&\hskip-12mm\big(B^{(s)}(t),F(0)\big)=\frac{1}{s!}\int\limits_{(\mathbb{R}^{3}\times\mathbb{R}^{3})^{s}}
      dx_{1}\ldots dx_{s}\,b_{s}^{\epsilon}(x_1,\ldots,x_s)F_s(t,x_1,\ldots,x_s\mid F_{1}(t)),
\end{eqnarray*}
where the reduced functionals of the state $F_{s}(t,x_1,\ldots,x_s \mid F_{1}(t))$ are determined
by series (\ref{f}).

Thus, for the initial state specified by a one-particle distribution function, the evolution
of all possible states of a system of many hard spheres can be described by means of the state
of a typical particle without any scaling approximations. We emphasize that reduced functionals
of the state(\ref{f}) describe all possible correlations created during the evolution of many
hard spheres in terms of the state of a typical hard sphere.

For $t\geq 0$ the one-particle distribution function (\ref{F(t)1}) is a solution of the following
Cauchy problem of the non-Markovian generalized Enskog kinetic equation \cite{GG22},\cite{GG12}:
\begin{eqnarray}
 \label{gke1}
   &&\hskip-12mm\frac{\partial}{\partial t}F_{1}(t,q_1,p_1)=
      -\langle p_1,\frac{\partial}{\partial q_1}\rangle F_{1}(t,q_1,p_1)+\\
   &&\hskip-5mm\epsilon^2\int\limits_{\mathbb{R}^3\times\mathbb{S}^2_+}d p_2 d\eta
      \langle\eta,(p_1-p_2)\rangle\Big(F_2(t,q_1,p_1^{*},q_1-\epsilon\eta,p_2^{*}\mid F_{1}(t))-\nonumber
\end{eqnarray}
\begin{eqnarray}
   &&\hskip-5mm F_2(t,q_1,p_1,q_1+\epsilon\eta,p_2\mid F_{1}(t))\Big),\nonumber\\ \nonumber\\
 \label{gkei}
   &&\hskip-12mm F_{1}(t)\big|_{t=0}= F_{1}^{\epsilon,0},
\end{eqnarray}
where the collision integral is determined by the reduced functional of the state (\ref{f})
in the case of $s=2$ and the expressions $p_1^{*}$ and $p_2^{*}$ are the pre-collision momenta
of hard spheres (\ref{momenta}). The series on the right-hand side of this equation converges
under the condition: $\|F_1(t)\|_{L^1(\mathbb{R}\times\mathbb{R})}<e^{-8}$ .

Hence in the case of the additive-type reduced observables the generalized Enskog kinetic equation
(\ref{gke1}) is dual to the dual BBGKY hierarchy of hard spheres (\ref{dh}) with respect to bilinear
form (\ref{B(t)}).

We observe that the structure of the collision integral of the generalized Enskog equation (\ref{gke1})
is such that the first term of its expansion is the collision integral of the Boltzman--Enskog kinetic
equation, and the next terms describe all possible correlations that are created by the dynamics of hard
spheres and by the propagation of initial correlations connected with the forbidden configurations, indeed
\begin{eqnarray*}\label{gke1e}
   &&\hskip-12mm\frac{\partial}{\partial t}F_{1}(t,x_1)=
                -\langle p_1,\frac{\partial}{\partial q_1}\rangle F_{1}(t,x_1)+\mathcal{I}_{GEE},
\end{eqnarray*}
where the collision integral is determined by the following series expansion:
\begin{eqnarray*}
   &&\hskip-12mm\mathcal{I}_{GEE}\doteq\epsilon^2\sum_{n=0}^{\infty}\frac{1}{n!}
      \int\limits_{\mathbb{R}^3\times\mathbb{S}^2_+}d p_2 d\eta
      \int\limits_{(\mathbb{R}^{3}\times\mathbb{R}^{3})^{n}}
      dx_3\ldots dx_{n+2}\langle\eta,(p_1-p_2)\rangle\times\nonumber\\
   &&\hskip-5mm\big(\mathfrak{V}_{1+n}(t,\{1^{*},2^{*}_{-}\},3,\ldots,n+2)
      F_1(t,q_1,p_1^{*})F_1(t,q_1-\epsilon\eta,p_2^{*})\prod_{i=3}^{n+2}F_{1}(t,x_i)- \nonumber\\
   &&\hskip-5mm\mathfrak{V}_{1+n}(t,\{1,2_{+}\},3,\ldots,n+2)F_1(t,x_1)
      F_1(t,q_1+\epsilon\eta,p_2)\prod_{i=3}^{n+2}F_{1}(t,x_i)\big),\nonumber
\end{eqnarray*}
and the notations adopted for the conventional notation of the Enskog collision integral
were used: indices $(1^{\sharp},2^{\sharp}_{\pm})$ denote that the evolution operator
$\mathfrak{V}_{1+n}(t)$ acts on the corresponding phase points $(q_1,p_1^{\sharp})$ and
$(q_1\pm\epsilon\eta,p_2^{\sharp})$, and the $(n+1)$th-order evolution operator
$\mathfrak{V}_{1+n}(t),\,n\geq0$, is determined by expansion (\ref{skrrc}) in the case
of $s=2$.

We note that in the work \cite{GG12} for the initial-value problem (\ref{gke1}),(\ref{gkei})
the existence theorem was proved in the space of integrable functions. The accordance of the
generalized Enskog equation (\ref{gke1}) and of the Markovian Enskog-type kinetic equation
was also established there.
By the point, we remark that in the paper \cite{T} the explicit soliton-like solutions of
kinetic equation (\ref{gke1}) were found.

Thus, if the initial state is specified by a one-particle distribution function on allowed
configurations, then the evolution of many hard spheres governed by the dual BBGKY hierarchy
(\ref{dh}) for reduced observables can be completely described by the generalized Enskog
kinetic equation (\ref{gke1}) and by a sequence of reduced functionals of the state (\ref{f}).

We remark also that in the case of the initial state that involves correlations (\ref{ins})
considered approach permits to take into consideration the initial correlations in the kinetic
equations \cite{GG21}.


Further, we sketch out the Boltzmann--Grad scaling behavior of the non-Markovian Enskog kinetic
equation (\ref{gke1}) and reduced state functional (\ref{f}).

Taking into account the validity of assumption (\ref{lh2}) for the initial one-particle
distribution function (\ref{h2}), in that case for a finite time interval, the Boltzmann--Grad
limit of dimensionless solution (\ref{F(t)1}) of the Cauchy problem of the non-Markovian Enskog
kinetic equation (\ref{gke1}),(\ref{gkei}) exists in the same sense, namely
\begin{eqnarray*}
   &&\lim_{\epsilon\rightarrow 0}\big(\epsilon^{2}F_{1}(t,x_1)-f_{1}(t,x_1)\big)=0,
\end{eqnarray*}
where the limit one-particle distribution function $f_{1}(t)$ is a weak solution of the
Cauchy problem of the Boltzmann kinetic equation (\ref{Bolz}),(\ref{Bolzi}).

Taking into consideration the fact of the existence of the Boltzmann--Grad scaling limit
of a solution of the non-Markovian Enskog kinetic equation (\ref{gke1}), for reduced
functionals of the state (\ref{f}) the following statement holds \cite{GG12}:
\begin{eqnarray*}
   &&\lim\limits_{\epsilon\rightarrow 0}\big(\epsilon^{2s}
      F_{s}\big(t,x_1,\ldots,x_s\mid F_{1}(t)\big)-\prod\limits_{j=1}^{s}f_{1}(t,x_j)\big)=0,
\end{eqnarray*}
where the limit one-particle distribution function $f_{1}(t)$ is governed by the Boltzmann kinetic
equation with hard sphere collisions (\ref{Bolz}). Because all possible correlations of many hard
spheres with elastic collisions are described by reduced functionals of the state (\ref{f}), as
noted above, this property means the propagation of the initial chaos in the Boltzmann--Grad limit.

The proof of these statements is based on the properties of cumulants of asymptotically perturbed
groups of operators (\ref{S*}) and the explicit structure (\ref{skrrc}) of the generating operators
of series expansions (\ref{f}) for reduced functional of state and of series (\ref{F(t)1}).

\bigskip
\subsection{Dynamics of correlations governed by kinetic equations}
Let the initial state be specified by a one-particle reduced correlation function, namely,
the initial state specified by a sequence of reduced correlation functions satisfying the
chaos property stated above, i.e., by the sequence $G^{(c)}=(G_0,G_1^{0},0,\ldots,0,\ldots)$.
We note that such an assumption about initial states is intrinsic to the contemporary kinetic
theory of many-particle systems \cite{CGP97},\cite{CIP}.

Since the initial data $G^{(c)}$ is completely specified only by a one-particle correlation
function, the Cauchy problem (\ref{gBigfromDFBa}),(\ref{gBigfromDFBai}) of the nonlinear
hierarchy for hard spheres is not a completely well-defined Cauchy problem because the initial
data is not independent for every unknown function determined of the hierarchy of mentioned
evolution equations.
As a consequence, it becomes possible to reformulate such a Cauchy problem as a new Cauchy
problem for a one-particle correlation function with independent initial data and explicitly
defined functionals of the solution of this Cauchy problem for the kinetic equation.


We formulate such a restated Cauchy problem and the sequence of the suitable functionals.
In the case under consideration, the reduced correlation functionals $G_s(t\mid G_{1}(t)),\,s\geq2$,
are represented with respect to the one-particle correlation function (\ref{mcc}), i.e.,
\begin{eqnarray}\label{ske}
   &&\hskip-12mm G_{1}(t,x_1)=\sum\limits_{n=0}^{\infty}\frac{1}{n!}\,
      \int\limits_{(\mathbb{R}^{3}\times\mathbb{R}^{3})^{n}}dx_{2}\ldots dx_{1+n}\,
      \mathfrak{A}^\ast_{1+n}(t,1,\ldots,n+1)\prod_{i=1}^{n+1}G_{1}^{0}(x_i)
      \mathcal{X}_{\mathbb{R}^{3(n+1)}\setminus \mathbb{W}_{n+1}},
\end{eqnarray}
as the following series expansions:
\begin{eqnarray}\label{fc}
     && \hskip-12mm G_{s}\bigl(t,x_1,\ldots,x_s\mid G_{1}(t)\bigr)=\\
     &&\hskip-5mm  \sum_{n=0}^{\infty }\frac{1}{n!}\,
         \int\limits_{(\mathbb{R}^{3}\times\mathbb{R}^{3})^{n}}dx_{s+1}\ldots dx_{s+n}\,
         \mathfrak{V}_{s+n}\bigl(t,1,\ldots,s+n\bigr)\prod_{i=1}^{s+n}G_{1}(t,x_i), \quad s\geq2.\nonumber
\end{eqnarray}
The generating operator $\mathfrak{V}_{s+n}(t),\,n\geq0$, of the $(s+n)th$-order of this series
is determined by the following expansion:
\begin{eqnarray}\label{skrrcc}
   &&\hskip-12mm\mathfrak{V}_{s+n}\bigl(t,1,\ldots,s,s+1,\ldots,s+n\bigr)= n!\,
     \sum_{k=0}^{n}\,(-1)^k\,\sum_{n_1=1}^{n}\ldots\\
   &&\hskip-8mm \sum_{n_k=1}^{n-n_1-\ldots-n_{k-1}}\frac{1}{(n-n_1-\ldots-n_k)!}\,
       \hat{\mathfrak{A}}_{s+n-n_1-\ldots-n_k}(t,1,\ldots,s+n-n_1-\ldots-n_k)\times\nonumber\\
   &&\hskip-8mm\prod_{j=1}^k\,\sum\limits_{\mbox{\scriptsize$\begin{array}{c}
       \mathrm{D}_{j}:Z_j=\bigcup_{l_j}X_{l_j},\\
       |\mathrm{D}_{j}|\leq s+n-n_1-\dots-n_j\end{array}$}}\frac{1}{|\mathrm{D}_{j}|!}
       \sum_{i_1\neq\ldots\neq i_{|\mathrm{D}_{j}|}=1}^{s+n-n_1-\ldots-n_j}\,
       \prod_{X_{l_j}\subset \mathrm{D}_{j}}\,\frac{1}{|X_{l_j}|!}
       \hat{\mathfrak{A}}_{1+|X_{l_j}|}(t,i_{l_j},X_{l_j}),\nonumber
\end{eqnarray}
where $\sum_{\mathrm{D}_{j}:Z_j=\bigcup_{l_j} X_{l_j}}$ is the sum over all possible dissections
of the linearly ordered set $Z_j\equiv(s+n-n_1-\ldots-n_j+1,\ldots,s+n-n_1-\ldots-n_{j-1})$ on no
more than $s+n-n_1-\ldots-n_j$ linearly ordered subsets, the $(s+n)th$-order scattering cumulant
is defined by the formula
\begin{eqnarray*}
    &&\hskip-12mm\hat{\mathfrak{A}}_{s+n}(t,1,\ldots,s+n)\doteq
      \mathfrak{A}^\ast_{s+n}(t,1,\ldots,s+n)\mathcal{X}_{\mathbb{R}^{3(s+n)}\setminus \mathbb{W}_{s+n}}
      \prod_{i=1}^{s+n}(\mathfrak{A}_{1}^\ast)^{-1}(t,i),
\end{eqnarray*}
and notations accepted above were used.

We adduce simplest examples of generating operators (\ref{skrrcc}):
\begin{eqnarray*}
   &&\hskip-12mm\mathfrak{V}_{s}(t,1,\ldots,s)=\mathfrak{A}_{s}(t,1,\ldots,s)\mathcal{X}_{\mathbb{R}^{3s}\setminus\mathbb{W}_{s}}
      \prod_{i=1}^{s}(\mathfrak{A}_{1}^\ast)^{-1}(t,i),\\
   &&\hskip-12mm\mathfrak{V}_{s+1}(t,1,\ldots,s,s+1)=
      \mathfrak{A}_{s+1}(t,1,\ldots,s+1)\mathcal{X}_{\mathbb{R}^{3(s+1)}\setminus \mathbb{W}_{s+1}}
      \prod_{i=1}^{s+1}(\mathfrak{A}_{1}^\ast)^{-1}(t,i)-\\
   &&\mathfrak{A}_{s}(t,1,\ldots,s)\mathcal{X}_{\mathbb{R}^{3s}\setminus \mathbb{W}_{s}}
      \prod_{i=1}^{s}(\mathfrak{A}_{1}^\ast)^{-1}(t,i)\sum_{j=1}^s\mathfrak{A}_{2}(t,j,s+1)\times\\
   &&\mathcal{X}_{\mathbb{R}^{6}\setminus \mathbb{W}_{2}}
      (\mathfrak{A}_{1}^\ast)^{-1}(t,j)(\mathfrak{A}_{1}^\ast)^{-1}(t,s+1).
\end{eqnarray*}

If $\|G_{1}(t)\|_{L^{1}(\mathbb{R}^{3}\times\mathbb{R}^{3})}<e^{-(3s+2)}$, for arbitrary
$t\in\mathbb{R}$ series (\ref{fc}) converges in the norm of the space $L^{1}_{s}$ \cite{GG12}.

We note that in the case of initial state specified by a one-particle correlation function
the reduced correlation functionals (\ref{fc}) describe all possible correlations generated
by the dynamics of many hard spheres in terms of a one-particle correlation function.

Thus, according to the representation \eqref{mcc} of reduced correlation functions, the cumulant
structure of their generating operators induces a generalized cumulant structure of the generating
operators for series \eqref{fc} of reduced correlation functionals.

In this case, the method of constructing reduced reduced correlation functionals \eqref{fc}
is based on the application of the variation of cluster expansions, the so-called kinetic
cluster expansions \cite{GG12}, to generating operators \eqref{cumcp} of series representing
reduced correlation functions \eqref{mcc}.


Indeed, taking into account relations of kinetic cluster expansions for scattering cumulants (\ref{scacu}):
\begin{eqnarray*}\label{rrrl2}
  &&\hskip-12mm \widehat{\mathfrak{A}}_{s+n}(t,1,\ldots,s+n)=\sum_{n_1=0}^{n}\frac{n!}{(n-n_1)!}
      \mathfrak{V}_{s+n-n_1}\big(t,1,\ldots, s+n-n_1\big)\times\nonumber\\
  &&\sum\limits_{\mbox{\scriptsize $\begin{array}{c}\mathrm{D}:Z=\bigcup_l X_l,\\|\mathrm{D}|\leq s+n-n_1\end{array}$}}
     \frac{1}{|\mathrm{D}|!}\sum_{i_1\neq\ldots\neq i_{|\mathrm{D}|}=1}^{s+n-n_1}\,\,
     \prod_{X_{l}\subset \mathrm{D}}\,\frac{1}{|X_l|!}\,\,\widehat{\mathfrak{A}}_{1+|X_{l}|}(t,i_l,X_{l})\nonumber,
\end{eqnarray*}
where $\sum_{\mathrm{D}:Z=\bigcup_l X_l,\,|\mathrm{D}|\leq s+n-n_1}$ is the sum over all possible dissections
$\mathrm{D}$ of the linearly ordered set $Z\equiv(s+n-n_1+1,\ldots,s+n)$ on no more than $s+n-n_1$ linearly ordered
subsets, we derive the expansions of the reduced correlation functionals $G_{s}(t,x_1,\ldots,x_s\mid G_{1}(t)),\,s\geq2$,
on the basis of solution expansions (\ref{mcc}) of the hierarchy of nonlinear evolution equations
(\ref{gBigfromDFBa}).

Note that the structure of kinetic cluster expansions of scattering cumulants of the groups of operators is similar
to the structure of virial expansions of equilibrium distribution functions, i.e., as a power series in the density.


If initial data $G_{1}^{0}\in L^{1}_1$, then for arbitrary $t\in\mathbb{R}$ one-particle
correlation function (\ref{ske}) is a weak solution of the Cauchy problem of the non-Markovian
Enskog kinetic equation
\begin{eqnarray}\label{gkec}
   &&\hskip-12mm\frac{\partial}{\partial t}G_{1}(t,x_1)=\mathcal{L}^{\ast}(1)G_{1}(t,x_1)+
      \int\limits_{\mathbb{R}^{3}\times\mathbb{R}^{3}}dx_{2}\,
      \mathcal{L}_{\mathrm{int}}^{\ast}(1,2)G_{1}(t,x_1)G_{1}(t,x_2)+\\
   &&\int\limits_{\mathbb{R}^{3}\times\mathbb{R}^{3}}dx_{2}\,
      \mathcal{L}_{\mathrm{int}}^{\ast}(1,2)G_{2}\bigl(t,x_1,x_2\mid G_{1}(t)\bigr),\nonumber\\
   \nonumber\\
 \label{gkeci}
   &&\hskip-12mm G_{1}(t,x_1)\big|_{t=0}=G_{1}^{0}(x_1),
\end{eqnarray}
where the first part of the collision integral in equation (\ref{gkec}) has the Boltzmann--Enskog
structure, and the second part of the collision integral is determined in terms of the two-particle
correlation functional represented by series expansion (\ref{fc}) which describes all possible
correlations that are created by hard-sphere dynamics and by the propagation of initial correlations
related to the forbidden configurations.

In the paper \cite{GG12}, similar statements were proved for the state evolution of a hard-sphere
system described in terms of reduced distribution functions governed by the BBGKY hierarchy.
We emphasize that the $nth$ term of expansions (\ref{fc}) of the reduced correlation functionals
are determined by the $(s+n)th$-order generating operator (\ref{skrrc}) in contradistinction to the
expansions of reduced distribution functionals of the state constructed in \cite{GG12} which are
determined by the $(1+n)th$-order generating operator (\ref{skrrc}).

Thus, for the initial state specified by a one-particle correlation function, the evolution
of all possible states of the system of hard spheres can be described without any approximations
within the framework of a one-particle correlation function governed by the non-Markovian Enskog-type
kinetic equation (\ref{gkec}), and by a sequence of explicitly defined functionals (\ref{fc}) of its
solution.

\vskip-5mm
\textcolor{blue!55!black}{\section{Conclusion}}

In conclusion, the challenges of the evolution of many hard spheres with inelastic collisions
will be reviewed, as will some applications of the methods outlined above to complex systems
of various natures.

\bigskip
\subsection{On the dynamics of inelastic collisions}
According to contemporary concept \cite{T04},\cite{V06} on a microscopic scale, the characteristic
properties of granular media are determined by dissipative collisional dynamics and can be described
as the evolution of a system of many hard spheres with inelastic collisions.

Since the characteristic features of the collective behavior of inelastically colliding particles
in one-dimensional space reflect the main properties of granular gases, an approach to the rigorous
derivation of the Boltzmann-type equation for one-dimensional granular gases will be presented below.
We note that, in contrast to the system of hard rods with inelastic collisions, in one-dimensional
space the evolution of hard rods with elastic collisions is trivial in the Boltzmann--Grad scaling
limit; it is known as so-called free molecular motion or the Knudsen flow \cite{CGP97}.

In the case of a one-dimensional granular gas for $t\geq0$ in dimensionless form the Cauchy problem
of the non-Markovian generalized Enskog kinetic equation (\ref{gke1}),(\ref{gkei}) takes the form
\cite{G20g},\cite{GB13},\cite{GB14}:
\begin{eqnarray}\label{GE1}
  &&\hskip-12mm\frac{\partial}{\partial t}F_{1}(t,q_1,p_1)=
      -p_1\frac{\partial}{\partial q_1}F_{1}(t,q_1,p_1)+\\
  &&\int\limits_0^\infty dP\,P\big(\frac{1}{(1-2\varepsilon)^2}\,
      F_{2}(t,q_1,p_1^\diamond(p_1,P),q_1-\epsilon,p_{2}^\diamond(p_1,P)\mid F_{1}(t))-\nonumber\\
  &&F_{2}(t,q_1,p_1,q_1-\epsilon,p_1+P\mid F_{1}(t))\big)+\nonumber\\
  &&\int\limits_0^\infty dP\,P \big(\frac{1}{(1-2\varepsilon)^2}\,F_{2}(t,q_1,\tilde p_1^\diamond(p_1,P),q_1+
      \epsilon,\tilde p_{2}^\diamond(p_1,P)\mid F_{1}(t))-\nonumber\\
  &&F_{2}(t,q_1,p_1,q_1+\epsilon,p_1-P\mid F_{1}(t))\big),\nonumber\\ 
\label{GKEi}
   &&\hskip-12mmF_{1}(t)|_{t=0}= F_{1}^{\epsilon,0},
\end{eqnarray}
where $\varepsilon=\frac{1-e}{2}\in [0,\frac{1}{2})$ and $e\in(0,1]$ is a restitution
coefficient, $\epsilon>0$ is a scaling parameter (the ratio of a hard sphere diameter (the length)
$\sigma>0$ to the mean free path), the collision integral is determined by reduced functional (\ref{f})
of the state $F_{1}(t)$ in the case of $s=2$ and the expressions:
\begin{eqnarray*}
   &&p_{1}^\diamond(p_1,P)=p_1-P+\frac{\varepsilon}{2\varepsilon -1}\,P,\\
   &&p_{2}^\diamond(p_1,P)=p_1-\frac{\varepsilon}{2\varepsilon -1}\,P\nonumber
\end{eqnarray*}
and the expressions
\begin{eqnarray*}
   &&\tilde p_{1}^\diamond(p_1,P)=p_1+P-\frac{\varepsilon}{2\varepsilon -1}\,P,\\
   &&\tilde p_{2}^\diamond(p_1,P)=p_1+\frac{\varepsilon}{2\varepsilon -1}\,P,\nonumber
\end{eqnarray*}
are transformed pre-collision momenta of inelastically colliding particles in a one-dimensional space.

The solution of the Cauchy problem \eqref{GE1},\eqref{GKEi} is represented by the following series:
\begin{eqnarray}\label{ske1}
   &&\hskip-12mm F_{1}^{\epsilon}(t,x_1)=\sum\limits_{n=0}^{\infty}\frac{1}{n!}
     \int\limits_{\mathbb{R}^n\times\mathbb{R}^n}dx_2\ldots dx_{n+1}\,\mathfrak{A}_{1+n}^\ast(t)
     \prod_{i=1}^{n+1}F_{1}^{\epsilon,0}(x_i)\mathcal{X}_{\mathbb{R}^{(1+n)}\setminus \mathbb{W}_{1+n}},
\end{eqnarray}
where the generating operator $\mathfrak{A}_{1+n}^\ast(t)$ is the $(1+n)$th-order cumulant
\eqref{cumulant} of the semigroups of operators (\ref{S*}) of inelastically colliding hard
rods in a one-dimensional space. Let the initial one-particle distribution function satisfy
the following condition: $|F_{1}^{\epsilon,0}(x_1)|\leq Ce^{\textstyle-\frac{\beta}{2}{p^{2}_1}},$
where $\textstyle{\beta}>0$ is a parameter and $C<\infty$ is some constant. Then every
term of series \eqref{ske1} exists; for a finite time interval it is the uniformly convergent
series with respect to $x_1$ from an arbitrary compact, and function \eqref{ske1} is a weak
solution of the Cauchy problem (\ref{GE1}),(\ref{GKEi}) of the non-Markovian Enskog-type equation
with inelastic collisions.

We assume that, in the sense of weak convergence, there exists a limit
\begin{eqnarray*}
   &&\mathrm{w-}\lim\limits_{\epsilon\rightarrow 0}\big(F_{1}^{\epsilon,0}(x_1)-f_{1}^0(x_1)\big)=0.
\end{eqnarray*}
Then, for a finite time interval, the Boltzmann--Grad limit of solution (\ref{ske1}) of the Cauchy problem
of the non-Markovian Enskog-type equation for a one-dimensional granular gas (\ref{GE1}) exists in the sense
of a weak convergence
\begin{eqnarray}\label{asymtin}
   &&\mathrm{w-}\lim\limits_{\epsilon\rightarrow 0}\big(F_{1}^{\epsilon}(t,x_1)-f_{1}(t,x_1)\big)=0,
\end{eqnarray}
where the limit of one-particle distribution function \eqref{ske1} is determined by the following series
uniformly convergent on an arbitrary compact set
\begin{eqnarray}\label{viterin}
  &&\hskip-9mm f_{1}(t,x_1)=\sum\limits_{n=0}^{\infty}\frac{1}{n!}
     \int\limits_{\mathbb{R}^n\times\mathbb{R}^n}dx_2\ldots dx_{n+1}\,\mathfrak{A}_{1+n}^0(t)
     \prod_{i=1}^{n+1}f_{1}^0(x_i),
\end{eqnarray}
and the generating operator $\mathfrak{A}_{1+n}^0(t)\equiv\mathfrak{A}_{1+n}^0(t,1,\ldots,n+1)$
is the $(n+1)th$-order cumulant of semigroups \eqref{S*} of point particles with inelastic
collisions. For $t\geq 0$ an infinitesimal generator of this semigroup of operators is determined
by the operator:
\begin{eqnarray*}
    &&\hskip-5mm (\mathcal{L}_{n}^{\ast,0}f_{n})(x_{1},\ldots,x_{n})=
       -\sum_{j=1}^{n}p_j\,\frac{\partial}{\partial q_j} f_{n}(x_{1},\ldots,x_{n})+\\
    &&\sum_{j_{1}<j_{2}=1}^{n}|p_{j_{2}}-p_{j_{1}}|\big(\frac{1}{(1-2\varepsilon)^{2}}
       f_{n}(x_{1},\ldots,x^{\diamond}_{j_{1}},\ldots,x^{\diamond}_{j_{2}},\ldots,x_{n})-
       f_{n}(x_{1},\ldots,x_{n})\big)\delta(q_{j_{1}}-q_{j_{2}}),
\end{eqnarray*}
where $x_j^\diamond\equiv(q_{j},p_{j}^\diamond)$ and the pre-collision momenta
$p^\diamond_{j_{1}},\,p^\diamond_{j_{2}}$ of inelastically colliding particles are determined by
the following expressions:
\begin{eqnarray*}
  &&p^{\diamond}_{j_{1}}=p_{j_{2}}+\frac{\varepsilon}{2\varepsilon -1}\,(p_{j_{1}}-p_{j_{2}}),\\
  &&p_{j_{2}}^{\diamond}=p_{j_{1}}-\frac{\varepsilon}{2\varepsilon-1}\,(p_{j_{1}}-p_{j_{2}}).\nonumber
\end{eqnarray*}

For $t\geq 0$ the limit one-particle distribution function represented by series (\ref{viterin})
is a weak solution of the Cauchy problem of the Boltzmann-type kinetic equation of point particles
with inelastic collisions \cite{GB14}
\begin{eqnarray}\label{Bolz1}
  &&\hskip-12mm\frac{\partial}{\partial t}f_1(t,q,p)=-p\,\frac{\partial}{\partial q}f_1(t,q,p)+\\
  &&\hskip-7mm\int\limits_{-\infty}^{+\infty}d p_1\,|p-p_1|
     \big(\frac{1}{(1-2\varepsilon)^2}\,f_1(t,q,p^\diamond)\,f_1(t,q,p^\diamond_1)
     -f_1(t,q,p)\,f_1(t,q,p_1)\big)+\sum_{n=1}^{\infty}\mathcal{I}^{(n)}_{0}.\nonumber
\end{eqnarray}
In kinetic equation (\ref{Bolz1}) the remainder $\sum_{n=1}^{\infty}\mathcal{I}^{(n)}_{0}$
of the collision integral is determined by the following expressions:
\begin{eqnarray*}
  &&\hskip-12mm\mathcal{I}^{(n)}_{0}\equiv\frac{1}{n!}\int\limits_0^\infty dP\,P\,
      \int\limits_{\mathbb{R}^n\times\mathbb{R}^n}dq_3dp_3\ldots \\
  && dq_{n+2}dp_{n+2}\mathfrak{V}_{1+n}(t)\big(\frac{1}{(1-2\varepsilon)^2}
      f_1(t,q,p_1^\diamond(p,P))f_1(t,q,p_{2}^\diamond(p,P))-\nonumber\\
  && f_1(t,q,p)f_1(t,q,p+P)\big)\prod_{i=3}^{n+2}f_{1}(t,q_i,p_i)+\nonumber\\
  && \int\limits_0^\infty dP\,P\,\int\limits_{\mathbb{R}^n\times\mathbb{R}^n}dq_3dp_3\ldots\nonumber\\
  &&  dq_{n+2}dp_{n+2}\mathfrak{V}_{1+n}(t)\big(\frac{1}{(1-2\varepsilon)^2}
      f_1(t,q,\tilde p_{1}^\diamond(p,P))f_1(t,q,\tilde p_{2}^\diamond(p,P))-\nonumber\\
  && f_1(t,q,p)f_1(t,q,p-P)\big)\prod_{i=3}^{n+2}F_{1}(t,q_i,p_i),\nonumber
\end{eqnarray*}
where the generating operators $\mathfrak{V}_{1+n}(t)\equiv\mathfrak{V}_{1+n}(t,\{1,2\},3,\ldots,n+2),\,n\geq0,$
of the series for a collision integral  are represented by expansions (\ref{skrrc}) with respect to the cumulants
of semigroups of scattering operators of point hard rods with inelastic collisions in a one-dimensional space
\begin{eqnarray*}\label{scat}
   &&\widehat{S}_{n}^{0}(t,1,\ldots,n)\doteq
       S_n^{\ast,0}(t,1,\ldots,s)\prod_{i=1}^{n}(S_1^{\ast,0})^{-1}(t,i).
\end{eqnarray*}

In fact, series expansions for the collision integral of the non-Markovian Enskog equation for
a granular gas (\ref{Bolz1}) or solution (\ref{ske1}) are represented as the power series over
the density, so that the terms $\mathcal{I}^{(n)}_{0},\,n\geq1,$ of the collision integral in
kinetic equation (\ref{Bolz1}) are corrections with respect to the density to the Boltzmann
collision integral of one-dimensional granular gases formulated in the paper \cite{T04}.

Since the scattering operator of point hard rods is an identity operator in the approximation
of elastic collisions, namely, in the limit $\varepsilon\rightarrow0$, the collision integral
of the Boltzmann kinetic equation (\ref{Bolz1}) in a one-dimensional space is identical to zero.
In the quasi-elastic approximation \cite{T04} the limit one-particle distribution function (\ref{viterin})
\begin{eqnarray*}
   &&\lim_{\varepsilon \rightarrow 0}\varepsilon f_1(t,q,p)=f^0(t,q,p),
\end{eqnarray*}
satisfies the nonlinear friction kinetic equation for one-dimensional granular gases \cite{T04}:
\begin{eqnarray*}
   &&\hskip-12mm\frac{\partial}{\partial t}f^0(t,q,p)=-p\,\frac{\partial}{\partial q} f^0(t,q,p)+
       \frac{\partial}{\partial p}\int\limits_{-\infty}^\infty dp_1\,
       |p_1-p|\,(p_1-p)\, f^0(t,q,p_1)f^0(t,q,p).
\end{eqnarray*}

Taking into consideration the result (\ref{asymtin}) on the Boltzmann--Grad asymptotic behavior of
the non-Markovian Enskog equation (\ref{GE1}), for reduced functionals of the state (\ref{f})
in a one-dimensional space, the following statement is true \cite{GB14}:
\begin{eqnarray}\label{lf}
  &&\hskip-12mm\mathrm{w-}\lim\limits_{\epsilon\rightarrow 0}
      \big(F_{s}\big(t,x_1,\ldots,x_s\mid F_{1}^{\epsilon}(t)\big)
      -f_{s}\big(t,x_1,\ldots,x_s\mid f_{1}(t)\big)\big)=0,\quad  s\geq2,
\end{eqnarray}
where in equality (\ref{lf}) the limit reduced functionals of the limit one-particle distribution
function (\ref{viterin}) are determined by the series expansions with a structure similar to series
(\ref{f}) and the generating operators represented by expansions (\ref{skrrc}) over the cumulants
of semigroups of scattering operators of point hard rods with inelastic collisions in a one-dimensional
space.

As mentioned above, in the case of a system of hard rods with elastic collisions, the limit reduced
functionals of the state are the product of the limit one-particle distribution functions, describing
the free motion of point particles.

Thus, the Boltzmann--Grad asymptotic behavior of solution (\ref{ske1}) of the non-Markovian
Enskog equation (\ref{GE1}) is governed by the Boltzmann kinetic equation (\ref{Bolz1}) for
a one-dimensional granular gas.

We emphasize that the Boltzmann-type equation (\ref{Bolz1}) describes the memory effects
in a one-dimensional granular gas. In addition, the limits
$f_{s}\big(t,x_1,\ldots,x_s\mid f_{1}(t)\big),\,s\geq2,$ of reduced functionals of the state
which are defined above, describe the process of the propagation of initial chaos in a
one-dimensional granular gas, or,in other words, the process of creating correlations in a
system of hard rods with inelastic collisions.

It should be noted that the Boltzmann--Grad asymptotic behavior of the non-Markovian Enskog
equation with inelastic collisions in a multidimensional space is analogous to the Boltzmann--Grad
asymptotic behavior of a hard sphere system with the elastic collisions \cite{GB14}, i.e., it is
governed by the Boltzmann equation for a granular gas \cite{V02},\cite{V06}, and the asymptotic
behavior of the reduced functionals of the state (\ref{f}) is described by the product of
one-particle distribution functions of its solution, i.e., describes the propagation of initial chaos.

\vskip+2.5mm
\subsection{Some bibliographic notes on collisional dynamics}
Above, it was studied systems of identical colliding particles, which are described by means
of functions of observables and distribution functions, which are symmetrical with respect
to arbitrary permutations of their arguments. In papers \cite{G82},\cite{G91p},\cite{GerRS},\cite{GS03},\cite{PG83},
the theory of the hierarchies of evolution equations for systems of many colliding particles
described by non-symmetric functions was developed. An example of such a system is a one-dimensional
system of particles interacting with their nearest neighbors, so-called non-symmetric systems of
particles \cite{G82}.

As is known, many-entity systems of active soft condensed matter are dynamic systems exhibiting
a collective behavior that differs from the statistical behavior of ordinary gases. To describe
the nature of entities (or self-propelled particles), in the paper \cite{ML}, collision dynamics
based on Markov jump processes, which should reflect the internal properties of living  creatures,
were proposed. In works \cite{Gsm},\cite{GF} an approach was developed to describe the collective
behavior of complex systems of mathematical biology within the framework of the evolution of
observables of many colliding stochastic processes, and the dual Vlasov hierarchy was constructed
in the mean field approximation. This representation of the kinetic evolution seems, in fact, to be
the direct mathematically fully consistent formulation modeling the kinetic evolution of biological
systems, since the notion of the state is more subtle and is an implicit characteristic of populations
of living creatures. In the paper \cite{Gsm} the processes of creation of correlations generated by
the dynamics of active soft matter and propagation of initial correlations have also been described
by means of the non-Markovian generalized kinetic equation with initial correlations, and, in particular,
in the mean-field scaling approximation, the Vlasov-type kinetic equation for many colliding stochastic
processes was constructed.

The study of systems of colliding particles in interaction with the environment, the so-called open
systems, involves a number of unsolved fundamental problems. One of them is related to the challenge
of the origin of stochastic behavior in dynamical systems of many particles.
In papers \cite{GGmb},\cite{GG14},\cite{GG15}, based on the approaches to the derivation of kinetic
equations outlined above, a generalization of the Fokker--Planck equation for open systems of
colliding particles was justified.

In previous decades, a lot of work has been performed on discrete-velocity models of the Boltzmann
equation, which are of significant conceptual interest for the kinetic theory of gases and, at the
same time, represent a fascinating mathematical subject \cite{PI}. In connection with this topic
of research, we note the works \cite{Git},\cite{GBLM},\cite{GBLM97},\cite{GG91}, in which the
discrete-velocity model was studied, related to the problem of deriving a model of the Enskog
discrete-velocity kinetic equation.

An overview of some modern applications of kinetic equations to the description of
non-equilibrium processes in complex systems of various natures is presented in the monograph \cite{DBK}.

\vskip+2.5mm
\subsection{Outlook}
The purpose of this review was to analyze the development and current advances of the
theory of evolution equations for systems of many colliding particles, in particular,
kinetic equations and their relations to the fundamental equations that describe the
laws of nature.

The problem of constructing a solution to the Cauchy problem for hierarchies of evolution
equations of observables \eqref{dh} and the state \eqref{h} of a system of hard spheres
with elastic collisions for initial data belonging to some functional spaces is considered.
As was established, solutions of hierarchies of evolution equations are determined by groups
of operators, which are represented by expansions over the groups of particles whose
evolution is described by cumulants of the corresponding order of the groups of operators
of the Liouville equations. Due to the fact that the cumulants of the groups of operators
are determined by cluster expansions of the groups of operators of the Liouville equations,
in the corresponding function spaces there are different representations for solutions
to the hierarchies of evolution equations. These cluster expansions of the groups of
operators underlie the classification of possible solution representations to the Cauchy
problem for the hierarchies of evolution equations of many colliding particles.

To describe the evolution of the state of a many-particle system, there is an alternative
approach that is based on the dynamics of correlations. In this approach, a state of finitely
many hard spheres is described with the employment of functions determined by the cluster
expansions of the probability distribution functions that are governed by the so-called
Liouville hierarchy \eqref{Lh}. It was above established that the constructed dynamics of
correlation underlie the description of the dynamics of infinitely many hard spheres governed
by the BBGKY hierarchy for reduced distribution functions \eqref{h} or the hierarchy of
nonlinear evolution equations for reduced correlation functions \eqref{gBigfromDFBa}, i.e.,
of the cumulants of reduced distribution functions. We emphasize the importance of the
mathematical description of the processes of the creation and propagation of correlations,
in particular, for numerous applications \cite{Bo2020},\cite{Bo2022},\cite{P62}.

To describe the evolution of many hard spheres within the framework of the evolution of states
for an initial state close to "kinetic," i.e., a state described in terms of the state of a
typical particle, there is another possibility: by means of the so-called non-Markovian Enskog
kinetic equation \eqref{gke1}. In other words, the origin of the collective behavior of a
hard-sphere system on a microscopic scale was examined above. As already mentioned, one of the
advantages of such an approach to the derivation of kinetic equations from underlying collisional
dynamics is the opportunity to construct the kinetic equations with initial correlations, which
makes it possible to describe the creation of correlations and propagation of initial correlations.
Another advantage of this approach is related to the rigorous derivation of the Boltzmann equation
\eqref{Bolz} with higher-order corrections to the canonical term of the collision integral.

Thus, the concept of cumulants of the groups of operators of Liouville equations underlies non-perturbative
expansions of solutions to hierarchies of fundamental evolution equations that describe the evolution
of observables and a state of many colliding particles, as well as underlies the kinetic description
of their collective behavior. We note that for quantum many-particle systems the concept of cumulants
of groups of operators is considered in review \cite{G20}.

In the paper, possible approaches to the rigorous derivation \cite{G13} and justification
\cite{GP97},\cite{GP98} of the kinetic equations  for many colliding particles were considered.
One of them is an approach to the description of the kinetic evolution within the framework of
the evolution of the observables of many colliding particles \cite{GG18}. The advances of the method
based on the dual Boltzmann hierarchy \eqref{vdh} are the opportunity to construct kinetic equations
\eqref{Bc}, taking into account the correlations of particles of the initial state, and the description
of the process of propagation of initial correlations in scaling approximations \eqref{cfmf}.

The paper \cite{GShZ} considered the challenge of deriving hydrodynamic equations from the dual BBGKY
hierarchy for reduced observed microscopic phase densities. We notice that the rigorous derivation of
hydrodynamic equations from the dynamics of many colliding particles is still an open problem.
Regarding the classical problem of rigorous derivation of the hydrodynamic equations from the Boltzmann
kinetic equation in scaling limits, we refer to the books \cite{GOL},\cite{S-R}.

\bigskip
\textcolor{blue!55!black}{\textbf{Acknowledgements.} Glory to Ukra\"{\i}ne!}


\bibliographystyle{plainurl}
\bibliography{aee}

\end{document}